%% file: paper.tex
\documentclass[floatfix]{revtex4}

\usepackage{epsfig}
\usepackage{graphicx}
\usepackage{dcolumn}
\usepackage{bm}
\usepackage{amsmath}
\usepackage{slashbox}
\usepackage{multirow}

\begin{document}

\def\be{\begin{equation}}
\def\ee{\end{equation}}
\def\bea{\begin{eqnarray}}
\def\eea{\end{eqnarray}}
\newcounter{Lcount}
\def\bl{\setcounter{Lcount}{0}
\begin{list}{\arabic{Lcount}.}{\usecounter{Lcount}\setlength{\leftmargin}{0.4cm}}}
\def\el{\end{list}}

\title{AKK Update: Improvements from New Theoretical Input and Experimental Data}
\author{S.\ Albino}
\affiliation{{II.} Institut f\"ur Theoretische Physik, Universit\"at Hamburg,\\
             Luruper Chaussee 149, 22761 Hamburg, Germany}
\author{B.\ A.\ Kniehl}
\affiliation{{II.} Institut f\"ur Theoretische Physik, Universit\"at Hamburg,\\
             Luruper Chaussee 149, 22761 Hamburg, Germany}
\author{G.\ Kramer}
\affiliation{{II.} Institut f\"ur Theoretische Physik, Universit\"at Hamburg,\\
             Luruper Chaussee 149, 22761 Hamburg, Germany}
\date{\today}
\begin{abstract}
We perform a number of improvements to 
the previous AKK extraction of fragmentation functions for $\pi^\pm$, $K^\pm$, $p/\bar{p}$, $K_S^0$ and 
$\Lambda/\overline{\Lambda}$ particles at next-to-leading order. 
Inclusive hadron production measurements from $pp(\bar{p})$ reactions at BRAHMS, CDF, PHENIX and STAR
are added to the data sample. We use the charge-sign asymmetry of the produced hadrons in 
$pp$ reactions to constrain the valence quark fragmentations.
Data from $e^+ e^-$ reactions in regions of smaller $x$ and lower $\sqrt{s}$ are added. 
Hadron mass effects are treated for all observables and, for each particle, the hadron mass 
used for the description of the $e^+ e^-$ reaction is fitted. 
The baryons' fitted masses are found to be only around
$1\%$ above their true masses, while the values of the mesons' fitted masses have the correct order of magnitude. 
Large $x$ resummation is applied in the coefficient functions of the
$e^+ e^-$ reactions, and also in the evolution of the fragmentation functions, which in most cases results in
a significant reduction of the minimized $\chi^2$.
To further exploit the data, all published normalization errors are incorporated via a correlation matrix.
\end{abstract}


\maketitle


\section{Introduction}
\label{Intro}

In perturbative QCD, fragmentation functions (FFs) $D_i^h (x,M_f^2)$, which can be interpreted
as the probability for a parton $i$ at the factorization scale $M_f$ to fragment to a hadron $h$ carrying away a fraction
$x$ of its momentum, are a necessary ingredient in the calculation of 
single hadron inclusive production in any reaction. Interest in FFs is widespread, to be found for example
in the study of the proposed hot quark gluon plasma (QGP) of the early universe currently
being sought in heavy ion collisions, in investigating the origin of proton spin,
and in tests of QCD such as theoretical calculations for recent measurements of 
inclusive production in $pp$ collisions at RHIC.

From the factorization theorem, the leading twist component of any single hadron inclusive production measurement
can be expressed as the convolution of FFs, being the universal soft parts containing the
final state, with the equivalent productions of real partons, which are perturbatively calculable,
up to possible parton distribution functions (PDFs) to account for any 
hadrons in the initial state. Thus, by using these data to constrain the FFs,
predictions for future measurements can be made from current data.
An exception to this possibility occurs when some new measurement
depends on any regions of the FFs' function space that has not yet been constrained by experiment.
For this reason, a failure to describe some data does not imply irrefutably
that there are relevant physics effects which have been neglected in calculations.
In particular, the apparent inconsistencies within this
framework of universality occurring between charge-sign unidentified hadron production
in $e^+ e^-$, for which the contribution from gluon fragmentation is much less
than from quark, and in $pp(\overline{p})$ reactions may
be attributed to the large experimental uncertainties on the gluon fragmentation 
determined from $e^+ e^-$ reaction data only, rather than to
neglected effects in the description of both reactions such as higher twist,
heavy quark masses, resummation at large and small $x$, 
and higher order terms in the perturbative approximation,
or to the less well understood hot QGP invoking parton energy loss.
A combined fit of FFs to data from $e^+ e^-$ and $pp(\overline{p})$ 
reactions would prove the optimum method of verifying consistency
between the two types of reactions, since a successful fit to both 
types of data would imply that these apparent inconsistencies
in fact lie within the experimental and theoretical uncertainties and so are not inconsistencies at all.
Success is expected for identified particles in general,
since good agreement is found \cite{Albino:2005me} for the theoretical calculation for $pp\rightarrow \pi +X$ data, 
where $\pi=\pi^0$ \cite{Adler:2003pb} and $\pi^\pm$ \cite{Adams:2006nd},
using FFs for $\pi^\pm$ constrained by data for $e^+ e^- \rightarrow \pi^\pm +X$ processes,
and data for the production of $\pi^\pm$ is generally more accurate and plentiful
than for the production of other particles due to the high abundance of $\pi^\pm$ in the particle sample.
In other words, the current theoretical state of the art is adequate in the kinematic regions studied.
The strongest caveat to this argument is the possible importance of hadron mass effects, which
are not so important for $\pi^\pm$, being the lightest hadrons, but which may be relevant for other particles.
Therefore, for the other particles it may be necessary to account for hadron mass effects in the theory. 
Furthermore, in this connection, it may also be necessary to account for contamination of the sample from 
decays of unstable particles. 

It is important to note in the discussion above that, due to insufficient information on the systematic effects, 
even a failure to fit certain data points in such a {\it global} fit
does not necessarily suggest other physics effects, unless the theory cannot describe
data from different experiments which are consistent with one another.

Since our previous fits \cite{Albino:2005me,Albino:2005mv}, a number of measurements have been published by collaborations
at RHIC and by the CDF collaboration at the Tevatron, which allow for a number of extensions in the knowledge of fragmentation: 
Because the gluon FF only appears at next-to-leading-order (NLO) in the 
$e^+ e^- \rightarrow h^\pm +X$ cross section but at leading order (LO) in the 
$pp(\overline{p}) \rightarrow h^\pm +X$ cross section, inclusion of these data in the purely $e^+ e^-$ sample
would significantly improve the constraints on gluon fragmentation, and may therefore give an FF set suitable 
for predictions of future measurements at e.g.\ the LHC and RHIC.
Furthermore, these {\it charge-sign unidentified} measurements provide much needed constraints
on the separation between the light quark flavour FFs
due to the differences between the light quark flavour PDFs.
Previously, the only data that could constrain these separations were 
the OPAL tagging probabilities \cite{Abbiendi:1999ry}, since these are the only data for which light quarks are
separately tagged. (Note that quark flavour untagged measurements from $e^+ e^-$ reactions cannot 
distinguish between quarks with similar electroweak couplings.)
These data are extracted from single and double inclusive
production measurements together with some reliable theoretical assumptions, the strongest being
SU(2) isospin symmetry between the $u$ and $d$ quark flavours and the standard model predictions
for the branching fractions of the $Z$-boson to each light quark flavour.
As required for any physical observable,
the theoretical definition of a quark flavour tagged measurement is trivially QCD scheme and scale independent
since it is obtained by setting to zero the electroweak couplings of 
all quark flavours except that of the tagged quark flavour.
Finally, valence quark FFs can be constrained by 
the difference between the production of a given species of charged 
hadron of one sign and the other, the {\it charge-sign asymmetry}, from $pp$ reactions at RHIC.
Since these data depend {\it only} on the valence quark FFs, in contrast 
to the charge-sign unidentified production in $e^+ e^-$ reactions which depend
only on the gluon, quark singlet and quark non-singlet FFs, fits to and predictions for the
charge-sign asymmetry and charge-sign unidentified production can be treated orthogonally. 
Since the total sample of the former data is much lower in quality than that of the latter,
the independence of the valence quark FFs from the other FFs
in this sense and also in the sense that it does not mix with the other FFs on evolution
should be reflected in the parameterization used for the FFs at the initial factorization scale.
Valence quark FFs will be useful for making predictions for other charge-sign separated 
data such as that at $ep$ colliders --- no such data at high negative photon virtuality exists 
at present, however such an extraction from HERA measurements is planned for the future. 
Such data in turn will be useful for improving the currently rather poor constraints on the valence quark FFs.
We note that there exists data from EMC for charged particle production \cite{Ashman:1991cj}.
However, the contamination by charged particles other than the ones we are interested in is unknown.
Furthermore, there exists data from HERMES for $\pi^+$, $\pi^-$, $K^+$ and $K^-$ production from $ep$ reactions
\cite{Hillenbrand:2005ke}, but for $Q\lesssim 2$ GeV where the validity of the fixed order (FO) approach comes into question.
However, since such processes are physically very similar to $\pi^\pm$ and $K^\pm$ production in $e^+ e^-$ reactions
at similar values for $\sqrt{s}$, inclusion of future measurements of the latter type at low $\sqrt{s}$
may help to average out these unknown theoretical systematic effects. 

In addition, theoretical developments in the calculation of inclusive production observables have occurred
since the analyses in Refs.\ \cite{Albino:2005me,Albino:2005mv}.
Small $x$ divergences can be resummed within the framework of DGLAP evolution \cite{Albino:2005gd}, 
which is crucial for improving the description at
small $x$. Since cross section measurements at small $x$ depend on the FFs at all
larger $x$ values, the inclusion of these measurements will also lead to improved constraints on the FFs at 
the $x$ values currently determined in global fits.
Unfortunately, while this procedure to any order is simple, explicit results 
for a full small $x$ resummed NLO calculation do not exist yet. 
However, effects of the produced hadron's mass, which must be treated first, can be incorporated. As we will see later,
this allows for $e^+ e^-$ data at smaller $x$ and lower centre-of-mass (c.m.) energy 
$\sqrt{s}$ to be added to the data sample to be fitted to,
and the results convincingly demonstrate both the accuracy of the fixed order approach and, particularly
for baryons, the fact that hadron mass effects are the most relevant small $x$, low $\sqrt{s}$ effects.

In this paper we repeat the AKK fits of FFs for $\pi^\pm$, $K^\pm$, $p/\overline{p}$ \cite{Albino:2005me},
$K_S^0$ and $\Lambda/\overline{\Lambda}$ \cite{Albino:2005mv}
to inclusive production measurements for these particles.
Those fits were intended to be conservative, in that we used only data for which the corresponding calculations
are reliable. In the fits of this paper we make a number of improvements.
Concerning the experimental information, we include all $e^+ e^-$ reaction data measured below the $Z$ pole. 
Of these, only TPC data were used in Ref.\ \cite{Albino:2005me} in order to constrain $\alpha_s(M_Z)$. 
The remaining data were excluded, due to unknown deviations from the
FO approach, and because the accuracy and number of data points was low. 
The former problem is handled by including hadron mass effects in our calculations and allowing the
hadron mass to be fitted.
Note that if $x$ and/or $\sqrt{s}$ were too low for the FO approach alone to be valid, 
this would also have the effect of subtracting out the
small $x$ and/or low $\sqrt{s}$ deviations, such as higher twist.
To meet the latter problem we exploit the data further than would be done if the 
normalization error were to be treated as a statistical error added in quadrature or as a normalization factor,
as is usually the case, by incorporating it instead as a systematic error in a covariance matrix.
Note that small $x$, low $\sqrt{s}$ data impose more constraints on 
the gluon FF than the larger $x$, higher $\sqrt{s}$ ones do.
We also resum large $x$ logarithms in the perturbative series for
the quark coefficient function for $e^+ e^-$ reactions, and in the DGLAP evolution of the 
FFs, since this affects all $x$ values in principle and, as we will see later, 
leads to an improvement in the fit over the unresummed case.
Furthermore, this modification is simple to apply to our Mellin space calculations.
The next important update to our fits is the inclusion of charge-sign unidentified hadron production data
from $pp(\overline{p})$ reactions at RHIC and the Tevatron to further constrain the gluon FF at large $x$. 
Similarly as for the $e^+ e^-$ reaction data,
we incorporate the systematic normalization errors of all RHIC data via a correlation matrix. 
The inclusion of these data also imposes further tests on universality, namely between
$pp$, $p\overline{p}$ and, because of the sizable gluon FF dependence, $e^+ e^-$ reaction data
at smaller $x$, lower $\sqrt{s}$.

In Ref.\ \cite{Albino:2005me}, only the charge-sign unidentified FFs were determined
since $e^+ e^-$ reactions are charge conjugate invariant, while the valence quark FFs were left completely unknown.
Therefore, it is not possible to obtain any information whatsoever on the
charge-sign asymmetry in fragmentation processes from the AKK FF sets, which are independent degrees of freedom.
Therefore, the most likely explanation
for the discrepancy between the charge-sign asymmetry in the BRAHMS data \cite{Arsene:2007jd}
and the corresponding calculation obtained
using charge-sign unidentified FFs together with certain assumptions relating them to the valence quark FFs is 
that there is a problem with these assumptions.
To resolve this issue, we perform a phenomenological extraction of the valence quark FFs
from the charge-sign asymmetry in the BRAHMS and STAR data, independently of the 
charge-sign unidentified fits.

For $pp$ reactions, a charge-sign asymmetry in the produced hadron sample
will be observed if the fragmentations from the initial protons' valence quarks 
were to dominate sufficiently over that from the proton's sea partons (gluons and sea quarks), since
the sea is charge conjugation invariant.
Furthermore, the contribution from the fragmentation to a particle of a given charge-sign from each of 
that particle's valence quarks must be positive.
We therefore study the relative contributions to the
charge-sign unidentified charged particle production from the 
fragmentation of the initial protons' valence quarks and sea partons,
and to the charge-sign asymmetry from the fragmentation of the produced hadron's valence quarks,
and look for deviations from our expectations which would signal a problem in the fit.

It is clear that any global fit is incomplete without a full error analysis, including 
correlation effects, on the fitted FFs. Since we wish to develop both the theory and 
develop and perform the technique
for doing this in detail, we postpone these analyses to a future publication,
while the present paper will be dedicated to the issues given above.

Since the last AKK analysis, two other analyses have been published in Refs.\
\cite{Hirai:2007cx} and \cite{deFlorian:2007aj}. 
The analyses of this paper differ in that we incorporate hadron mass effects
to improve the small $x$ and low $\sqrt{s}$ regions, resum large $x$ logarithms,
perform a complete treatment of the normalization errors on the experimental data
via a correlation matrix, and use $pp(\overline{p})$ reaction data including that for
$K_0^S$ and $\Lambda/\overline{\Lambda}$. 
However, we avoid the HERMES data from $ep$ reactions for $\pi^\pm$ and $K^\pm$ \cite{Hillenbrand:2005ke}
for the reasons given earlier.
(Data from $pp(\overline{p})$ reactions with transverse 
momentum values as low as $\simeq 2$ GeV are used, however the particle energy is 
somewhat higher because of the non-zero average rapidity, the measurements are for
c.m.\ energy $\geq 200$ GeV, and the cross section only depends on the FFs at large $x$.)
For these reasons our analyses provide a complement to the other analyses.

The paper is organized as follows. In Section \ref{expinput} we explain our choice of data sets
used to constrain the FFs. In Section \ref{theoryinput} we describe the theoretical input for our calculations,
and derive new tools for our analyses. 
Note that Sections \ref{expinput} and \ref{theoryinput} are intended to complement rather than repeat the 
discussions in Refs.\ \cite{Albino:2005me,Albino:2005mv}.
All results are presented in Section \ref{results},
and finally the work is summarized in Section \ref{conc}.

\section{Treatment of experimental input}
\label{expinput}

In this section we discuss the motivation for and application of the data used to constrain the FF sets.
For further details concerning the $e^+ e^-$ reaction data, we refer the reader to
Refs.\ \cite{Albino:2005me,Albino:2005mv}.

\subsection{Experimental data}

The data used for the extraction of FFs for $\pi^\pm$, $K^\pm$ and $p/\overline{p}$ 
particles in Ref.\ \cite{Albino:2005me} and of
FFs for $K_S^0$ and $\Lambda/\overline{\Lambda}$ particles in Ref.\ \cite{Albino:2005mv} 
were limited to measurements from $e^+ e^-$
reactions for which $x_p>0.1$, where $x_p$ is the fraction of available spatial momentum taken 
away by the produced hadron, given in terms of the hadron's momentum ${\mathbf p_h}$ in the c.m.\ frame
and the c.m.\ energy $\sqrt{s}$ by $x_p=2|{\mathbf p_h}|/\sqrt{s}$. 
In the analysis of this paper, we make a number of extensions to the experimental input. 
Firstly, to our sample of data from $e^+ e^-$ reactions we add 
all available data for which $\sqrt{s}<M_Z$. Such lower $\sqrt{s}$
data suffer from larger theoretical errors and were not considered in Ref.\ \cite{Albino:2005me}
(apart from data from the TPC Collaboration, which were necessary for a determination of $\alpha_s(M_Z)$).
For this reason, we account for the hadron mass effects \cite{Albino:2005gd}, which requires only a simple
modification to the calculation.
Furthermore, these older data have larger errors.
However, in this analysis we exploit the given statistical information concerning the 
normalization, i.e.\ the normalization error is treated as a systematic error by
incorporating it in a correlation matrix, instead of as a
random error by adding it in quadrature to the statistical error as is usually done. 
The general treatment of systematic errors and its justification
is outlined in Appendix \ref{correrr}.

We also include the $pp(\overline{p})$ reaction data from inclusive production measurements:
in $pp$ reactions at $\sqrt{s}=200$ GeV from the BRAHMS \cite{Arsene:2007jd}, 
PHENIX \cite{Adler:2003pb} and STAR \cite{Adams:2006nd,Adams:2006uz,Abelev:2006cs} collaborations at RHIC, and 
in $p\overline{p}$ reactions at $\sqrt{s}=630$ GeV from 
the CDF collaboration \cite{Acosta:2005pk} at the Tevatron (the data at $\sqrt{s}=1800$ GeV are excluded
due to their small effective $x=2p_T/\sqrt{s}$). 
We impose a lower bound $p_T > 2$ GeV to avoid large 
theoretical errors at small $p_T$. Any errors, systematic and statistical, 
are added in quadrature, except for the normalization errors which are treated, as for the $e^+ e^-$ reaction data
discussed above, as systematic errors by using a correlation matrix. 
For each of $K^0_S$ and $\Lambda/\overline{\Lambda}$, the CDF data are normalized by the unknown total cross section, 
which we therefore leave as a free parameter in the fit. 
Therefore, the CDF data only constrain the shape and relative normalizations of the FFs but not
the overall normalization.

We include $\pi^0$ production data from the STAR collaboration \cite{Adams:2006uz} in the
sample of data for constraining $\pi^\pm$ FFs 
(note that, as in the previous AKK fits, each of our charge-sign unidentified FFs
for charged particles and $\Lambda/\overline{\Lambda}$
is defined to be the FF of a parton of given species and
charge for the particle in question added to that for its antiparticle),
and $K^0_S$ production data from STAR \cite{Abelev:2006cs} and CDF \cite{Acosta:2005pk} 
and $K^\pm$ production data from BRAHMS \cite{Arsene:2007jd} in
the sample of data used to constrain both $K^\pm$ and $K^0_S$ FFs, by respectively imposing the relations
\be
D_i^{\pi^0}(x,M_f^2)=\frac{1}{2}D_i^{\pi^{\pm}}(x,M_f^2)
\label{ffspi0tochargedpi}
\ee
and
\be
D_i^{K_S^0}(x,M_f^2)=\frac{1}{2}D_j^{K^\pm}(x,M_f^2),
\label{ffforks0tokpm}
\ee
where $j=u,d$ if $i=d,u$, otherwise $i=j$, which follow from the highly reliable assumption of
SU(2) isospin symmetry between $u$ and $d$ quarks. 
Due to the nature of the DGLAP evolution,
these constraints are independent of $M_f^2$, as physical constraints should be.

For each light charged hadron species, we perform two mutually exclusive fits, one fit to the
charge-sign unidentified data (from $e^+ e^-$ 
and  $pp(\overline{p})$ reactions) to constrain the charge-sign unidentified gluon, 
$u$, $d$, $s$, $c$ and $b$ quark FFs,
and the other fit to the charge-sign asymmetries from $pp$ reactions at RHIC
to constrain the valence quark FFs. The only measurements distinguishing $\Lambda$ from $\overline{\Lambda}$ that exist
come from STAR \cite{Abelev:2006cs}, which are too inaccurate for a reasonable fit of valence
quark FFs for $\Lambda/\overline{\Lambda}$,
while there are no valence quarks in $K^0_S$ and its orthogonal state $K^0_L$ is not measured.

For the readers' convenience, we have listed all data sets used in our analyses and their properties 
in Tables \ref{PionResults}--\ref{K0SResults}, to the left of the vertical line.

\section{Theoretical formalisms used for the calculations}
\label{theoryinput}

In this Section we outline some known theoretical tools that we
use for our calculations, not already discussed in Ref.\ \cite{Albino:2005me},
and derive some improvements to these tools.

\subsection{Production in hadron-hadron reactions}
\label{hhreac}

We briefly outline the theoretical calculation that is used for the $pp(\overline{p})$ reactions.
A generic reaction $h_1 h_2 \rightarrow h+X$, where $h_{(i)}$ is a hadron, can be described by the quantity
\be
F^h_{h_1 h_2}(\chi,y,s)=s^2 E \frac{d^3 \sigma^h_{h_1 h_2}}{d{\mathbf p}^3}(p_T,y,s)
= s^2 \frac{1}{2\pi p_T} \frac{d^2 \sigma^h_{h_1 h_2}}{dp_T dy}(p_T,y,s)
\ee
(exploiting azimuthal symmetry in the second line),
where $\sqrt{s}$ is the c.m.\ energy, $E$ and ${\mathbf p}$ the energy and
spatial momentum respectively of the detected particle $h$,
$p_T$ its transverse momentum (relative to the spatial momenta of $h_1$ and $h_2$, which are antiparallel) 
and $y$ its rapidity, given by
\be
y=\frac{1}{2}\ln \frac{E+p_L}{E-p_L},
\ee
where $p_L$ is the longitudinal momentum (in the direction of the spatial momentum of $h_1$) of $h$.
The dimensionless variable $\chi$, which
will be convenient for later discussions, is given by 
\be
\chi=1-V+VW,
\label{chifromVW}
\ee
where $V$ and $W$ are the variables typically used in perturbative calculations,
related to the usual Mandelstam variables $S$, $T$ and $U$ of $h$ through
\be
\begin{split}
V=&1+\frac{T}{S},\\
W=&-\frac{U}{S+T}.
\label{VWfromMandelstam}
\end{split}
\ee
Working now in the c.m.\ frame and assuming 
that the mass of $h$ can be neglected so that $E=p_T \cosh y$, we find that
\be
\chi=(2p_T /\sqrt{s})\cosh y=2E/\sqrt{s}
\ee
is the fraction of available energy or momentum taken away by the hadron
and is therefore the scaling variable of the factorization theorem.

In the factorization theorem at leading twist, the following picture emerges which
allows the process to be partially calculated perturbatively: 
In any frame related to the c.m.\ frame via a boost 
(anti-)parallel to the beam direction (for massless hadrons the precise
choice of frame is irrelevant in the study below --- later when we treat the mass effects of the 
produced hadron we will need to specify a frame),
a parton in each initial state hadron $h_k$, $k=1,2$, moving parallel to it and carrying away a momentum fraction $x_k$,
interacts with the other. The interaction results in the inclusive production of a third parton, which subsequently 
fragments to a hadron moving in the same direction and carrying away a fraction $x$ of the parton's momentum. 
This fragmentation is expressed by writing
\be
F^h_{h_1 h_2}(\chi,y,s)
=\sum_i \int_{\chi}^1 dx \widehat{F}^i_{h_1 h_2}\left(\frac{\chi}{x},y,s,M_f^2\right)D_i^h\left(x,M_f^2\right).
\label{ptdistfromff}
\ee
The equivalent partonic production cross section has been denoted $\widehat{F}^i_{h_1 h_2}$. Note that the partonic rapidity
is the same as the hadronic rapidity, since for massless hadrons $y$ can be approximated by the pseudorapidity
\be
\eta=-\ln\left(\tan\frac{\theta}{2}\right),
\ee
where $\theta$ is the angle which both the produced hadron and the 
massless fragmenting parton make with the beam in the c.m.\ frame.

Due to the smallness of $\widehat{F}^i_{h_1 h_2}$ at sufficiently large $\chi/x$, the FFs at small
$x$ do not contribute significantly to $F^h_{h_1 h_2}$. To see that (large $x$)
gluon fragmentation contributes significantly to $F^h_{h_1 h_2}$, so that measurements of 
such observables are ideal for improving the constraints on gluon fragmentation, consider the
dependence of the $\widehat{F}^i_{h_1 h_2}$ on
the parton distribution functions (PDFs) $f^i_{h_j}$, where $i$ labels the parton species and $h_j=h_1,h_2$ 
labels the initial hadrons,
\be
\widehat{F}^i_{h_1 h_2}\left(x,y,s,M_f^2\right)
=\sum_{i_1 i_2}\int_x^1 dx_1 \int_{\frac{x}{x_1}}^1 dx_2 
\widetilde{F}^i_{i_1 i_2}\left(\frac{x}{x_1 x_2},y, \frac{x_2}{x_1},x_1 x_2 s,M_f^2\right)
f^{i_1}_{h_1}(x_1,M_f^2)f^{i_2}_{h_2}(x_2,M_f^2).
\label{frompdfs}
\ee
The $\widetilde{F}^i_{i_1 i_2}$, which are perturbatively calculable \cite{ACGG},
are the equivalent partonic cross sections when the hadrons $h$, $h_1$ and $h_2$ in $F^h_{h_1 h_2}$
are replaced by the partons $i$, $i_1$ and $i_2$
respectively, with c.m.\ energy squared $x_1 x_2s$.
For simplicity, the same factorization scale $M_f$ for all 3 partons has been used.
Since the gluon PDF $f^g_{h_j}$ dominates over the quark PDFs at small $x_i$
and since $\widetilde{F}^i_{gg}$ at LO is only non zero when $i=g$, 
we expect Eq.\ (\ref{frompdfs}) to be largest when $i=g$.

We use the Mellin transform approach as discussed in detail in Ref.\ \cite{Albino:2005me} for evaluating
the cross sections for the $e^+ e^-$ reactions, since this is fast and numerically accurate.
For these reasons we evaluate the $pp(\overline{p})$ reaction cross section in Mellin space.
This can be found by rewriting Eq.\ (\ref{ptdistfromff}) as a proper convolution
(using an obvious simplified notation),
\be
F(\chi)=\int_{\chi}^1 \frac{dx}{x} \widehat{F}\left(\frac{\chi}{x}\right)xD\left(x\right),
\label{ptdistfromff_brief}
\ee
so that the Mellin transform, defined by
\be
f(N)=\int_0^1 dx x^{N-1} f(x)
\label{defofmelltrans}
\ee
for any function $f(x)$, of Eq.\ (\ref{ptdistfromff_brief}) is the simple product of the Mellin transform
of $\widehat{F}(x)$ and $xD(x)$,
\be
F(N)=\widehat{F}(N)D(N+1).
\label{FfromFhatandDinMell}
\ee
Note from Eq.\ (\ref{ptdistfromff_brief}) that $\widehat{F}(x)$ is defined only in the range
\be
\chi < x < 1.
\label{rangeforhatF}
\ee
It will be chosen to vanish outside this range, which is reached for example by the integration region of 
the Mellin transform, Eq.\ (\ref{defofmelltrans}).
The Mellin transform is inverted according to
\be
f(x)=\frac{1}{2\pi i}\int_C dN x^{-N} f(N),
\ee
where $C$ is any contour which lies to the right of all poles in the complex $N$ plane and 
which extends to ${\rm Im} (N)=\pm \infty$, so that $F(\chi)$ can be obtained via
\be
F(\chi)=\frac{1}{2\pi i}\int_C dN \chi^{-N} \widehat{F}(N)D(N+1).
\label{FchifrominvMelltrans}
\ee
Since an analytical determination of the Mellin transform $\widehat{F}(N)$ of $\widehat{F}(x)$ itself cannot be determined
(one reason being that the evolved PDFs on which it depends may be extracted numerically from, 
e.g., a grid of values in $x$ and $M_f^2$),
it is necessary to obtain an approximate form for $\widehat{F}(x)$ whose analytic Mellin transform can be obtained.
The expansion of a function as a weighted sum of Chebyshev polynomials of the first kind \cite{numrec},
whose weights are determined by demanding that the expansion and the function agree exactly
at the zeros of the Chebyshev polynomials,
proves to be the most accurate representation of a continuous function as a polynomial of a given degree. 
In particular, it is close to the 
much harder to calculate minimax polynomial expansion, the polynomial with the smallest maximum deviation
from the function which is being approximated.
In detail, any continuous function $f(w)$ defined over the interval $-1 \leq w \leq 1$ can be approximated by 
\be
f(w)\approx \left[ \sum_{k=1}^M c_k T_{k-1}(w) \right]-\frac{c_1}{2},
\label{fexpaninchebpoly}
\ee
where $M-1$ is the degree of the polynomial, the coefficients $c_n$ are given by
\be
c_n=\frac{2}{M}\sum_{k=1}^M f\left(\cos\left(\frac{\pi \left(k-\frac{1}{2}\right)}{M}\right)\right)
\cos\left(\frac{\pi (j-1)\left(k-\frac{1}{2}\right)}{M}\right)
\label{coeffpolynomfromf}
\ee
and the Chebyshev polynomials of the first kind, $T_n(w)$, are defined by
\be
\begin{split}
T_0(w)=&1\\
T_1(w)=&w\\
T_{n+1}(w)=&2wT_n(w)-T_{n-1}(w).
\end{split}
\ee
That Eq.\ (\ref{fexpaninchebpoly}) is exact at the zeroes of the $M$th Chebyshev polynomial,
i.e.\ for $w$ values for which $T_M(w)=0$, may
be verified by directly substituting Eq.\ (\ref{coeffpolynomfromf}) 
into Eq.\ (\ref{fexpaninchebpoly}) and using the property
$T_n(\cos \theta)=\cos (n\theta)$. 

It is not possible to expand $\widehat{F}(x)$ as a Chebyshev polynomial directly,
due to the limiting discontinuous behaviour
\be
\lim_{x \rightarrow \chi}\widehat{F}(x)\propto \ln \left(1-\frac{\chi}{x}\right),
\ee
so we instead apply the expansion to $\widehat{F}(x)/\ln \left(1-\chi/x\right)$,
\be
\widehat{F}(x)=\ln \left(1-\frac{\chi}{x}\right)\theta \left(x-\chi\right)
\left(\left[\sum_{k=1}^M c_k T_{k-1}\left(\frac{2x-\left(1+\chi\right)}{1-\chi}\right)\right]-\frac{c_1}{2}\right)
\label{Fcheb}
\ee
where we have used the linear map from the Chebyshev polynomial argument $w$ to 
$x$ in the range of Eq.\ (\ref{rangeforhatF}),
\be
x=\frac{1}{2}\left(1-\chi\right)w+\frac{1}{2}\left(1+\chi\right).
\ee
In order to be able to apply the Mellin transform to the right hand side of Eq.\ (\ref{Fcheb}),
the weighted sum over Chebyshev polynomials must be
rewritten as a weighted sum over the $x^a$, where $a=0,...,M-1$. However, this expansion in $x$
is numerically very sensitive \cite{numrec5.10}: Although the magnitude of a Chebyshev polynomial is less than one, 
the coefficients of its expansion in powers of $w$ (and therefore $x$) are typically much greater than one, growing
in order of magnitude with $M$. Therefore, in our calculations, for which $M$ typically needs to be as large as 35 for
the data points with $\chi$ as small as $\simeq 0.02$, delicate cancellations among large numbers occur which 
require somewhat more accuracy than that of the 8-byte (``double'') precision provided by most computer systems. Various packages 
for increasing the precision of numbers exist; we found the DDFUN90 \cite{DDFUN90} package sufficient
to achieve both a sufficiently reasonable speed for fast fitting 
and an accuracy to a few parts per mil. The Mellin transform of each term 
$\ln \left(1-\chi/x \right) x^a$ on the right hand side of Eq.\ (\ref{Fcheb}) is calculated as
\be
\int_0^1 dx x^{N-1}
\ln \left(1-\frac{\chi}{x}\right) x^a \theta \left(x-\chi\right)
=\int_{\chi}^1 \frac{dx}{x} x^{N'} \ln \left(1-\frac{\chi}{x}\right)
=\chi^{N'} \sum_{n=1}^\infty \frac{1-\chi^{n-N'}}{n (N'-n)}
=\chi^{N'} \frac{S_1(-N')}{N'}- G(\chi,N')
\label{MellofeachterminexpanforFchi}
\ee
with the shorthand $N'=N+a$. The series
\be
G(x,N)=\sum_{n=1}^\infty \frac{x^n}{n(N-n)}
\ee
is quickly converging for $x<1$. $S_1(N)$ is given by
\be
S_1(N-1)=\sum_{k=1}^{N-1}\frac{1}{k}
=\ln (N) +\gamma_E-\frac{1}{2N}-\frac{1}{12N^2}+\frac{1}{120 N^4}
-\frac{1}{252 N^6}+\frac{1}{240 N^8}
-\frac{1}{132 N^{10}}+O\left(\frac{1}{N^{12}}\right),
\label{defsofSiN}
\ee
where the second equality is the analytic continuation to complex $N$ whose magnitude is large.
If $|N|$ is not large, this approximation is still applicable by repeated use of the result 
\be
S_1(-N-1)=S_1(-N)+\frac{1}{N},
\ee
which follows from the first equality in Eq.\ (\ref{defsofSiN}). This last result also
allows the $S_1(-(N+a))$ for all the values of $a$ to be evaluated quickly from $S_1(-N)$.

Poles created by the denominator in the third line of Eq.\ (\ref{MellofeachterminexpanforFchi})
at $N=n-a$ for $n=1,...,\infty$ would render the inverse Mellin
transform inapplicable, however inspection of the numerator reveals that 
these poles are canceled.

In the cases where the rapidity $y$ is integrated over, the above procedure can be applied
by replacing the value for $y$ used in $\chi$ by the minimum bound on $y$.

\subsection{Hadron mass effects}

In this subsection we show how to incorporate the effects of the mass $m_h$ of the produced hadron $h$ 
into inclusive hadron production in
$e^+ e^-$ and hadron-hadron reactions and into their relevant kinematic variables.

\subsubsection{$e^+ e^-$ reactions}

We repeat the study of Ref.\ \cite{Albino:2005gd}, including now some discussion on the produced hadrons' energy and on
cross sections averaged over a range of the scaling variable. The theoretical calculation, which follows from
the factorization theorem, takes the form
\be
\frac{d\sigma}{d\xi}(\xi,s)=\int_\xi^1 \frac{dz}{z}\frac{d\sigma}{dz}(z,s,M_f^2)D\left(\frac{\xi}{z},M_f^2\right),
\ee
where $z$ is the light cone momentum fraction and $\xi$ the light cone scaling variable.
Specifically, in the c.m.\ frame in which we will work from now on,
using light cone coordinates $V=(V^+=(V^0+V^3)/\sqrt{2},V^-=(V^0-V^3)/\sqrt{2},{\mathbf V}_T=(V^1,V^2))$, 
these variables are defined to be the ratio of the produced hadron's ``$+$'' component of light cone 
momentum to the partons' and to the intermediate vector boson's respectively.
The energy and momentum fractions measured in experiments, given
by $x_E=2E_h/\sqrt{s}$ and $x_p=2|{\mathbf p_h}|/\sqrt{s}$ respectively,
are determined from the momentum of the intermediate vector boson,
\be
q=\left(\frac{\sqrt{s}}{\sqrt{2}},\frac{\sqrt{s}}{\sqrt{2}},{\mathbf 0}\right),
\ee
of the produced parton,
\be
k=\left(\frac{p_h^+}{z},0,{\mathbf 0}\right),
\ee
and of the produced hadron,
\be
p_h=\left(\frac{\xi \sqrt{s}}{\sqrt{2}},\frac{m_h^2}{\sqrt{2}\xi\sqrt{s}},{\mathbf 0}\right).
\ee
Note that the hadron mass effects are accounted for simply by introducing a non-zero ``$-$'' component
into the hadron's momentum.
From these results we get immediately our desired relations,
\be
\begin{split}
x_p=&\xi \left(1-\frac{m_h^2}{s \xi^2}\right),\\
x_E=&\xi \left(1+\frac{m_h^2}{s \xi^2}\right).
\end{split}
\ee
A cross section which is measured at a fixed value for some scaling variable $\overline{x}$
and which is differential in some other scaling variable $x$ is calculated from the 
theoretical result $d\sigma/d\xi (\xi,s)$ using
\be
\frac{d\sigma}{dx}(\overline{x},s)=\frac{d\xi}{dx}(x,s)\frac{d\sigma}{d\xi}(\xi(\overline{x},s),s).
\ee
Averaged cross sections are calculated according to
\be
\bigg{\langle} \frac{d\sigma}{dx} \bigg{\rangle}_{\overline{x}^0 < \overline{x} < \overline{x}^1}
=\frac{1}{x(\overline{x}^1,s)-x(\overline{x}^0,s)}\int_{x(\overline{x}^0,s)}^{x(\overline{x}^1,s)} dx \frac{d\sigma}{dx} 
=\frac{1}{x(\overline{x}^1,s)-x(\overline{x}^0,s)}\int_{\xi(\overline{x}^0,s)}^{\xi(\overline{x}^1,s)} 
d\xi \frac{d\sigma}{d\xi} (\xi,s).
\ee

\subsubsection{Hadron-hadron reactions}

To incorporate hadron mass effects in $pp(\overline{p})$ reactions, we start with the general result from the factorization theorem,
\be
d\sigma^h_{h_1 h_2}(p,\sqrt{s})=\sum_{ii_1 i_2}\int dx_1 \int dx_2 
f_{i_1}^{h_1}(x_1,M_f^2) f_{i_2}^{h_2}(x_2,M_f^2) \int dx D_i^h(x,M_f^2)d \sigma^i_{i_1 i_2}(l,\sqrt{s}),
\label{starteqforptspec}
\ee
where $d\sigma^i_{i_1 i_2}$ is the equivalent partonic differential cross section for the 
production of the massless parton $i$ (which will fragment to the produced hadron $h$),
with only the two partons $i_1$ and $i_2$ in the initial state.
We now work in the partonic c.m.\ frame where ${\mathbf l}$ is parallel to ${\mathbf p}$,
since this leads to various simplifications: Firstly,
\be
x=\frac{p^0 +|{\mathbf p}|}{l^0+|{\mathbf l}|},
\ee
being the ratio of the produced hadron's $+$ component of light cone momentum to the parton's
when the 3-axis is aligned with their spatial momenta.
Eliminating $l^0=|{\mathbf l}|$ and $p^0=\sqrt{|{\mathbf p}|^2+m_h^2}$ gives
\be 
2x|{\mathbf l}|=|{\mathbf p}|+\sqrt{|{\mathbf p}|^2 +m_h^2}.
\label{kplustopplus2}
\ee
Secondly, the relation between the experimentally measured observable $E d^3 \sigma^h_{h_1 h_2}/d{\mathbf p}^3$
from the equivalent partonic ``observable'' $|{\mathbf l}| d^3\sigma^i_{jk}/d {\mathbf l}^3$
via Eq.\ (\ref{starteqforptspec}) can be obtained from the simultaneous results
\be
\begin{split}
\frac{d {\mathbf p}^3}{E}=&\frac{|{\mathbf p}|^2}{\sqrt{|{\mathbf p}|^2+m_h^2}}d|{\mathbf p}| d\Omega,\\
\frac{d {\mathbf l}^3}{|{\mathbf l}|}=&|{\mathbf l}|d|{\mathbf l}| d\Omega,
\end{split}
\ee
where $\Omega$ is the solid angle. Therefore, in the presence of hadron mass,
\be
E\frac{d^3\sigma^h_{h_1 h_2}}{d{\mathbf p}^3}=\sum_{ijk} \int dx_1 \int dx_2 
f_j^{h_1}(x_1,M_f^2) f_k^{h_2}(x_2,M_f^2) 
\int dx D_i^h(x,M_f^2) |{\mathbf l}| \frac{d^3\sigma^i_{jk}}{d{\mathbf l}^3}
\frac{1}{x^2 R^2},
\ee
where the divisor
\be
R=1-\frac{m_h^2}{(|{\mathbf p}|+\sqrt{|{\mathbf p}|^2 +m_h^2})^2}
\ee
in the integrand is the only modification required to incorporate hadron mass effects.
However, we must still obtain the relation between $|{\mathbf p}|$ of the partonic c.m.\ frame
and the quantities $p_T$ and $y$ of the c.m.\ frame used to define the kinematics of the produced hadron in experiments.
Since $p_T$ has the same value in both frames,
the dependence of $|{\mathbf p}|$ on $p_T$ is easily found to be
\be
|{\mathbf p}|=m_T \cosh y',
\label{pfrompt}
\ee
where
\be
m_T=\sqrt{p_T^2+m_h^2}
\ee
and $y'$ is the rapidity in the partonic c.m.\ frame. Then
\be
R=\left(1-\frac{m_h^2}{\left(m_T \cosh y'
+\sqrt{m_T^2 \cosh^2 y'-m_h^2}\right)^2}\right)^2.
\ee
The relation between $y'$ and $y$ is 
\be
y'=y+\phi,
\ee
where $\phi$ is the boost rapidity between the partonic c.m.\ and lab frames. 
To evaluate $\phi$, note that, in the partonic c.m.\ frame,
$x_{1,2}$ are each replaced by a common momentum fraction $x_1 e^\phi =x_2 e^{-\phi}$, which implies
\be
\phi=\ln \sqrt{\frac{x_2}{x_1}}.
\ee

Finally, the variables $V$ and $W$ defined in Eq.\ (\ref{VWfromMandelstam})
which are used to define the kinematics of the produced parton in the theoretical
calculation \cite{ACGG} are obtained from $p_T$ and $y$ by
\be
\begin{split}
V=&1+\frac{m_h^2}{s}-\frac{m_T}{\sqrt{s}}e^{-y},\\
W=&\frac{\frac{m_T}{\sqrt{s}}e^y -\frac{m_h^2}{s}}{1+\frac{m_h^2}{s}-\frac{m_T}{\sqrt{s}}e^{-y}}.
\end{split}
\ee
For completeness, we note from Eq.\ (\ref{chifromVW}) that the above results imply that
\be
\chi=\frac{2m_T}{\sqrt{s}}\cosh y-\frac{2m_h^2}{s}.
\ee

The relation between $y$ and the pseudorapidity $\eta$ is then
\be
y=\sinh^{-1} \left(\frac{p_T}{m_T}\sinh \eta\right).
\ee
For experimental data where a range of $\eta$ values is specified, we use the operator
\be
\frac{1}{\eta_2-\eta_1}\int_{\eta_1}^{\eta_2} d\eta
=\frac{1}{\eta_2-\eta_1} \int_{y_1}^{y_2} dy \frac{m_T \cosh y}{\sqrt{m_T^2 \cosh^2 y-m_h^2}}.
\ee

\subsection{Large $x$ resummation}

When the transferred momentum (or energy) fraction $x$, being equal to $x_p$ for $e^+ e^-$ reactions and
$\chi$ for $pp(\overline{p})$, is large, the accuracy of the FO perturbative calculation for the hard part of 
a cross section and the DGLAP evolution in the soft parts is worsened by unresummed
divergences occurring in the formal limit $x\rightarrow 1$. The accuracy and abundance of data 
may be sufficiently limited at large $x$ that the resummation of these divergences is unnecessary.
However, since the cross section depends on the hard part in the range $x<z<1$, where $z$ is the 
partonic momentum fraction, the true ``large $x$'' region, being the region where resummation could make a difference,
may include significantly lower values of $x$ than just those for which $x\simeq 1$.
In addition, resummation should reduce theoretical errors. These effects of resummation
should become more significant at lower $\sqrt{s}$. Since we use $e^+ e^-$ 
reaction data for $\sqrt{s}$ as low as $\simeq 10$ GeV, and since large $x$
resummation results are easily implemented in FO calculations in Mellin space, 
we implement this resummation in our fits.

In general, in the series expansion of the hard part ${\mathcal W}(a_s,x)$ of some cross section, these
large $x$ divergences take the form $a_s^n [\ln^{n-r} (1-x)/(1-x)]_+$, where $r=0,...,n$ labels the class of divergence.
In Mellin space these divergences take the form $a_s^n \ln^{n+1-r} N$. 
These divergences may be factored out, which results in
the calculation of ${\mathcal W}$ taking the form
\be
{\mathcal W}(a_s,N)={\mathcal W}_{\rm res}(a_s,N)\left(\sum_n a_s^n {\mathcal W}_{\rm FO}^{(n)}(N)\right),
\label{genresproc}
\ee
where the FO series in parenthesis on the right hand side is free of
these divergences since they are all contained in ${\mathcal W}_{\rm res}$. ${\mathcal W}$ at large $N$ is approximated
by ${\mathcal W}_{\rm res}$ when the divergences in ${\mathcal W}_{\rm res}$ are resummed, which involves
writing ${\mathcal W}_{\rm res}$ as an exponential and expanding the exponent in $a_s$ keeping $a_s \ln N$ fixed.
We will resum the divergences in the quark coefficient function $C_q$ for inclusive quark production
in $e^+ e^-$ reactions in the manner of Eq.\ (\ref{genresproc}), 
once we have obtained to all orders its leading (class $r=0$) and next-to-leading (class $r=1$) divergences
from the results of Ref.\ \cite{Cacciari:2001cw}. The result, Eq.\ (\ref{howtoresumCq}) below, 
is used in our calculations instead of the FO expression for $C_q$.

As we will see, resummation of large $x$ divergences in the $e^+ e^-$ reactions typically improves fits
of fragmentation to charge-sign unidentified hadrons. 
Formal resummed results also exist for the hard parts of $pp(\overline{p})$ reactions
integrated over all rapidity values \cite{deFlorian:2005yj}. 
However, while the generalization to a given rapidity range can
be determined approximately, no formal results exist at the time of writing.
Therefore, it is not possible at present to perform 
large $x$ resummation in the hard parts of $pp(\overline{p})$ reactions.

The divergences of the $e^+ e^-$ reactions (in the $\overline{\rm MS}$ scheme in which we are working)
can be reorganized according to the formula \cite{Cacciari:2001cw}
\be
\ln C_q(N,a_s(s))=\int_0^1 dz \frac{z^{N-1}-1}{1-z}\Bigg[\int_{s}^{(1-z)s} \frac{dq^2}{q^2} A(a_s(s))
+B(a_s((1-z)s))\Bigg] +O(1),
\label{genresumforCq}
\ee
where
\be
(A,B)(a_s)=\sum_{n=1}^\infty (A,B)^{(n)} a_s^n.
\ee
The perturbative series for the quark coefficient function to NLO contains a leading (class $r=0$) 
and next-to-leading (class $r=1$) divergence,
which will be replaced just now by the sum of all divergences belonging to these two classes from every order.
All of these divergences can be explicitly extracted from Eq.\ (\ref{genresumforCq}) by taking
\be
\begin{split}
A^{(1)}=&2 C_F\\
A^{(2)}=&-C_F\left(C_A \left(\frac{\pi^2}{3}-\frac{67}{9}\right)+\frac{20}{9}T_R n_f \right)\\
B^{(1)}=&-\frac{3}{2}C_F.
\end{split}
\ee
Since Eq.\ (\ref{genresumforCq}) is algebraically similar to the resummed quark coefficient function
of deeply inelastic scattering \cite{Catani:1989ne}, we may obtain the divergences of classes $r=0,1$ 
directly from the $\overline{\rm MS}$ result in Ref.\ \cite{Albino:2000cp},
\be
\begin{split}
\ln C_q^{r=0,1}(N,a_s)=&\frac{A^{(1)}}{a_s \beta_0^2}\left[(1-\lambda_s)\ln (1-\lambda_s)+\lambda_s\right]
+\left(\frac{B^{(1)}}{\beta_0}-\frac{A^{(1)}\gamma_E}{\beta_0}+\frac{A^{(1)} \beta_1}
{\beta_0^3}-\frac{A^{(2)}}{\beta_0^2}\right) \ln (1-\lambda_s)\\
&+\frac{A^{(1)}\beta_1}{2\beta_0^3}\ln^2 (1-\lambda_s)
-\left(\frac{A^{(2)}}{\beta_0^2}-\frac{A^{(1)}\beta_1}{\beta_0^3}\right)\lambda_s,
\label{lnCqforr1andr0}
\end{split}
\ee
where $\lambda_s=a_s \beta_0 \ln N$. The constant $\beta_0$ is that which appears in the expansion 
of the evolution of $a_s$,
\be
\frac{da_s(\mu^2)}{d\ln \mu^2}=\beta(a_s(\mu^2))=-\sum_{n=0}^\infty \beta_n a_s^{n+2}.
\ee
According to the general form of Eq.\ (\ref{genresproc}),
the resulting resummed quark coefficient function that we seek is finally
\be
C_q=C_q^{r=0,1}\left(1+a_s(C_q^{(1)}-C_q^{r=0,1\ (1)})\right).
\label{howtoresumCq}
\ee
The presence in the above equation of $C_q^{r=0,1\ (1)}$, the coefficient of the $O(a_s)$ 
term in the expansion of $C_q^{r=0,1}$ in $a_s$, given by
\be
C_q^{r=0,1\ (1)}=\frac{A^{(1)}}{2}\ln^2 N+\left(A^{(1)}\gamma_E-B^{(1)}\right)\ln N,
\ee
ensures that the original NLO result is obtained when the
whole of the right hand side of Eq.\ (\ref{howtoresumCq}) is expanded in $a_s$, i.e.\ when the resummation
is ``undone'', and therefore also prevents double counting of the divergences.
Note that there are an infinite number of other schemes which are consistent with this criteria and give the
large $N$ behaviour of Eq.\ (\ref{lnCqforr1andr0}), a typical feature of perturbation theory.

Equation (\ref{lnCqforr1andr0}) contains a Landau pole when $\lambda_s=1$, for which $N$ is real and $\gg 1$. 
However, in the inverse Mellin transform it is not necessary for the contour 
in the complex $N$ plane to run to the right of this pole, as it should for the other poles, 
because it is unphysical, created by the ambiguity of the asymptotic series, which in $x$ space is 
essentially a higher twist contribution and therefore negligible. We will keep the 
contour to the left of this pole, which is known as the {\it minimal prescription} \cite{Catani:1996yz}, since this is the
most efficient choice for our numerical evaluation of the inverse Mellin transform.

The DGLAP evolution of the FFs also contains large $x$ logarithms, which must also be resummed.
We do this according to the approach of Ref.\ \cite{Albino:2007ns}, which
uses the fact that the FO splitting functions are already resummed \cite{Albino:2000cp,Korchemsky:1988si} to
perform the resummation in the Mellin space evolution analytically.

\subsection{Evolution with heavy quarks}

NLO matching conditions at the heavy quark flavour thresholds for FF evolution have been derived \cite{Cacciari:2005ry}, 
and occur for the gluon FF (if the thresholds differ from the heavy quark pole masses $m_q$) and the 
extrinsic heavy quark FF (for all choices of threshold). ``Extrinsic'' here 
means that the fragmentation from the heavy quark proceeds via perturbative fragmentation of the heavy quark
to a gluon, followed by non perturbative fragmentation of the gluon to the detected hadron. 
Following the results of spacelike factorization, the intrinsic (non perturbative) component of each heavy quark FF, which
must be added to the extrinsic one, may be negligible since it is expected
to be of $O((\Lambda_{\rm QCD}/m_q)^p)$, where $p$ is a positive integer.
However, the intrinsic component may be important at large $x$ \cite{Pumplin:2007wg}, and 
therefore we will not neglect intrinsic heavy quark fragmentation here. Since this intrinsic component
is unknown, we simply parameterize the whole heavy quark FF at threshold and neglect
its contribution to the calculation below threshold, as in the previous AKK studies.
A study of the relative contributions of the extrinsic and intrinsic components of 
heavy quark fragmentation would be interesting, but is not relevant to the goals
of this paper, and will therefore be left to a future publication.
To ensure a continuous gluon FF at each threshold for simplicity, 
we set the thresholds to the heavy quark pole masses $m_c=1.65$ GeV and 
$m_b=4.85$ GeV \footnote{These are the centres of the current best ranges,
$m_c=$1.47 -- 1.83 GeV and $m_b=$4.7 -- 5 GeV \cite{Yao:2006px}.},
as opposed to twice these values as practiced in previous fits.

\subsection{Other theoretical choices}

We use NLO results for the calculations of all data. To account for the initial (anti)protons, 
we use the CTEQ6.5S0 PDFs \cite{Lai:2007dq}. Consequently, for consistency, $\Lambda_{\rm QCD}$ for 5 active
quark flavours is fixed in our fits to the value at which these PDFs are determined, $226$ MeV.

Other than the improvements that have been discussed in this section,
the theoretical calculations and choices used here for the $e^+ e^-$ reaction data
are identical to those used in Refs.\ \cite{Albino:2005me,Albino:2005mv}, and are detailed therein
\footnote{We point out that Eq.\ (7) of Ref.\ \cite{Albino:2005me} is only for the transverse component of the gluon
coefficient function. The addition of the longitudinal component has the effect of removing the
last term, $-2(1-z)/z$. The full gluon coefficient function was used in the actual calculations.}.
There we considered only charge-sign unidentified quantities (cross sections and FFs) for 
which a particle is not distinguished from its antiparticle, which are given the
label $H=h^\pm$ or $h/\bar{h}$ and are given by
$O^{h^\pm}=O^{h^+}+O^{h^-}$ (or $O^{h/\bar{h}}=O^{h}+O^{\bar{h}}$).
These particles were 
$H=\pi^\pm$, $K^\pm$, $p/\bar{p}$ \cite{Albino:2005me}, $K^0_S$ and $\Lambda/\overline{\Lambda}$ \cite{Albino:2005mv},
requiring 5 independent fits.
Since particles are distinguished from their antiparticles in the BRAHMS and STAR measurements
for $\pi^\pm$, $p/\bar{p}$ (BRAHMS and STAR) and $K^\pm$ (BRAHMS), we may now also perform fits
for the charge-sign asymmetries $H=\Delta_c h^\pm$, for which quantities take the form
\be
O^{\Delta_c h^\pm}=O^{h^+}-O^{h^-}.
\ee
Therefore, for this paper there will be a total of 8 separate fits. Note that these combinations are not the only possibility.
For $\pi^\pm$, $K^\pm$ and $p/\bar{p}$ we could instead consider quantities for which
$H=h^+$ and $H=h^-$. However, in that case
the corresponding quantities $O^{h^+ (h^-)}=O^{h^\pm}/2+(-)O^{\Delta_c h^\pm}/2$ 
are statistically contaminated by the highly uncertain quantities $O^{\Delta_c h^\pm}$,
so that the more reliable information provided by the quantities $O^{h^\pm}$ is lost.
Secondly, the theoretical calculations for $O^{h^+}$ and $O^{h^-}$ receive contributions
from the same FFs through the mixing occurring in the evolution 
(see Eq.\ (\ref{valrelbetqandqbar}) below), so that all data must be fitted simultaneously.
The combinations $H=h^\pm$ and $H=\Delta_c h^\pm$ that we use avoid all this entanglement.
For comparison, we also perform fits without the large $x$ resummation, which 
therefore leads to a total of 16 independent fits.
We allow a total of 11 parton species to produce the observed hadron through fragmentation, being
the gluon, the quarks $i=u$, $d$, $s$, $c$ and $b$ and their antiquarks.
The FFs are parameterized at a starting scale $M_0=\sqrt{2}$ GeV in the form
\be
\begin{split}
D^{h^\pm}_i(x,M_0^2)=&N_i^{h^\pm} x^{a_i^{h^\pm}} (1-x)^{b_i^{h^\pm}} (1+c_i^{h^\pm} (1-x)^{d_i^{h^\pm}}),\\
D^{\Delta_c h^\pm}_i(x,M_0^2)=&N_i^{\Delta_c h^\pm} x^{a_i^{\Delta_c h^\pm}} (1-x)^{b_i^{\Delta_c h^\pm}}.
\end{split}
\ee
This is the strongest non perturbative assumption that we will make.
(Eqs.\ (\ref{ffspi0tochargedpi}) and (\ref{ffforks0tokpm}) and Eq.\ (\ref{SU2forpicharged}) below 
have more physical justification, namely SU(2) isospin invariance between $u$ and $d$ quarks, than the
above parameterization.) The term proportional to $c_i^H$ was fixed to zero in the previous AKK
analysis, and is found to significantly improve the fit to the data when freed. The constraints
\be
D^{h^\pm}_{\bar{q}}(x,M_0^2)=D^{h^\pm}_q (x,M_0^2)
\ee
and
\be
D^{\Delta_c h^\pm}_{\bar{q}}(x,M_0^2)=-D^{\Delta_c h^\pm}_q (x,M_0^2)
\label{valrelbetqandqbar}
\ee
are exact in QCD and are applied in the fits. For the charge-sign unidentified combinations and the neutral particles,
the first constraint implies 6 independent FFs, being those for
the gluon and the $u$, $d$, $s$, $c$ and $b$ quarks.
The second constraint implies that charge-sign asymmetry FFs vanish when $q$ is a unfavoured quark (i.e.\ a sea quark of the
produced hadron), since
the equality also holds without the minus sign, leaving only the FFs for the favoured quarks 
(i.e.\ the valence quarks of the produced hadron) for charge-sign asymmetry data: 
for $i=u$ and $d$ when $H=\Delta_c \pi^\pm$ and $H=\Delta_c p/\bar{p}$, and
for $i=u$ and $s$ when $H=\Delta_c K^\pm$. 
For $\pi^\pm$ we impose the SU(2) $u$ and $d$ isospin symmetry relations
\be
D^{\pi^\pm/\Delta_c \pi^\pm}_u=D^{\pi^\pm/\Delta_c \pi^\pm}_{\bar{d}}.
\label{SU2forpicharged}
\ee
In principle, this relatively precise assumption could also be applied to use 
$K^\pm$ FFs to calculate $K_S^0$ production, so that
a fit to $K^\pm$ and $K_S^0$ data simultaneously could be performed to obtain a single set of FFs for $K^\pm/K_S^0$ particles.
However, the differences in the details of these particles' production mechanisms, including
large time scale effects, may be too significant.

The hadron mass appearing in the calculation of the $pp(\overline{p})$ reaction data
must be assigned before fitting, since the precalculation of subsection \ref{hhreac} depends on it.
Each particle's mass will therefore be fixed to its known value.
On the other hand, since the hadron mass appearing in the calculation 
of the $e^+ e^-$ reaction data is not limited in this way, it will
be freed in the fit. 

\subsection{Contributions of the partonic fragmentations to $pp\rightarrow (\Delta_c) h^\pm +X$}

When we come to present our results, we will also verify whether the relative contributions of the 
fragmentation from valence quarks and sea partons to the $pp$ cross sections is consistent with our expectations,
to be given below. This will also help in determining the sources of the charge-sign asymmetry.
Writing out the indices labeling parton and hadron production in Eq.\ (\ref{FfromFhatandDinMell}) gives
\be
F^H=\sum_{i=-n_f}^{n_f} \widehat{F}^i D_i^H.
\label{FfromFhatandDwithindices}
\ee
This can also be taken as the $x$ space result, by defining the product in this equation to be the $x$ space
convolution. The cross section for $H=h^\pm$ can be written
\be
F^{h^\pm}=\widehat{F}^{u_v} D_u^{h^\pm} +\widehat{F}^{d_v} D_d^{h^\pm} 
+\sum_{i=g,q_s}\widehat{F}^i D_i^{h^\pm},
\label{FfromFhatandDwithindicesdecomp}
\ee
where $\widehat{F}^{q_v}=\widehat{F}^q-\widehat{F}^{\bar{q}}$ and $q_s$ refers to sea quarks.
(In this subsection, ``quark'' means a quark or antiquark of the same flavour.)
The first two terms give respectively the contribution to the
production of $h^\pm$ from the protons' valence $u$ and $d$ quarks. 
These quarks are the source of the charge-sign asymmetry, to be discussed in more detail
around Eq.\ (\ref{FfromFhatandDwithindicesforval}) below.
Since there are more valence $u$ than $d$ quarks
in the initial protons, the first term is expected to dominate over the second for $\pi^\pm$ production,
since the $u$ and $d$ quark fragmentations are equal, and even more so for
$p/\bar{p}$ production, because then $u$ is larger than $d$ quark fragmentation.
For $K^\pm$, $d$ quark fragmentation is unfavoured, and therefore 
the contribution from the protons' valence $d$ quark is expected to be much smaller than from their valence $u$ quark. 

The third term or remainder in Eq.\ (\ref{FfromFhatandDwithindicesdecomp}) can be regarded as the contribution
from the collective sea of the initial protons. This term does not
contribute to the charge-sign asymmetry because it is
charge conjugation invariant. Therefore, the larger the third term is relative to the first two,
the smaller the charge-sign asymmetry relative to the charge-sign unidentified cross section.
For hadrons which have non-zero strangeness, or which are superpositions of hadrons with non-zero strangeness,
which in our set are $K^\pm$ and $\Lambda/\overline{\Lambda}$ in the first case and $K^0_S$ in the second,
the third term is expected to dominate: Here, the favoured 
$s$ quark and $u$ and/or $d$ quark fragmentation from the abundant sea occurs, while in the first two terms only
the fragmentation from the protons' valence $u$ and/or $d$ quark contributes which necessarily 
involves the production of a heavier $s$ quark. 
For $\pi^\pm$ and $p/\bar{p}$ production, for which $u$ and $d$ quark fragmentations are favoured,
it is not clear whether the first two terms are more important due to the FFs there being an order
of magnitude larger than the rest, or the third which accounts for fragmentation from the 
protons' abundant partonic sea. In fact,
while the decomposition in Eq.\ (\ref{FfromFhatandDwithindicesdecomp}) provides a simple and direct
method to determine the underlying partonic physics processes involved, it is ``non-physical'', or
the physical meaning is arbitrary, since each term is factorization scheme and scale dependent.
An unambiguous decomposition into factorization scheme and scale independent terms is
\be
F_{pp}=\left(F_{p\overline{p}}-F_{\overline{p}\overline{p}}\right)\big{|}_{u_v}
+\left(F_{p\overline{p}}-F_{\overline{p}\overline{p}}\right)\big{|}_{d_v}
+(F_{pp}+F_{\overline{p}\overline{p}}-F_{p\overline{p}}),
\label{FfromFhatandDwithindicesdecompphys}
\ee
where now the final state superscript ``$h^\pm$'' is omitted for brevity, while the initial state is
explicitly indicated by the subscript. (Note that $F_{\overline{p}\overline{p}}=F_{pp}$.
However, we have not made this replacement in order to emphasize that
the final states differ, by the interchange $\pi^+ \leftrightarrow \pi^-$.)
The charge-sign asymmetry originates from the protons' valence $u$ and $d$ quarks, represented by the first and second term
respectively, where e.g.\ ``$\big{|}_{u_v}$'' means that all PDFs except 
that for the valence $u$ quark are set to zero. 
In other words, in Mellin space the first (second) term is proportional to the square of the $u$ ($d$) valence quark PDF.
Note that these asymmetry generating terms
neither depend on the protons' sea partons nor receive contributions from interactions between valence $u$ and $d$ quarks.
The third term corresponds to the charge-sign symmetric contribution,
becoming equal to $F_{pp}$ when the protons' valence quarks vanish. The arguments
which apply to Eq.\ (\ref{FfromFhatandDwithindicesdecomp}) that were given 
above apply also for Eq.\ (\ref{FfromFhatandDwithindicesdecompphys}).
It turns out that, qualitatively, the relative sizes of the first two terms in Eq.\ (\ref{FfromFhatandDwithindicesdecompphys})
are similar to those in Eq.\ (\ref{FfromFhatandDwithindicesdecomp}).
However, for graphical purposes we will study the contributions in Eq.\ (\ref{FfromFhatandDwithindicesdecomp}),
since the valence quark terms in Eq.\ (\ref{FfromFhatandDwithindicesdecompphys})
are typically a few orders of magnitude lower than the third term.
In other words, the fragmentation from the initial protons' sea partons always dominates, even if
the charge-sign asymmetry is very significant.

The charge-sign asymmetry is determined from the FFs according to
\be
F^{\Delta_c h^\pm}=\widehat{F}^{u_v} D_u^{\Delta_c h^\pm} +\widehat{F}^{d_v} D_d^{\Delta_c h^\pm}. 
\label{FfromFhatandDwithindicesforval}
\ee
Both terms are factorization scheme and scale independent.
Since $D_d^{\Delta_c \pi^\pm}$ is negative and $D_u^{\Delta_c \pi^\pm}$ positive, e.g.\ an excess of $\pi^+$ over $\pi^-$ 
requires a sufficiently large excess of fragmenting $u$ over 
$\bar{u}$ quarks relative to $d$ over $\bar{d}$.
For $h^\pm =p/\bar{p}$, all 4 quantities in 
Eq.\ (\ref{FfromFhatandDwithindicesforval}) are positive so that a definite excess
of $p$ over $\bar{p}$ is predicted. Likewise, an excess of $K^+$ over $K^-$ is predicted, although
in this case the second term vanishes since $D_d^{\Delta_c K^\pm}=0$. Note that there is no dependence
on $D_s^{\Delta_c K^\pm}$ since the $s$ quark can only be found in the protons' sea. 
In the limit that the light quark masses are equal, the resulting 
SU(3) invariance would imply that this FF is equal to $D_{\bar{u}}^{\Delta_c K^\pm}$. 
However, this symmetry is rather poor for the low energy processes which make up the bulk of these FFs. 
(SU(3) invariance would be useful for the high energy processes,
where the light quark masses can be neglected.) 
Therefore this symmetry should be used with care.

\section{Results}
\label{results}

The minimized $\chi^2$ values for the main fits and the fits in which no large $x$ resummation
is used are compared in Table \ref{chi2summary}.
(No comparison is made for the charge-sign asymmetry fits.
For these fits only $pp$ reaction data is used, so that the difference between the resummed and unresummed fit results only 
from the differences in the FF evolution, which is not as large as that in the quark coefficient function of $e^+ e^-$ reactions.)  
For $K^\pm$, $p/\overline{p}$ and $\Lambda/\overline{\Lambda}$, 
the resummation significantly improves the fit.
In this sense one can say that the data are sufficiently ``large $x$'' to warrant the resummation.
\begin{table}[h]
\caption{The minimized $\chi^2$ values in each of the charge-sign unidentified fits. For comparison, 
the $\chi^2$ values for the unresummed fit are shown (under ``Unres.\ fit'').
\label{chi2summary}}
\begin{center}
\begin{tabular}{|c|c|c|}
\hline
\multirow{2}{*}{$H$} & \multicolumn{2}{c|}{\multirow{2}{*}{\vspace{0.3cm} $\chi^2$}} \\
\cline{2-3}
    & Main fit & Unres.\ fit \\
\hline \hline
\input{figs_chi2summary}
\end{tabular}
\end{center}
\end{table}

From now on we consider only our main fits. 
Table \ref{hadmasses} shows the results for the hadron masses that are constrained by the $e^+ e^-$ data.
Excellent agreement is found for the baryons, which suggests that hadron mass effects comprise 
almost all the deviation from the standard calculation at the smaller $x$, lower $\sqrt{s}$ values of the data considered. 
While the excess in each of the fitted baryon masses is only about 1\%, it is somewhat larger for $\pi^\pm$.
This is expected because the $\pi^\pm$ sample receives large contributions from the decay of the much heavier 
$\rho$(770) to $\pi^+ +\pi^-$, while those baryons not produced from direct fragmentation will mostly come from
decays of their slightly heavier resonances.
While the fitted masses for $K^\pm$ and $K^0_S$ have the correct order of magnitude,
there is clearly a large undershoot.
A possible explanation is the fact that these particles' production mechanisms are so involved
that they cannot be even partly accounted for by mass effects of the heavier parent particles alone.
For example, a kaon (charged or neutral) resonance can decay to a pion and kaon simultaneously.
Interestingly, the undershoots of the fitted charged and neutral kaon masses are very similar, i.e. both are around 154.3 MeV,
which is consistent with SU(2) isospin symmetry between $u$ and $d$ quarks after all.
\begin{table}[h]
\caption{Fitted particle masses used in the calculation of the hadron production from $e^+ e^-$ reactions. For comparison,
the true particle masses are also shown.
\label{hadmasses}}
\begin{center}
\begin{tabular}{|c|c|c|}
\hline
Particle & Fitted mass (MeV) & True mass (MeV) \\
\hline
\input{figs_hadmasssummary}
\end{tabular}
\end{center}
\end{table}

The minimized reduced $\chi^2$ ($\chi^2_{\rm DF}$) values are shown in the last
row of Tables \ref{PionResults}--\ref{LambdaResults}. 
In each case, the total $\chi^2_{\rm DF}$ is in the range 1--2, indicating an overall good quality of fit. 
However, the $\chi^2_{\rm DF}$ value for each data set can sometimes be large due to unknown systematic effects.
For this reason some data sets were excluded in Refs.\ \cite{Albino:2005me,Albino:2005mv}.
In the fits of this paper we have taken the other extreme and have included all data sets in order to enhance the 
mutual cancellation of these unknown systematic effects.
In Ref.\ \cite{Albino:2005me}, OPAL quark tagging probabilities were included
to improve the quark flavour separation of the FFs, particularly in the light quark sector which no other
data could constrain at the time those fits were performed. 
In the fits of this paper, the RHIC data also help to separate out the light quark flavours
due to the different weighting provided by the PDFs.
This may be the main cause of the slightly larger  $\chi^2_{\rm DF}$ values for the individual light 
quark tagged OPAL probabilities relative to those of the previous AKK fits.
Note that the $\chi^2_{\rm DF}$ values of the individual light quark tagged OPAL probabilities for $K^\pm$, $K^0_S$ and
$\Lambda/\overline{\Lambda}$ are reasonable, as they were for the previous AKK fit.
This is not surprising since the OPAL data give the predicted strange quark suppression.
For $\pi^\pm$ and $K^\pm$, the fits to the heavy quark tagging probabilities are unsuccessful,
which may be due to the large angle gluon emission effects that we discussed in Ref.\ \cite{Albino:2005me}.

Turning now to the known systematic effects, the magnitudes of the fitted $\lambda_K$ values,
defined in Appendix \ref{correrr} to be the shift, upwards or downwards, 
of the measurements resulting from and relative to the $K$th source of systematic error, are 
typically less than around 1--2, being the upper limit for a reasonable fit. 
The most serious exceptions are the $\lambda_K$ values for the 
$p/\overline{p}$ data from BRAHMS for which $3.25<y<3.35$
and for the $\Lambda/\overline{\Lambda}$ data from CDF and STAR.
For example, the global $\pi^\pm$ production data imply that the central values from TASSO at $\sqrt{s}=$34 GeV
overshoot the best fit calculation by 3\%. All central values from RHIC and Tevatron (CDF) overshoot the best fit
calculation, as do most $e^+ e^-$ reaction data for which a 
systematic error is given. This may result from contamination
of the measured sample in each case by other particles, or something else.

Finally, for completeness and to 
clarify our choice of parameterizations, the central values of 
all parameters are listed in Tables \ref{paramtable} and \ref{paramValtable}. 
Due to the low factorization scale at which these FFs are parameterized, 
extractions of our FFs at arbitrary momentum fraction and factorization scale should be performed using
the grids and FORTRAN routines referred to in Section \ref{conc}. They should not be extracted by using these parameters
directly in a NLO evolution routine, which will in general differ from our routine by a potentially large NNLO error 
at $M_f =O(1)$ GeV. For the same reason, any physical interpretation of these parameters should be avoided.
However, we note for the charge-sign unidentifed FFs in Table \ref{paramtable}
that whenever a large value ($>1000$) for $N_i$ is obtained, 
it is compensated for by a large value ($>9$) for $b_i$, which signifies a large statistical correlation
between these parameters in these cases, and consequently poor constraints at large $x$. 
Conversely, for the charge-sign asymmetries in Table \ref{paramValtable}, poor constraints at small $x$ were signified in 
previous fits by a large correlation between the $N_u^{\Delta_c K^\pm}$ and $a_u^{\Delta_c K^\pm}$, which is why the 
$a_u^{\Delta_c K^\pm}$ were fixed to zero for these FFs in our final main fits.
Note that SU(2) isospin symmetry between $u$ and $d$ was used to constrain $D^{\Delta_c \pi^\pm}_d
=D^{\Delta_c \pi^\pm}_u$, while due to poor experimental constraints we fixed $D^{\Delta_c p/\bar{p}}_d
=0.5 D^{\Delta_c p/\bar{p}}_u$.
\begin{table}[h]
\caption{Values, to 2 decimal places, 
of the parameters of the charge-sign unidentified FFs at $M_f =\sqrt{2}$ GeV for which the $\chi^2$ values are minimal.
We fix $D_d^{\pi^\pm} (x,M_f^2)=D_u^{\pi^\pm} (x,M_f^2)$ as dictated by SU(2) isospin symmetry between
$u$ and $d$ quarks, and also fix $c_s^{\pi^\pm}=d_s^{\pi^\pm}=0$ 
since they were rather independent of $\chi^2$. \label{paramtable}}
\begin{center}
\begin{tabular}{|c||c|c|c|c|c|}
\hline
Parameter & $\pi^\pm$ & $K^\pm$ & $p/\bar{p}$ & $K^0_S$ & $\Lambda/\overline{\Lambda}$ \\ \hline \hline
\input{figs_paramtable}
\end{tabular}
\end{center}
\end{table}

\begin{table}[h]
\caption{As in Table \ref{paramtable}, but for the charge-sign asymmetry FFs at $M_f =\sqrt{2}$ GeV.
We fix $D_d^{\Delta_c \pi^\pm} (x,M_f^2)=D_u^{\Delta_c \pi^\pm} (x,M_f^2)$ as dictated by SU(2) isospin symmetry between
$u$ and $d$ quarks, and also fix $a_i^{\Delta_c K^\pm}=a_i^{\Delta_c p/\bar{p}}=0$ 
since they were rather independent of $\chi^2$. In addition, we fix $D_d^{\Delta_c K^\pm}=0$ as dictated by charge-sign
symmetry.
\label{paramValtable}}
\begin{center}
\begin{tabular}{|c||c|c|c|c|c|}
\hline
Parameter & $\Delta_c \pi^\pm$ & $\Delta_c K^\pm$ & $\Delta_c p/\bar{p}$ \\ \hline \hline
\input{figs_paramValtable}
\end{tabular}
\end{center}
\end{table}

\begin{figure}
\begin{center}
\includegraphics[width=17cm]{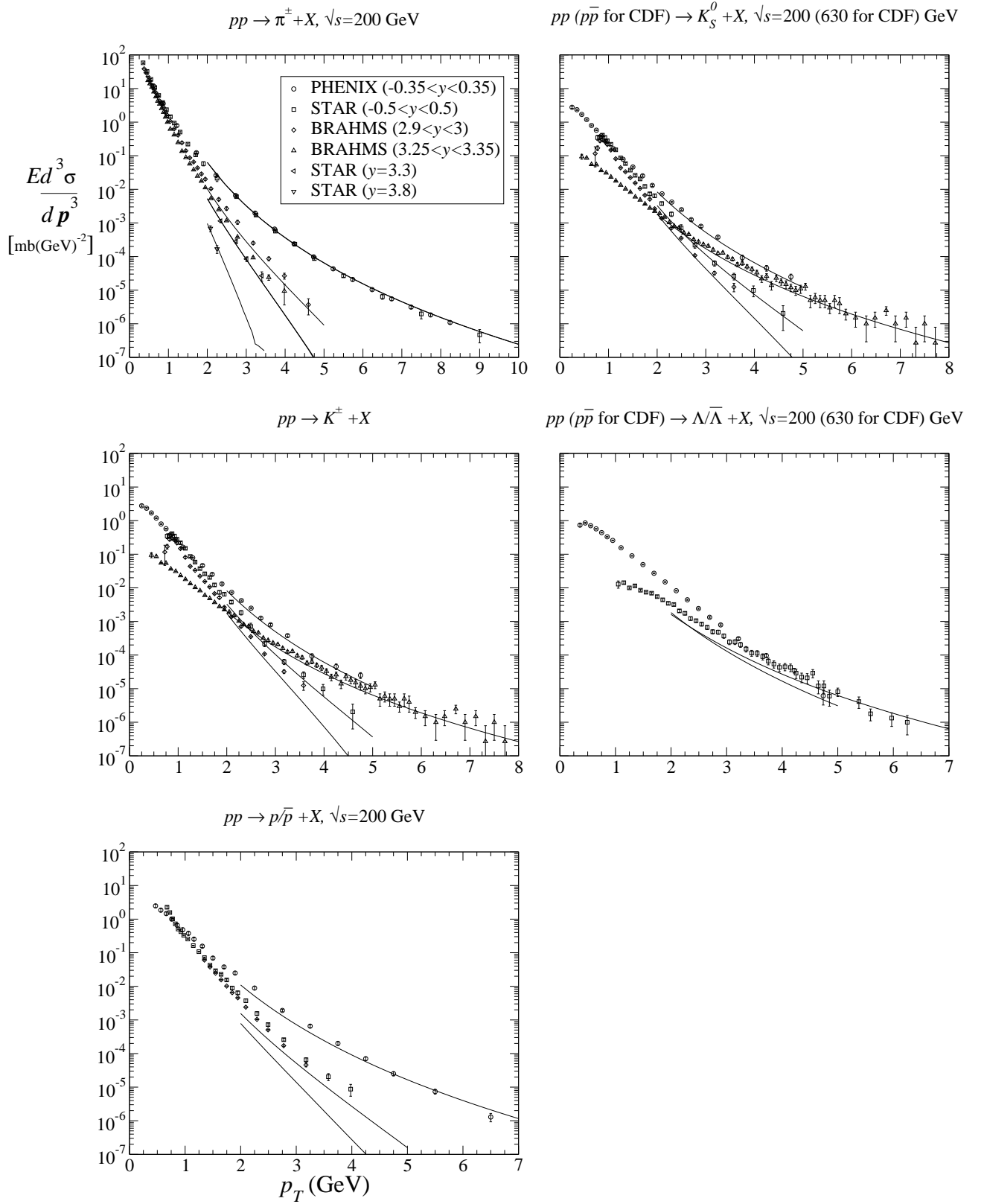}
\caption{Comparison of the calculation with measurements of the invariant 
differential cross sections for inclusive production in $pp$
collisions at $\sqrt{s}=200$ (630 for CDF) GeV. 
Note that $E d^3\sigma /d {\mathbf p}^3$ is averaged over rapidity 
using the operation $(1/(y_{\rm max}-y_{\rm min}))\int^{y_{\rm max}}_{y_{\rm min}}dy$.
For $\pi^\pm$, the curves for which $-0.35<y<0.35$ and $-0.5<y<0.5$ overlap,
as do those for which $3.25<y<3.35$ and $y=3.3$. \label{pp}}
\end{center}
\end{figure}

\begin{figure}
\begin{center}
\includegraphics[width=8.5cm]{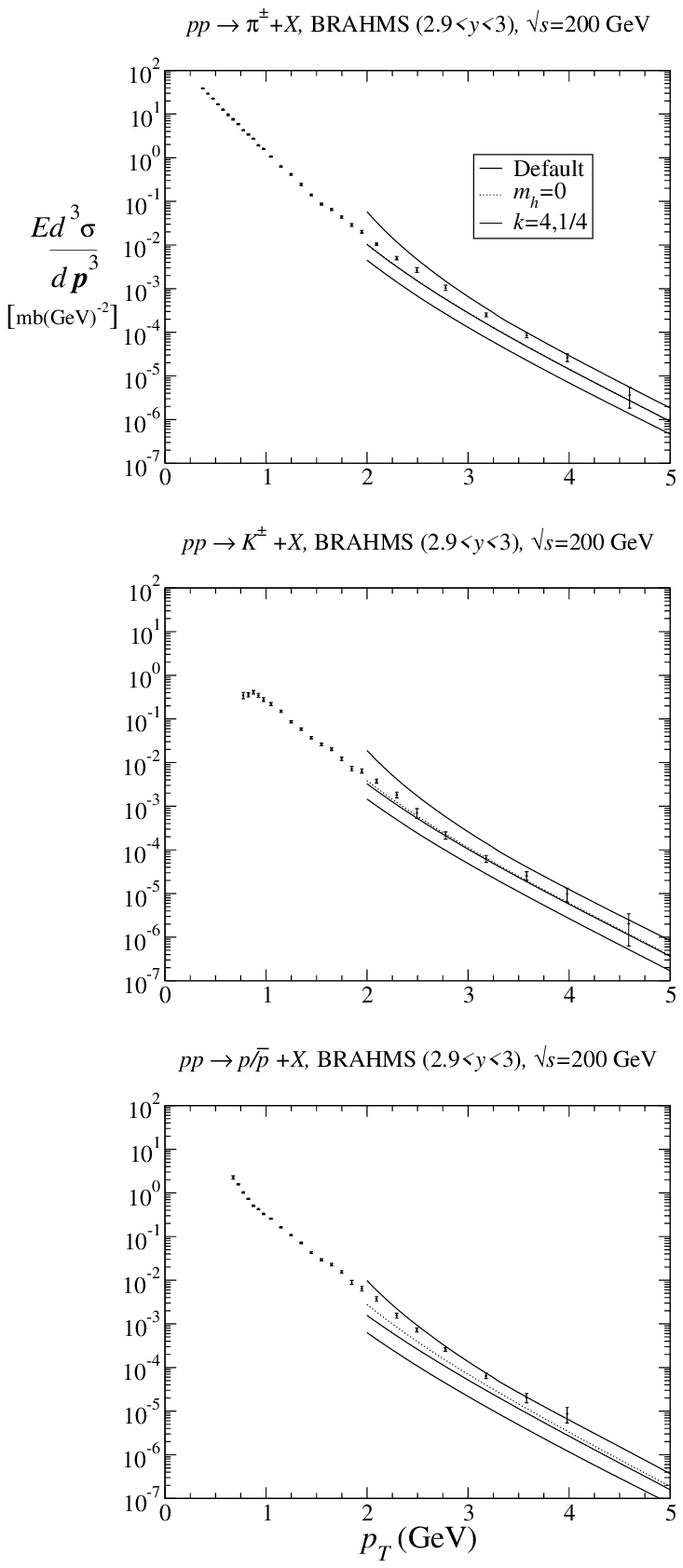}
\caption{As in Fig.\ \ref{pp} for the BRAHMS data for which $2.9<y<3$ (labeled ``Default'').
Also shown is the calculation when mass effects are
neglected (the dotted curve labeled ``$m_h=0$'') and when the ratio $k=M_f^2 /s$ is increased to 4 (lower solid curve) and decreased
to $1/4$ (upper solid curve). In the case of $\pi^\pm$, the $m_h=0$ curve cannot be seen because it overlaps 
with the ``Default'' curve. \label{ppBRAHMS}}
\end{center}
\end{figure}

\begin{figure}
\begin{center}
\includegraphics[width=17cm]{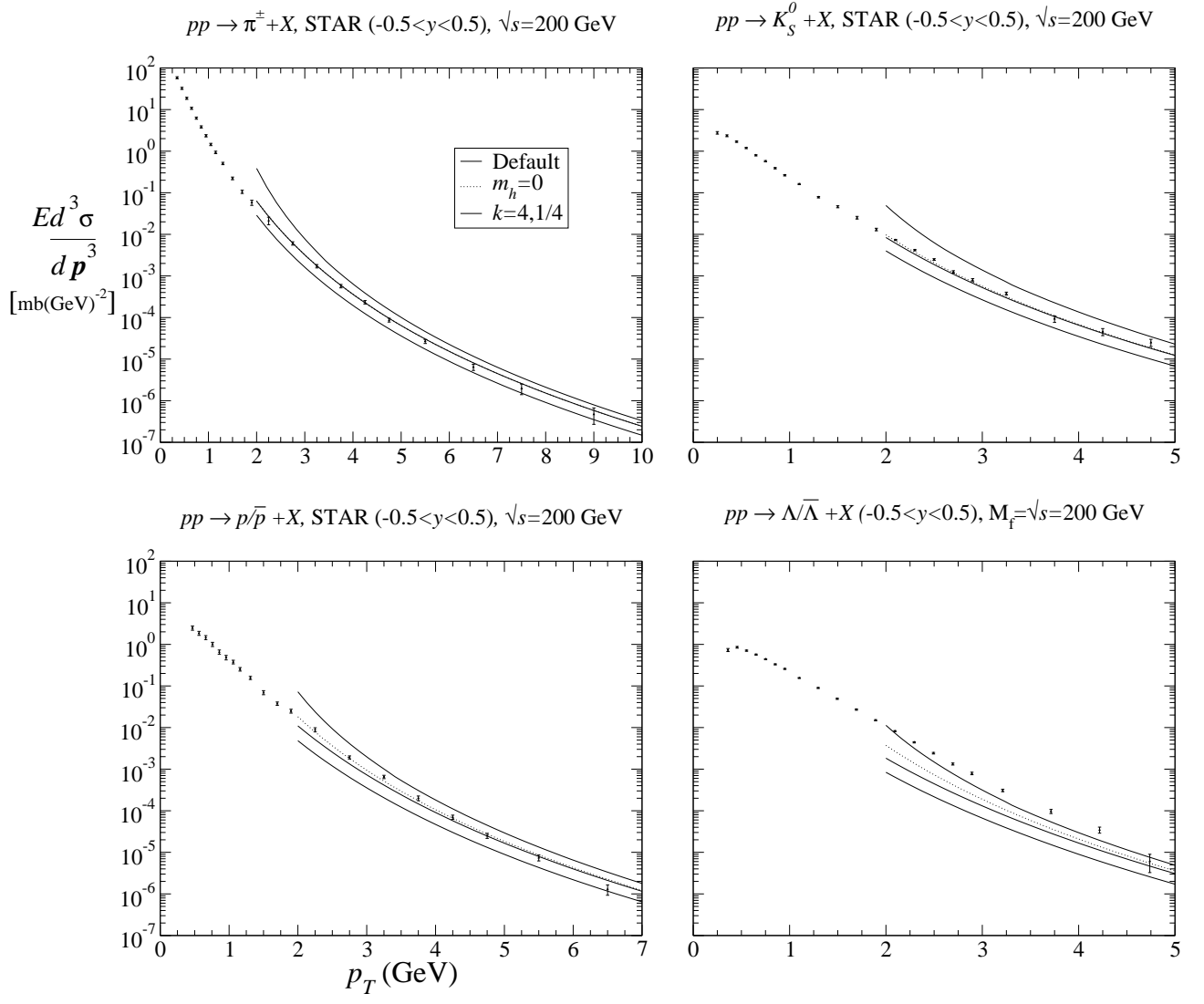}
\caption{As in Fig.\ \ref{ppBRAHMS}, but for the STAR data.\label{ppSTAR}}
\end{center}
\end{figure}

\begin{figure}
\begin{center}
\includegraphics[width=8.5cm]{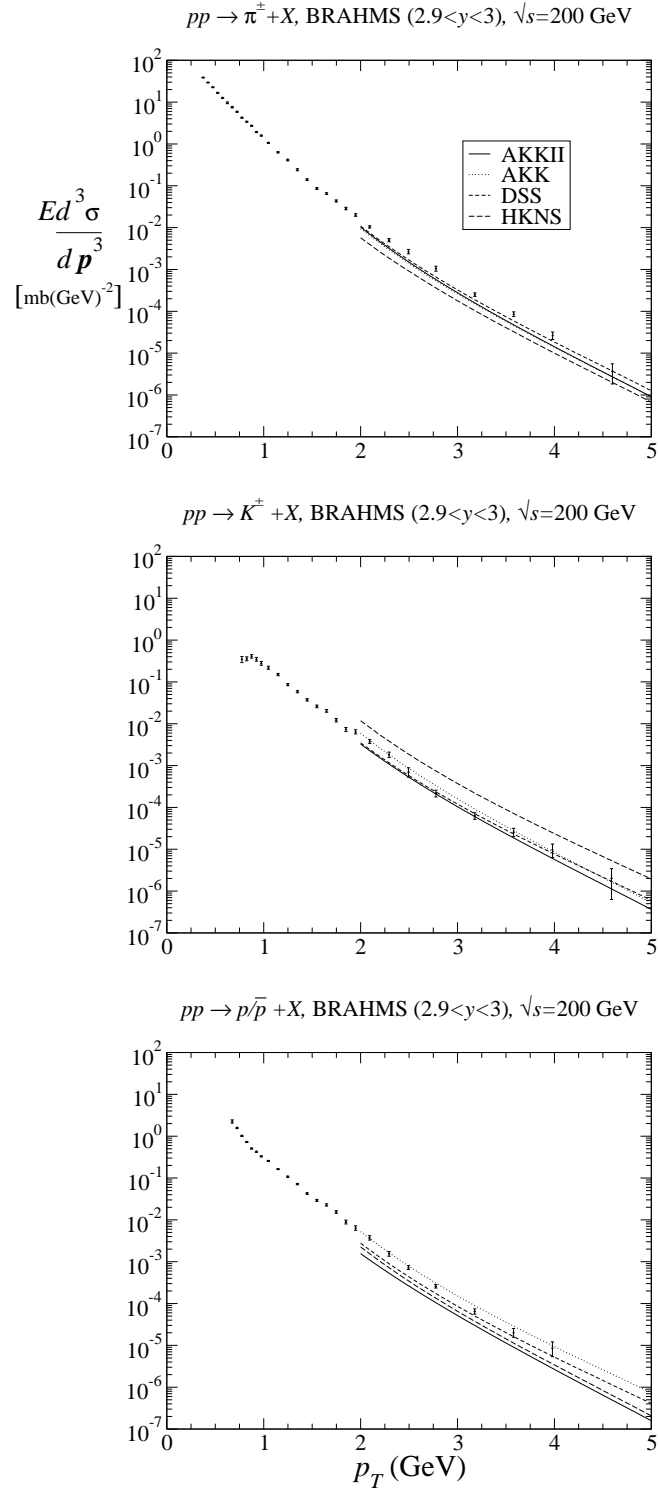}
\caption{As for the ``Default'' cases in Fig.\ \ref{ppBRAHMS} (now labeled ``AKK08''), including also the calculations
using the AKK \cite{Albino:2005me,Albino:2005mv}, DSS \cite{deFlorian:2007aj} and 
HKNS \cite{Hirai:2007cx} FF sets with the hadron mass set to zero. \label{ppBRAHMS_ffc}}
\end{center}
\end{figure}

\begin{figure}
\begin{center}
\includegraphics[width=17cm]{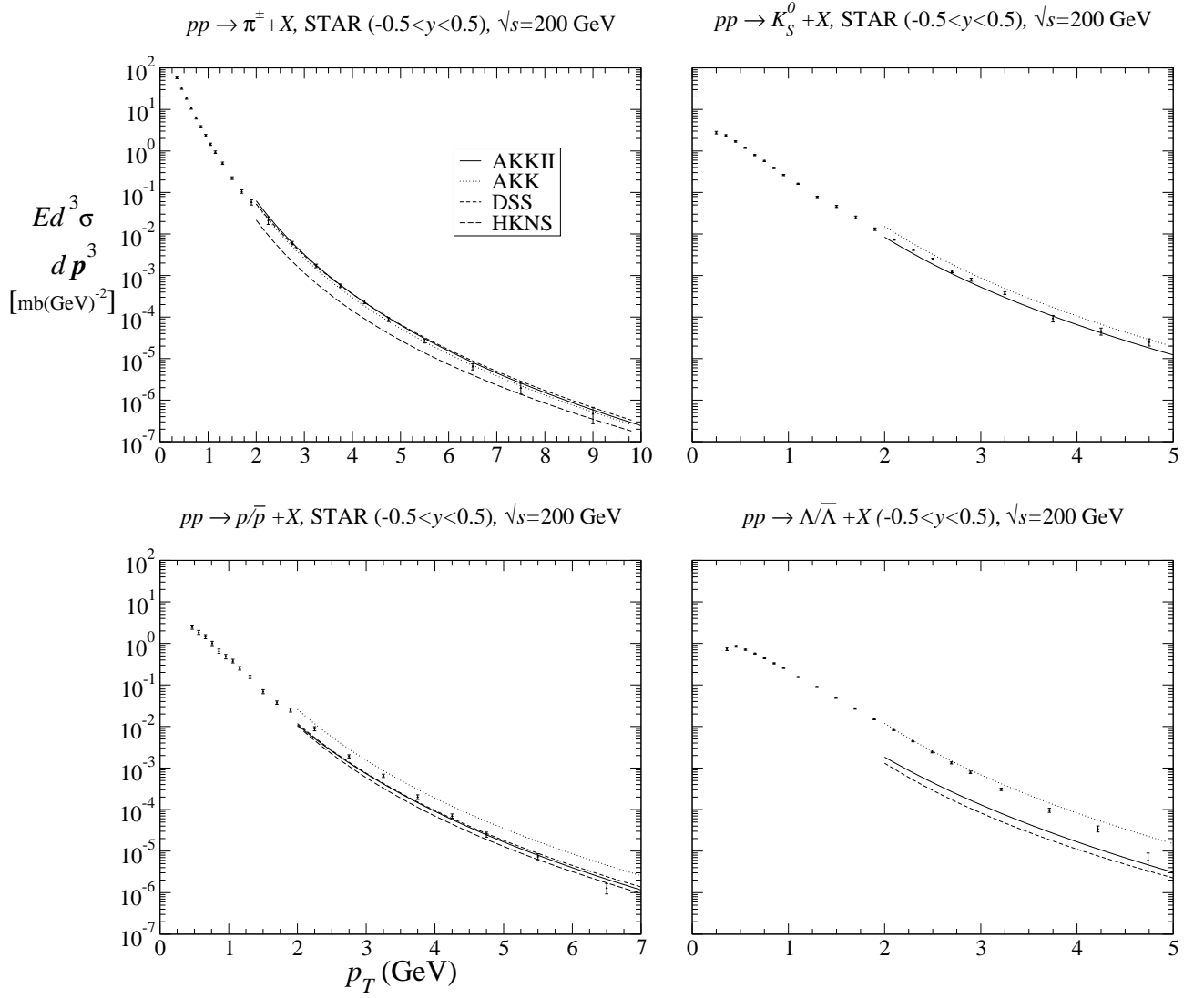}
\caption{As in Fig.\ \ref{ppBRAHMS_ffc}, but for the STAR data. Also shown, for $\Lambda/\overline{\Lambda}$, 
is the calculation using the DSV \cite{deFlorian:1997zj} FF set. \label{ppSTAR_ffc}}
\end{center}
\end{figure}

\begin{figure}
\begin{center}
\includegraphics[width=8.5cm]{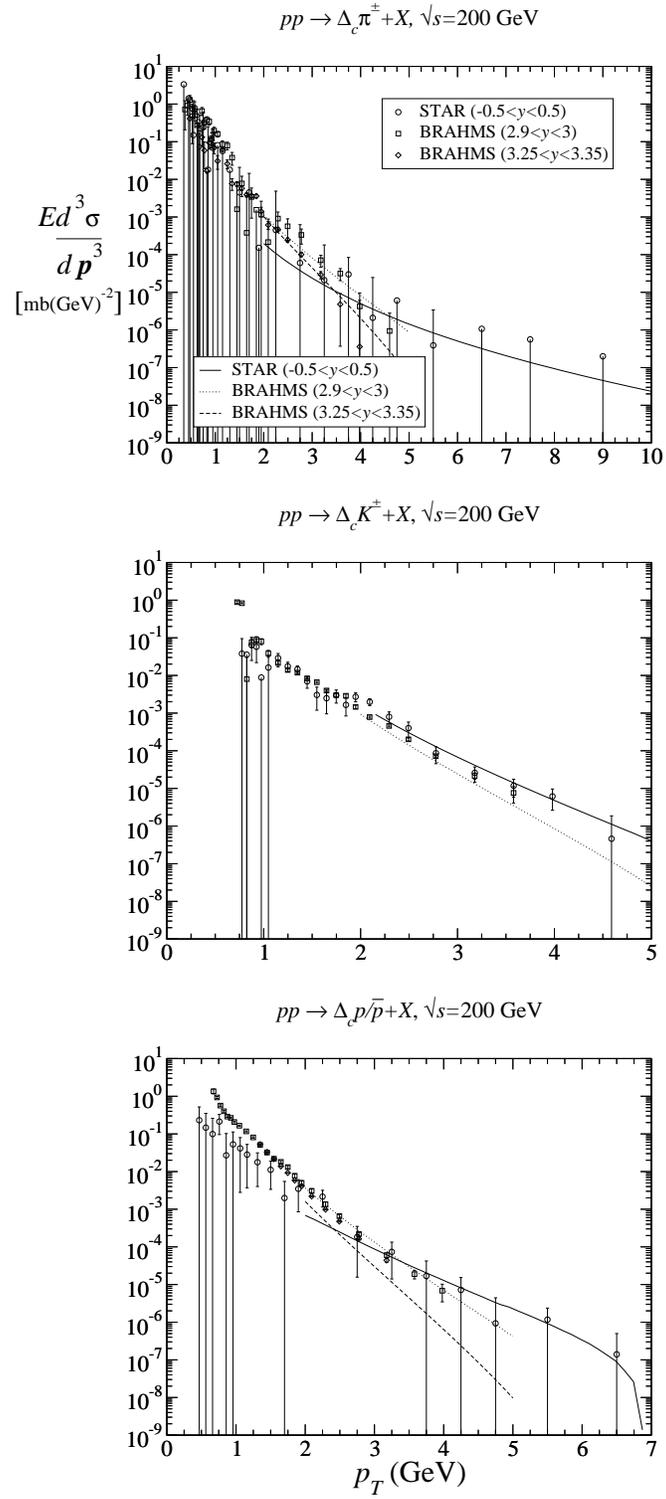}
\caption{As in Fig.\ \ref{pp}, but for the charge-sign
assymetry $\Delta_c h^\pm$. \label{ppVal}}
\end{center}
\end{figure}

\begin{figure}
\begin{center}
\includegraphics[width=8.5cm]{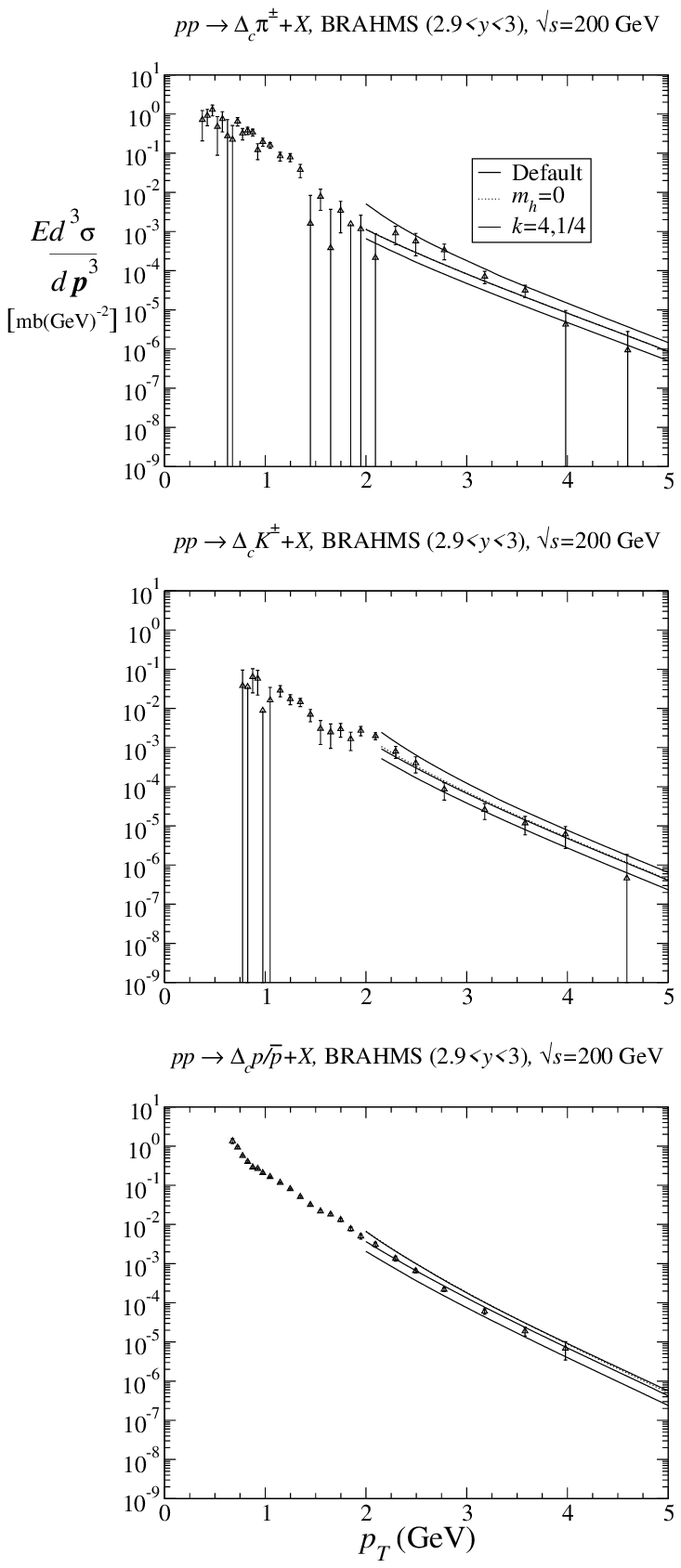}
\caption{As in Fig.\ \ref{ppVal} for the BRAHMS data for which $2.9<y<3$ (labeled ``Default'').
Also shown is the calculation when mass effects are
neglected (the dotted curve labeled ``$m_h=0$'') and when the ratio $k=M_f^2 /s$ is increased to 4 (lower solid curve) and decreased
to $1/4$ (upper solid curve). In the case of $\Delta_c \pi^\pm$, the $m_h=0$ curve cannot be seen because it overlaps 
with the ``Default'' curve. \label{ppValBRAHMS}}
\end{center}
\end{figure}

\begin{figure}
\begin{center}
\includegraphics[width=8.5cm]{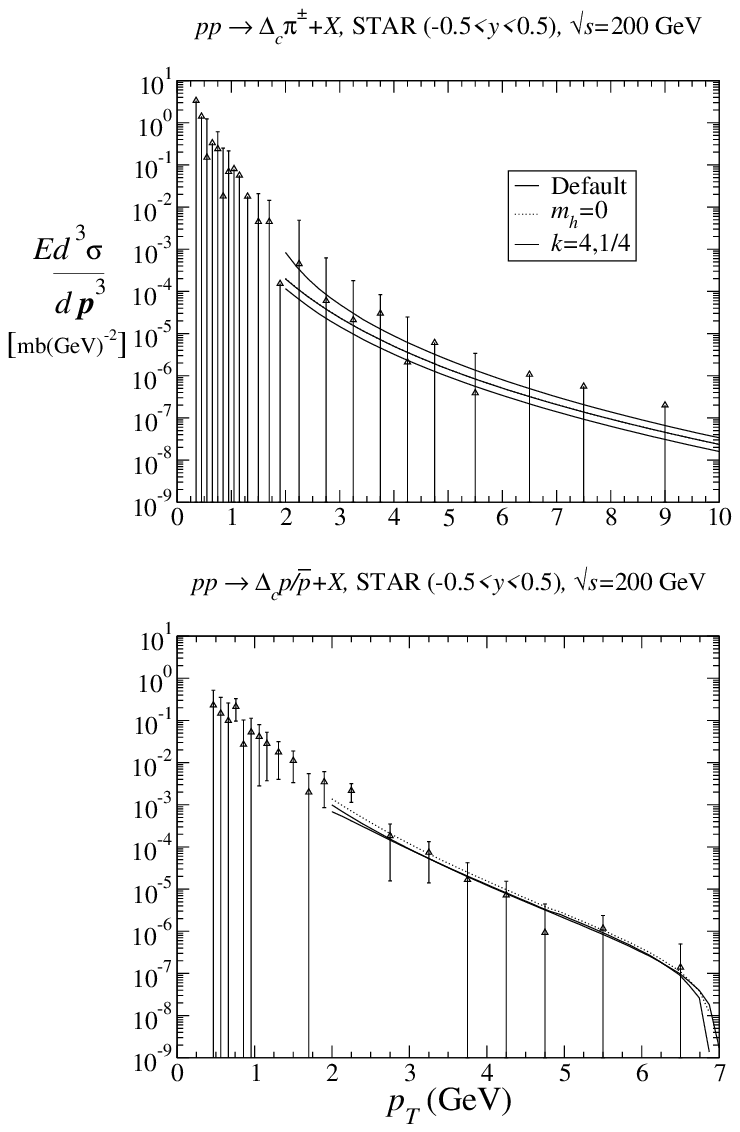}
\caption{As in Fig.\ \ref{ppValBRAHMS}, but for STAR kinematics. \label{ppValSTAR}}
\end{center}
\end{figure}

\begin{figure}
\begin{center}
\includegraphics[width=8.5cm]{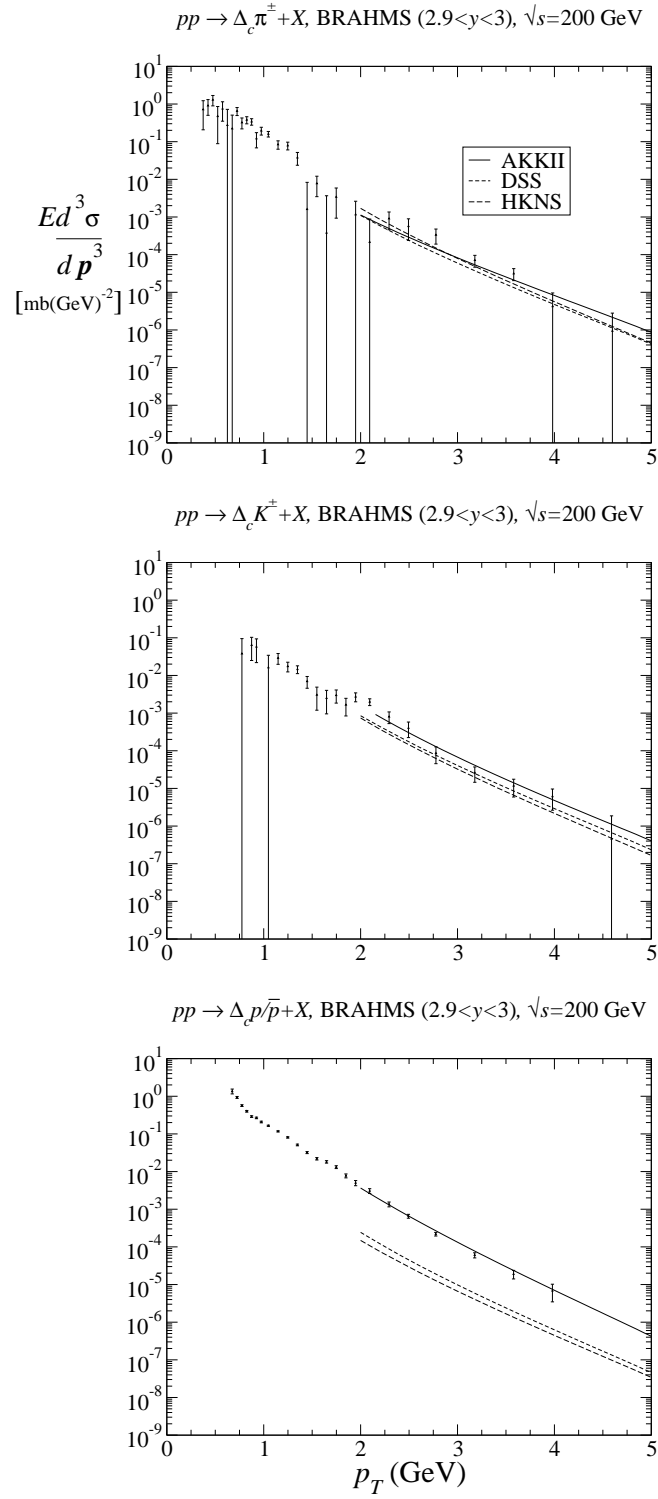}
\caption{As in Fig.\ \ref{ppBRAHMS_ffc}, but for $\Delta_c h^\pm$. \label{ppValBRAHMS_ffc}}
\end{center}
\end{figure}

\begin{figure}
\begin{center}
\includegraphics[width=8.5cm]{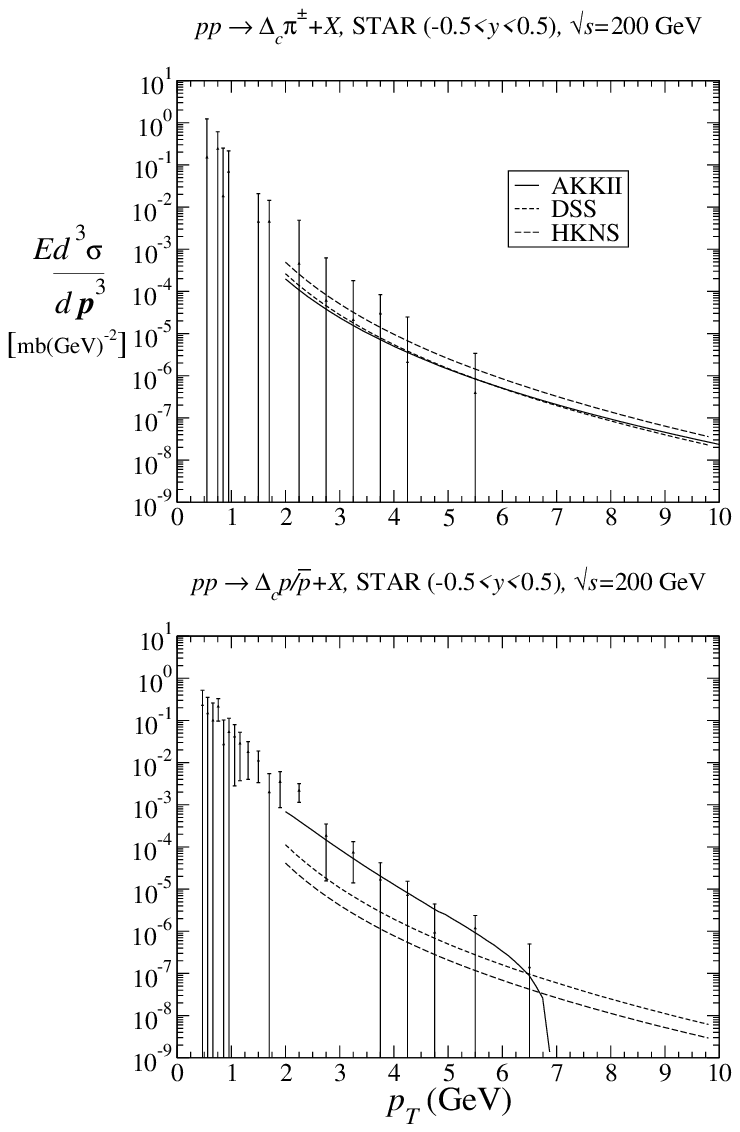}
\caption{As in Fig.\ \ref{ppValBRAHMS_ffc}, but for STAR kinematics. \label{ppValSTAR_ffc}}
\end{center}
\end{figure}

\begin{figure}
\begin{center}
\includegraphics[width=8.5cm]{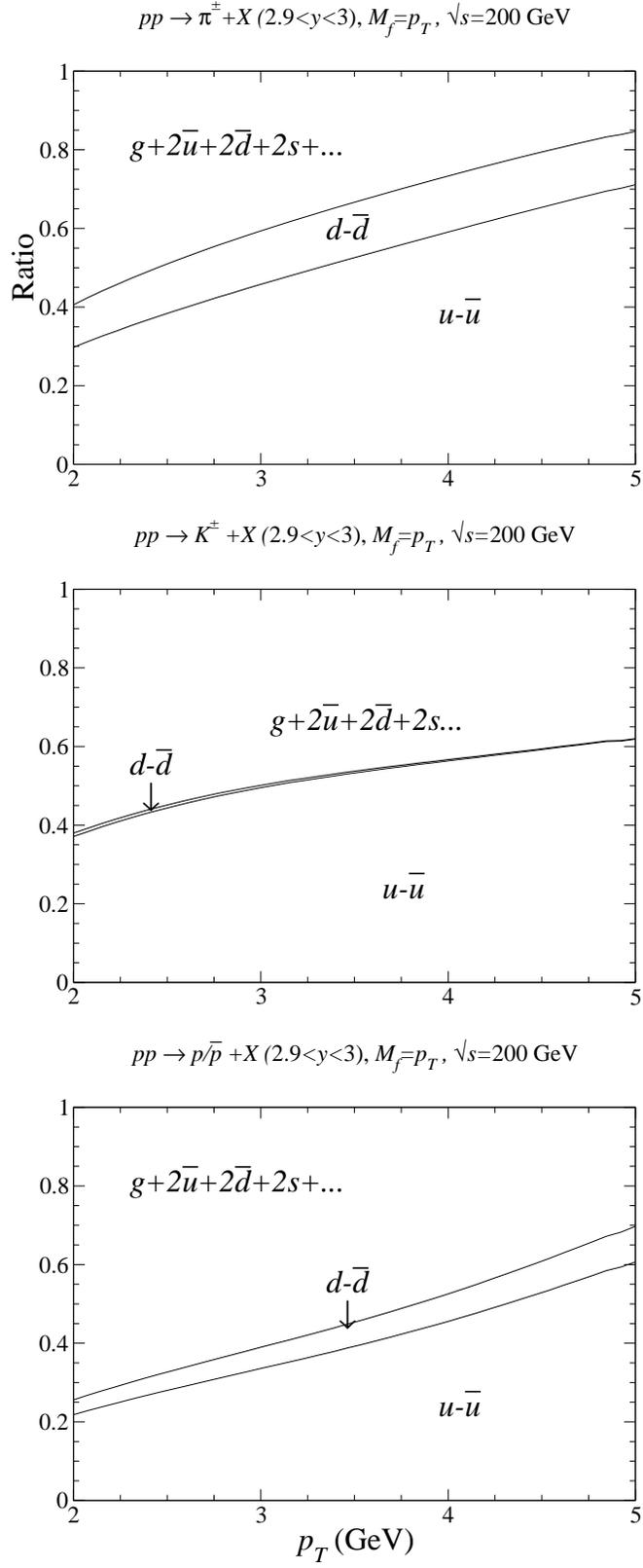}
\caption{Contributions to the production from fragmentation of the initial
state protons' valence quarks and sea partons, for BRAHMS kinematics with $2.9<y<3$. 
In the notation of Eq.\ (\ref{FfromFhatandDwithindices}), ``$u-\bar{u}$''
refers to the contribution from $u$ valence fragmentation
$\widehat{F}_u D_u^{h^\pm}-\widehat{F}_{\bar{u}}D_{\bar{u}}^{h^\pm} $ etc.
The ratios are stacked, i.e.\ for a given $p_T$ value, the distance on the $y$-axis
from zero to the first curve gives the $u-\bar{u}$ contribution, from the first to the second curve
the $d-\bar{d}$ contribution, and from the second curve to $1$ the sea parton contribution.
\label{ppValenceRatioBRAHMS}}
\end{center}
\end{figure}

\begin{figure}
\begin{center}
\includegraphics[width=8.5cm]{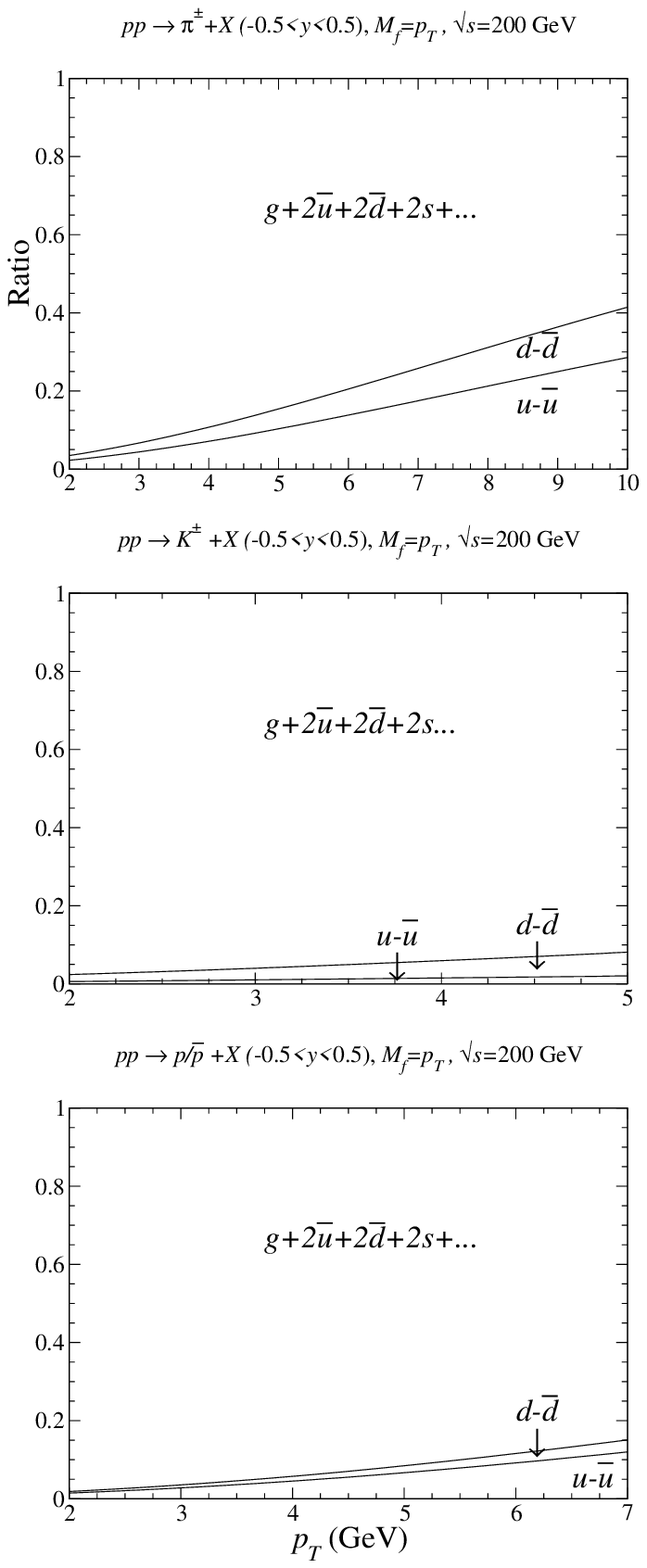}
\caption{As in Fig.\ \ref{ppValenceRatioBRAHMS}, but for STAR kinematics. \label{ppValenceRatioSTAR}}
\end{center}
\end{figure}

\begin{figure}
\begin{center}
\includegraphics[width=8.5cm]{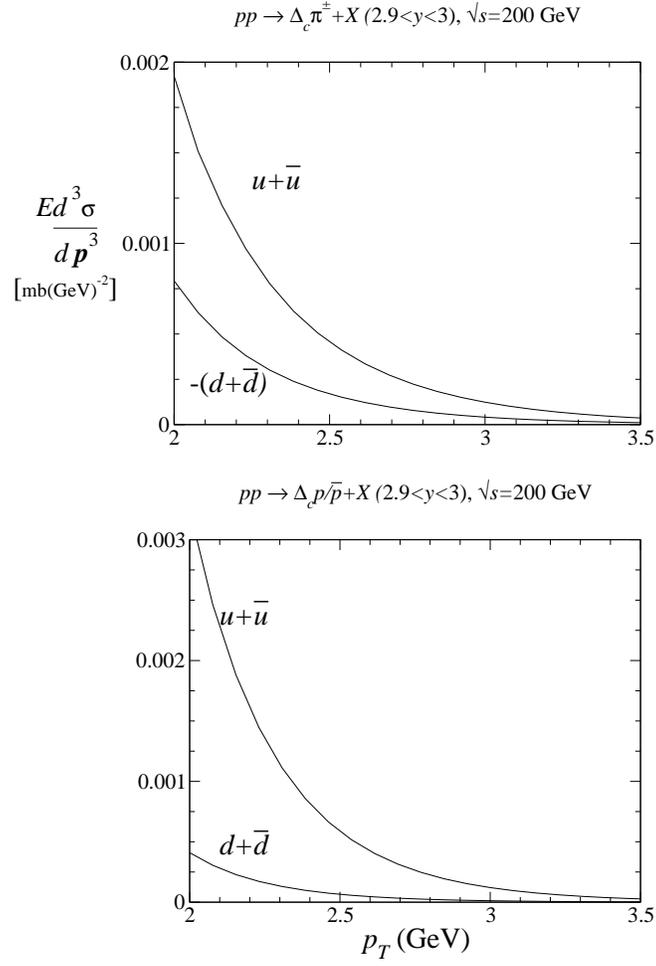}
\caption{Contributions to the charge-sign assymetry $\Delta_c h^\pm$ for BRAHMS kinematics for which $2.9<y<3$
from fragmentation of the initial state protons' valence quarks. 
In the notation of Eq.\ (\ref{FfromFhatandDwithindices}), ``$u+\bar{u}$''
means $\widehat{F}_u D_u^{\Delta_c \pi^\pm}+\widehat{F}_{\bar{u}}D_{\bar{u}}^{\Delta_c \pi^\pm} $ etc.
The cross section is given by the difference between the two curves.
\label{ppValenceRatioVal_BRAHMS}}
\end{center}
\end{figure}

\begin{figure}
\begin{center}
\includegraphics[width=8.5cm]{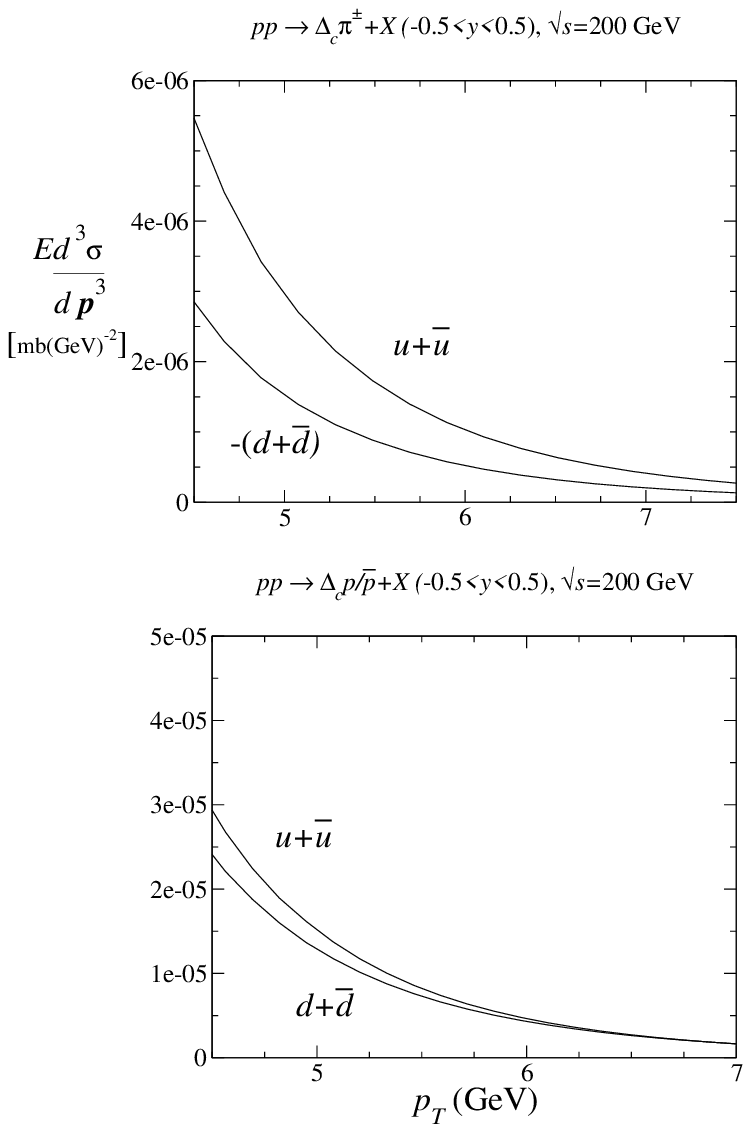}
\caption{As in Fig.\ \ref{ppValenceRatioVal_BRAHMS}, but for STAR kinematics. \label{ppValenceRatioVal_STAR}}
\end{center}
\end{figure}

\begin{figure}
\begin{center}
\includegraphics[width=17cm]{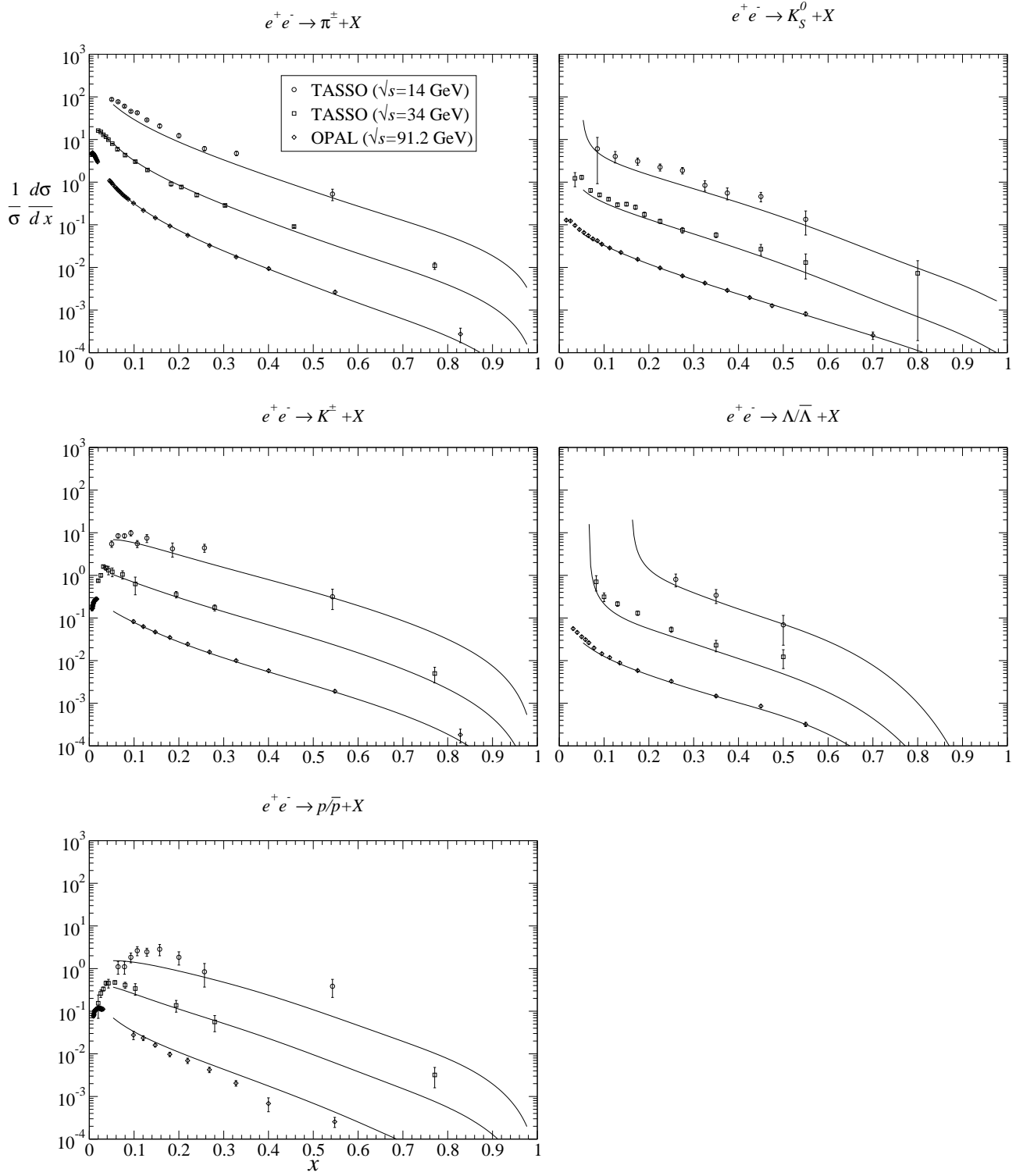}
\caption{Comparison of the calculation with measurements of the invariant 
differential cross sections for inclusive $h^\pm$ production in $e^+ e^-$
collisions. For clarity, the TASSO data at $\sqrt{s}=34$ GeV and the OPAL data at $\sqrt{s}=91.2$ GeV
have been multiplied by 0.1 and 0.01 respectively, as well as their corresponding calculations. \label{ee}}
\end{center}
\end{figure}

\begin{figure}
\begin{center}
\includegraphics[width=8.5cm]{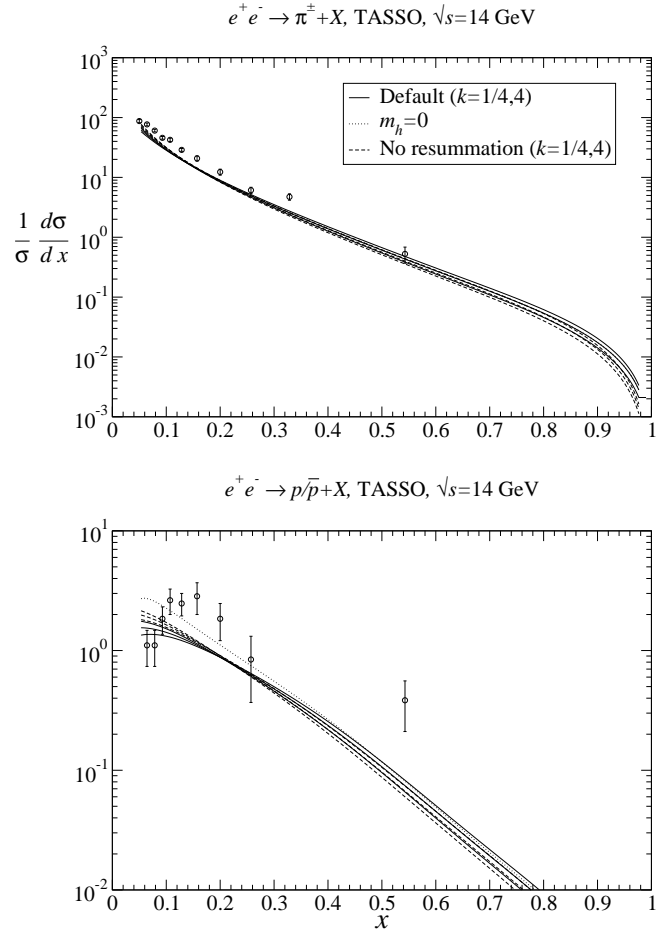}
\caption{As in Fig.\ \ref{ee} for the TASSO data at $\sqrt{s}=14$ GeV (labeled ``Default''). 
Also shown is the calculation when mass effects are 
neglected (the dotted curve labeled ``$m_h=0$'') and when the ratio $k=M_f^2 /s$ is increased to 4 (lower solid curve) and decreased
to $1/4$ (upper solid curve). \label{TASSO14_Z-qq}}
\end{center}
\end{figure}

\begin{figure}
\begin{center}
\includegraphics[width=8.5cm]{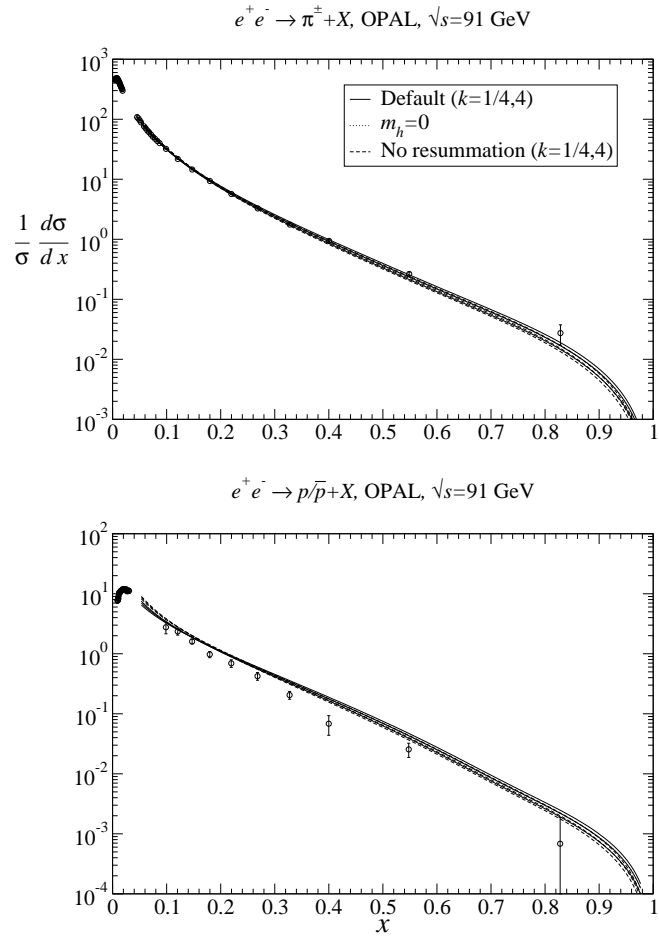}
\caption{As in Fig.\ \ref{TASSO14_Z-qq}, but for the OPAL data at $\sqrt{s}=91.2$ GeV. \label{OPAL_Z-qq}}
\end{center}
\end{figure}

\begin{figure}
\begin{center}
\includegraphics[width=17cm]{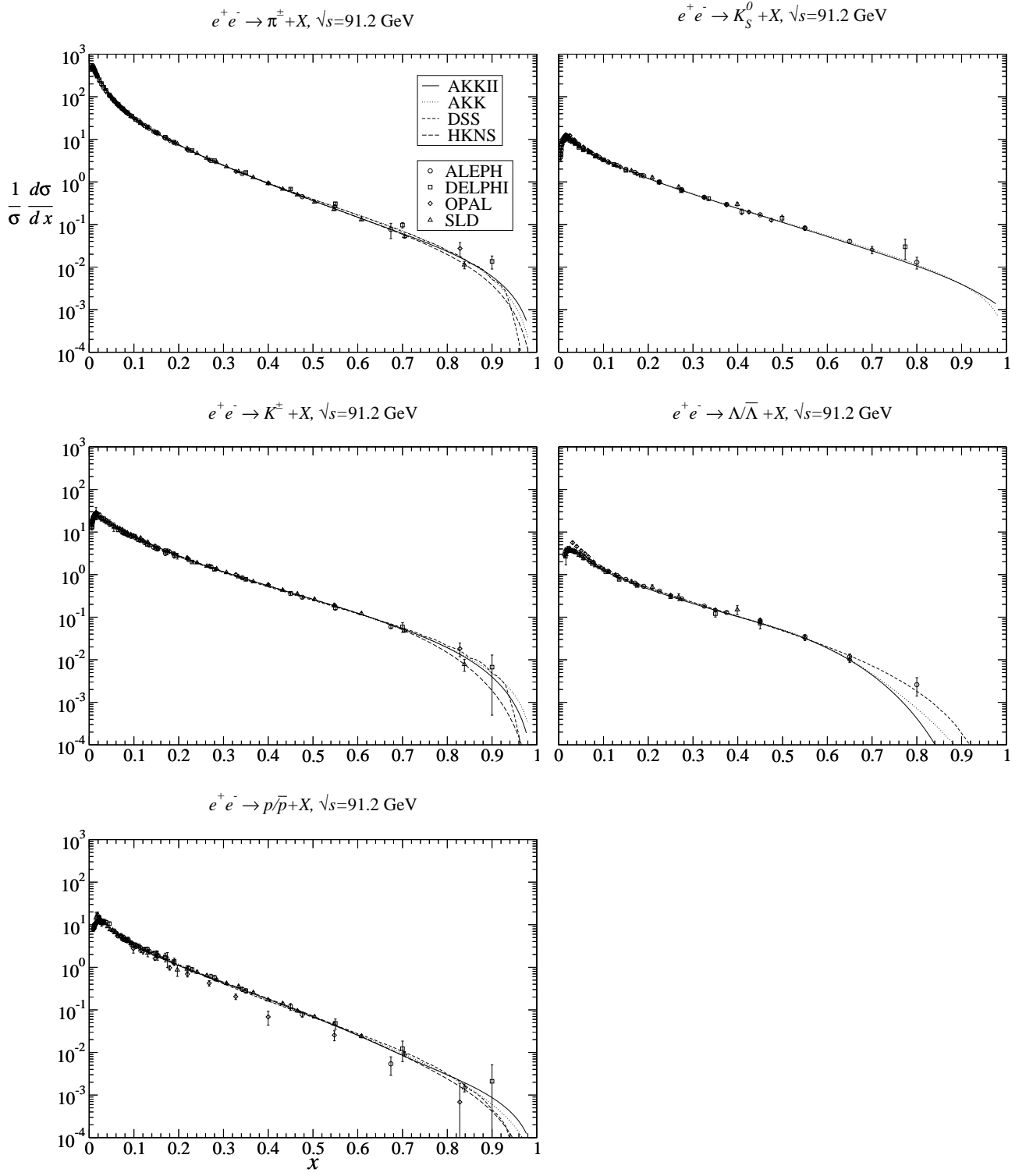}
\caption{As in Fig.\ \ref{ee}, but for all fitted $\sqrt{s}=91$ GeV data and with 
the corresponding calculations from other FF sets. \label{ee_ffc}}
\end{center}
\end{figure}

\begin{figure}
\begin{center}
\includegraphics[width=17cm]{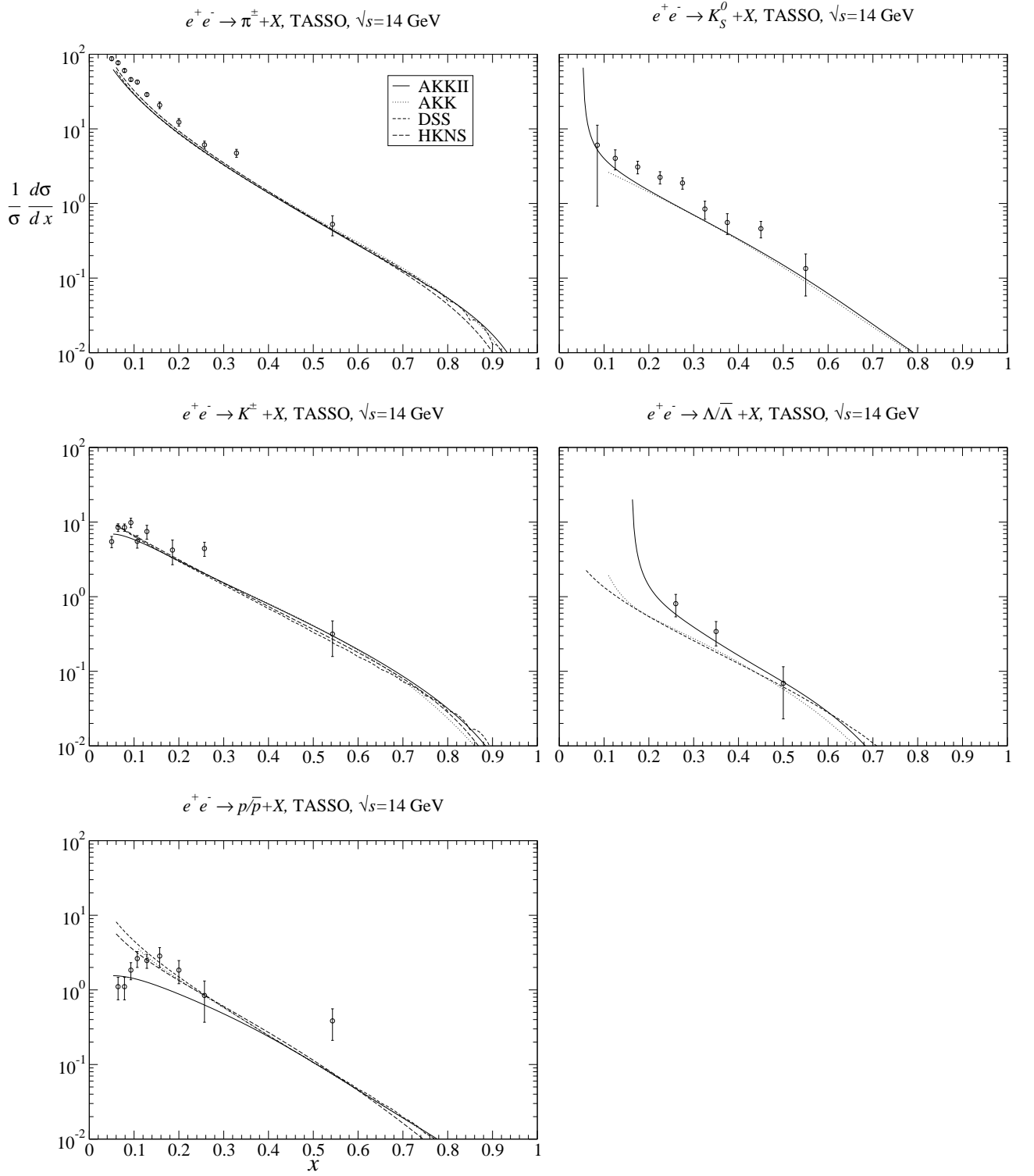}
\caption{As in Fig.\ \ref{ee_ffc}, but for the TASSO data at $\sqrt{s}=14$ GeV data and with 
the corresponding calculations from other FF sets. \label{TASSO14_Z-qq_ffc}}
\end{center}
\end{figure}

\begin{figure}
\begin{center}
\includegraphics[width=8.5cm]{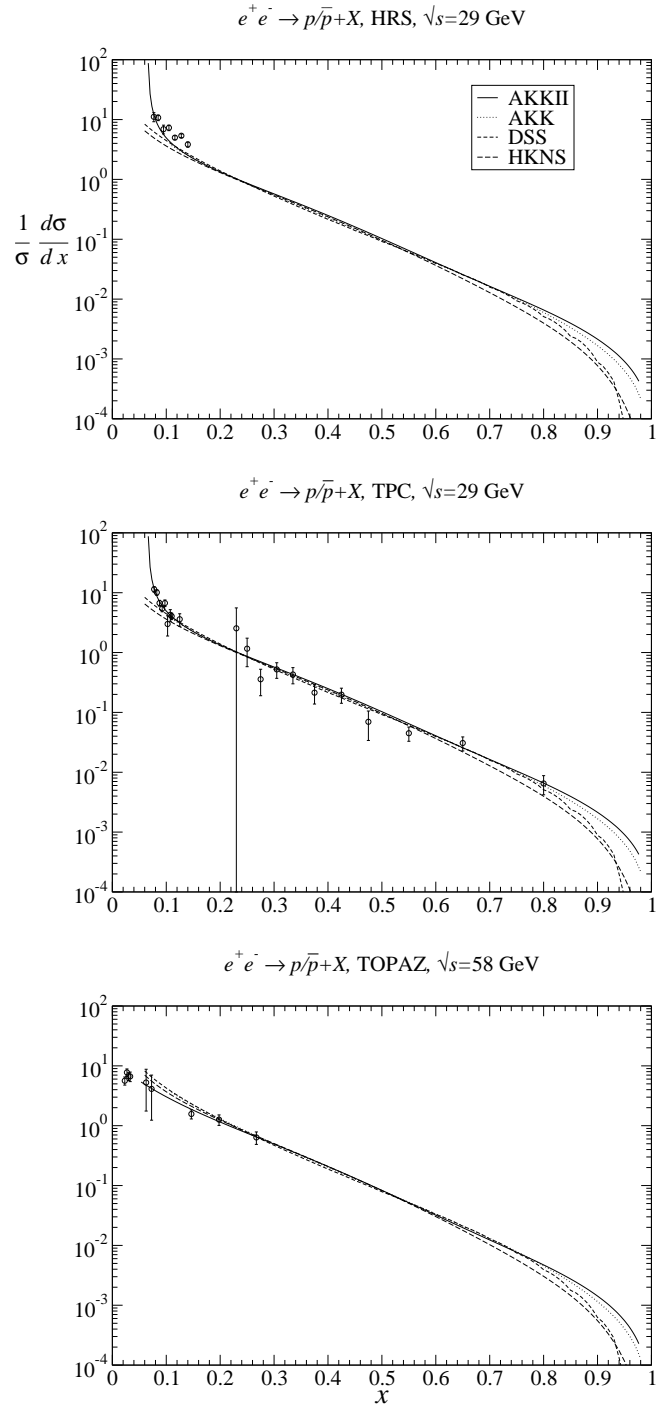}
\caption{As in Fig.\ \ref{TASSO14_Z-qq_ffc}, but for the HRS and TPC data at $\sqrt{s}=29$ GeV 
and the TOPAZ data at $\sqrt{s}=58$ GeV. \label{qqProton_ffc}}
\end{center}
\end{figure}

\begin{figure}
\begin{center}
\includegraphics[width=17cm]{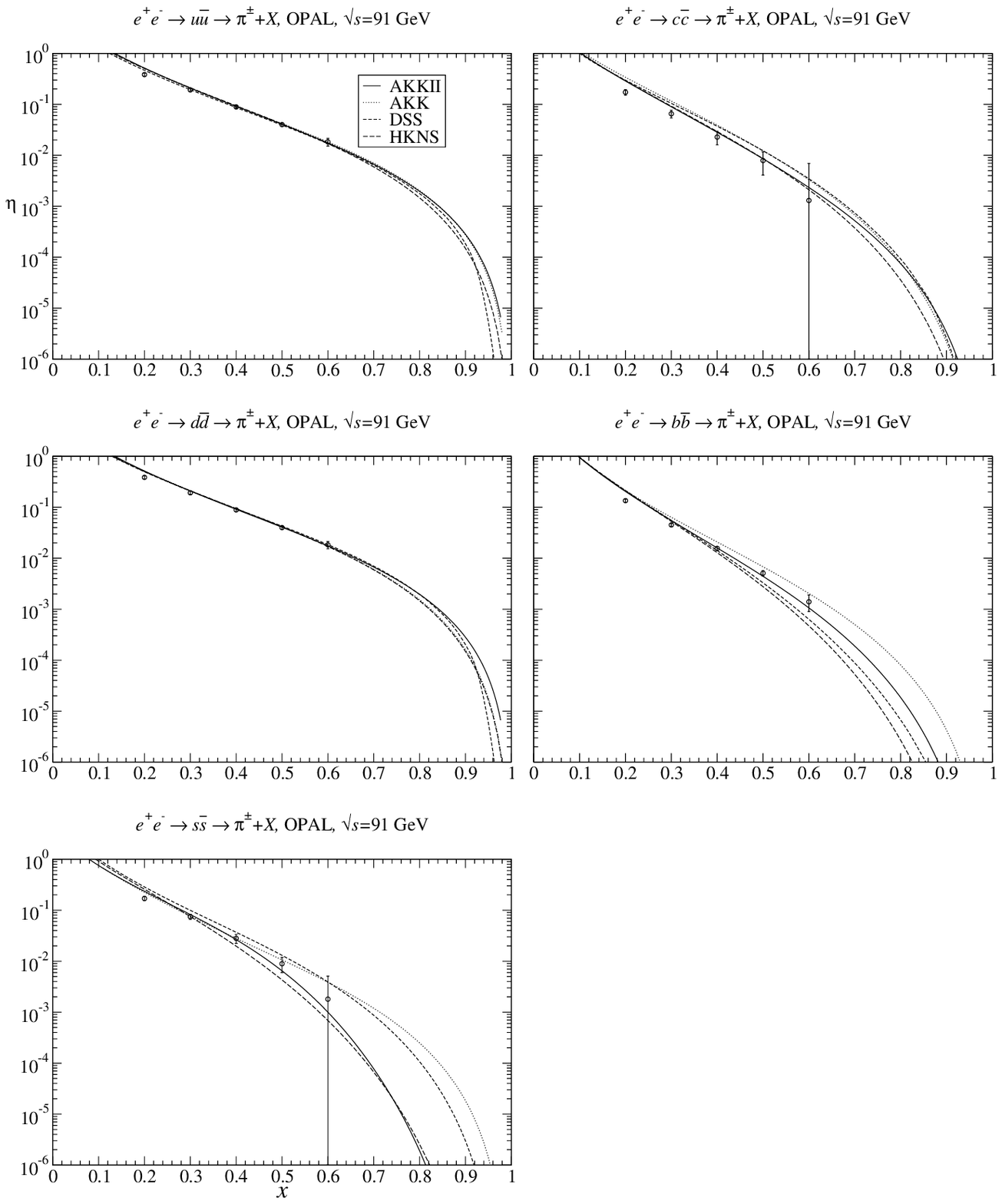}
\caption{OPAL quark tagging probabilities for $\pi^\pm$. \label{OPAL-eta_eta-Pion_ffc}}
\end{center}
\end{figure}

\begin{figure}
\begin{center}
\includegraphics[width=17cm]{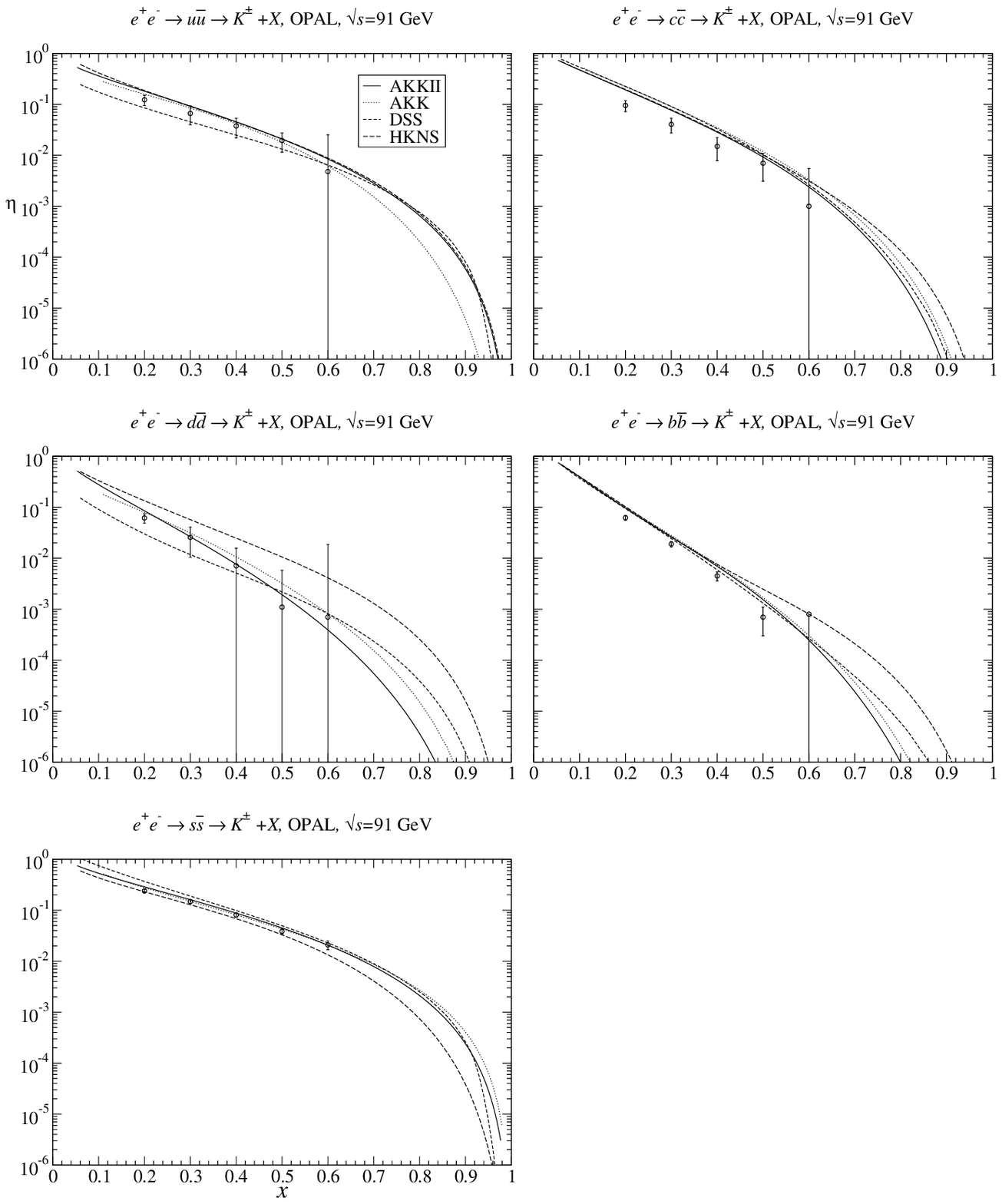}
\caption{OPAL quark tagging probabilities for $K^\pm$. \label{OPAL-eta_eta-Kaon_ffc}}
\end{center}
\end{figure}

\begin{figure}
\begin{center}
\includegraphics[width=17cm]{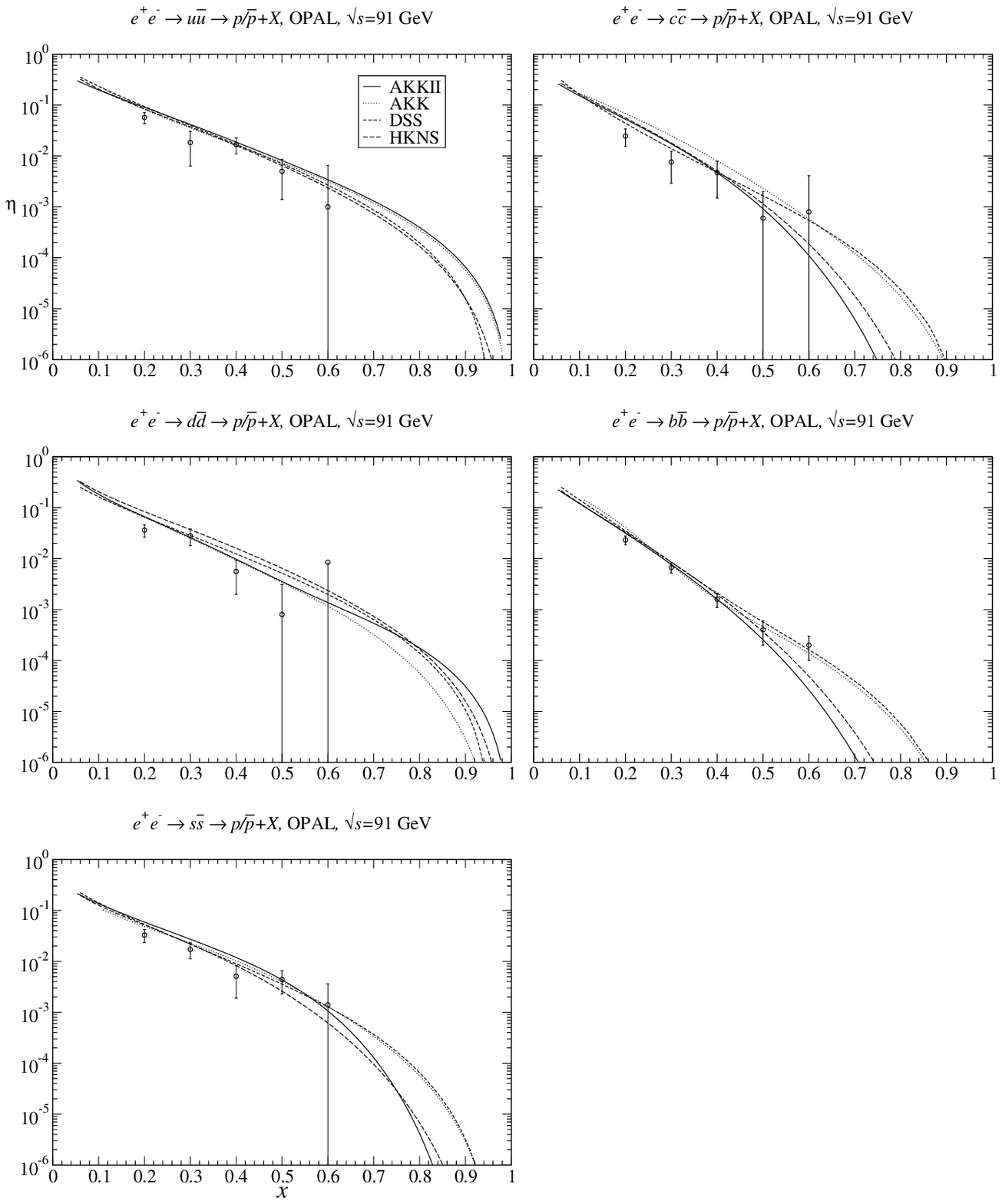}
\caption{OPAL quark tagging probabilities for $p/\overline{p}$. \label{OPAL-eta_eta-Proton_ffc}}
\end{center}
\end{figure}

\begin{figure}
\begin{center}
\includegraphics[width=17cm]{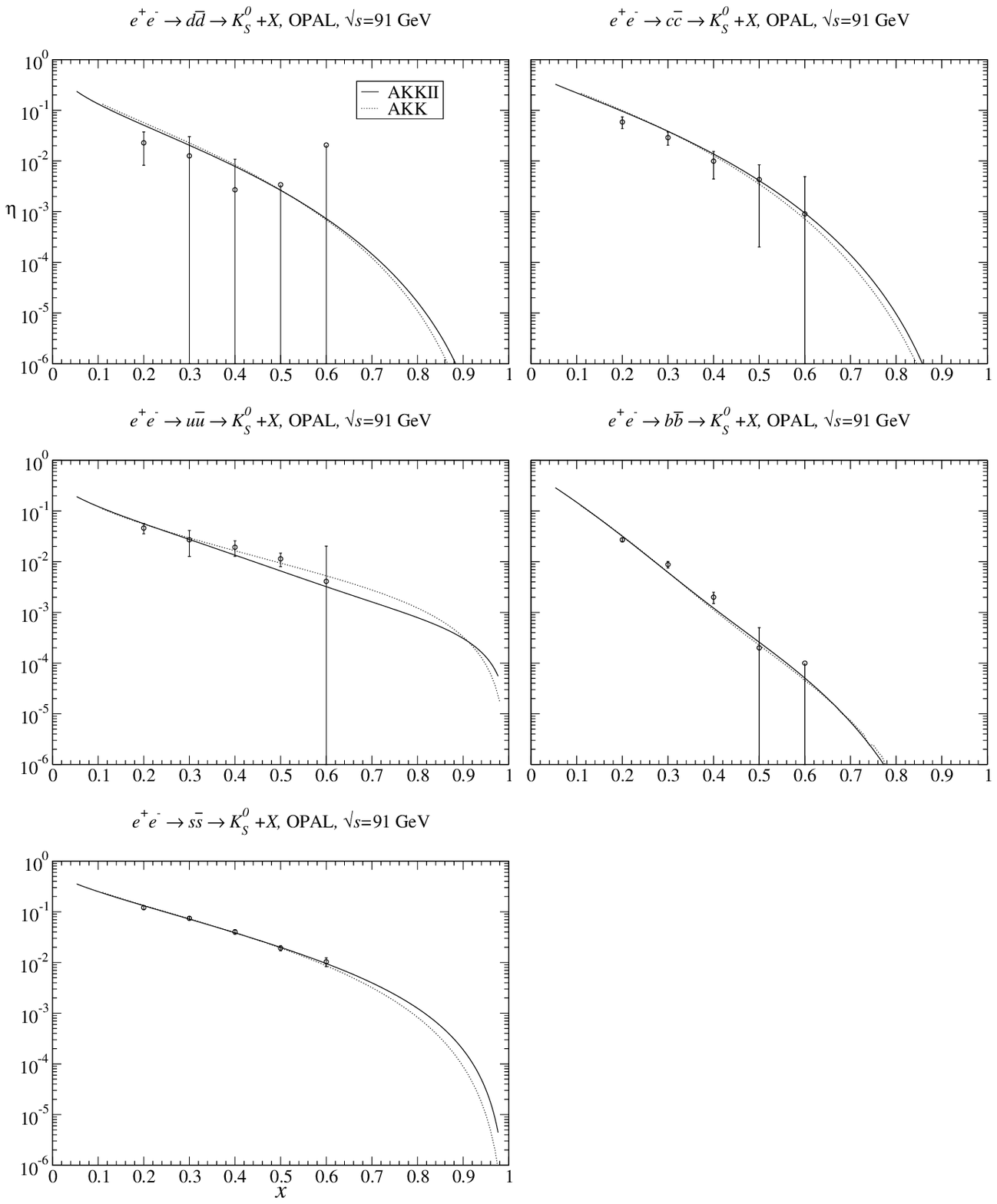}
\caption{OPAL quark tagging probabilities for $K^0_S$. \label{OPAL91-eta-K0S_ffc}}
\end{center}
\end{figure}

\begin{figure}
\begin{center}
\includegraphics[width=17cm]{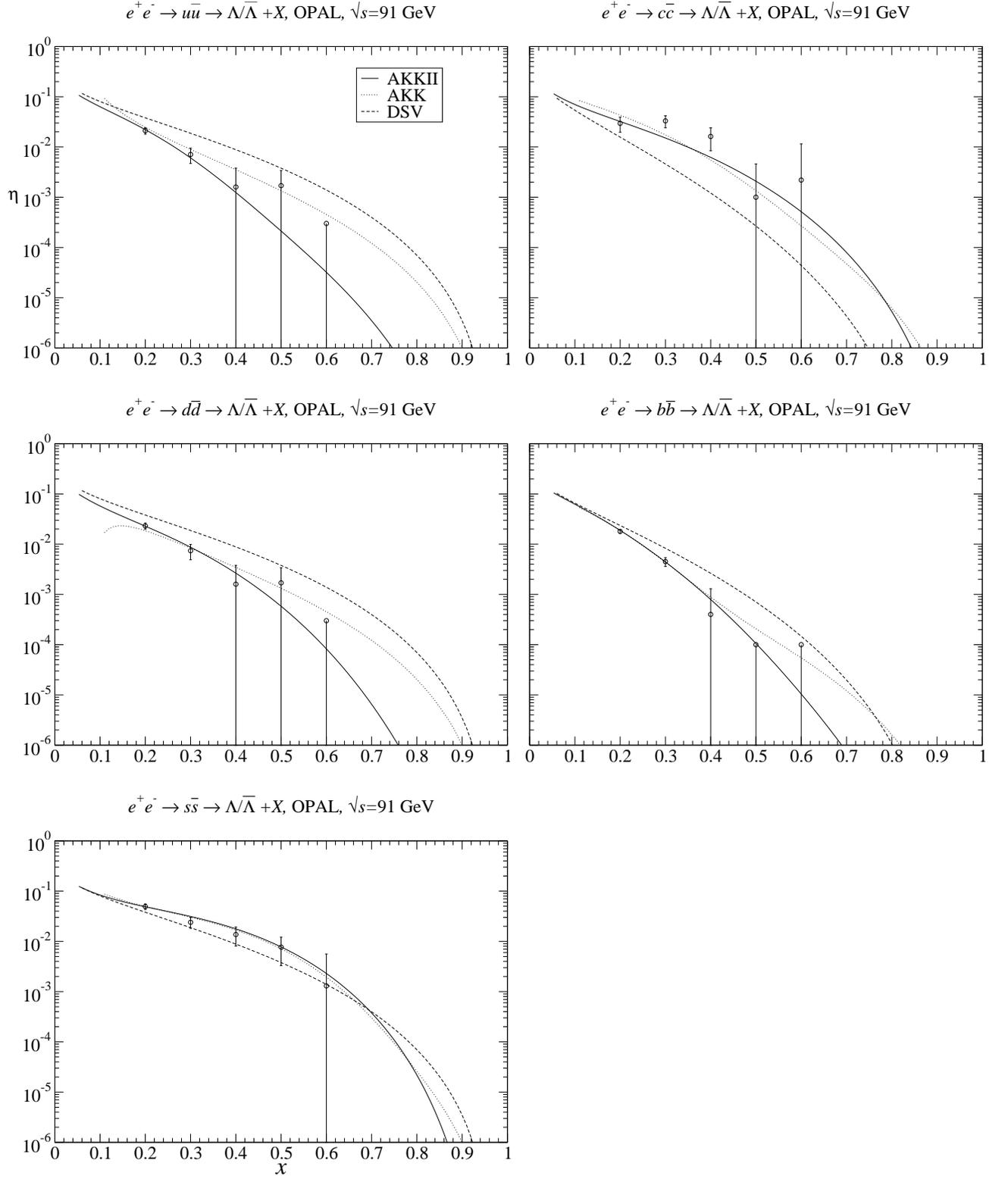}
\caption{OPAL quark tagging probabilities for $\Lambda/\overline{\Lambda}$. \label{OPAL91-eta-qqLambda_ffc}}
\end{center}
\end{figure}

\begin{figure}
\begin{center}
\includegraphics[width=17cm]{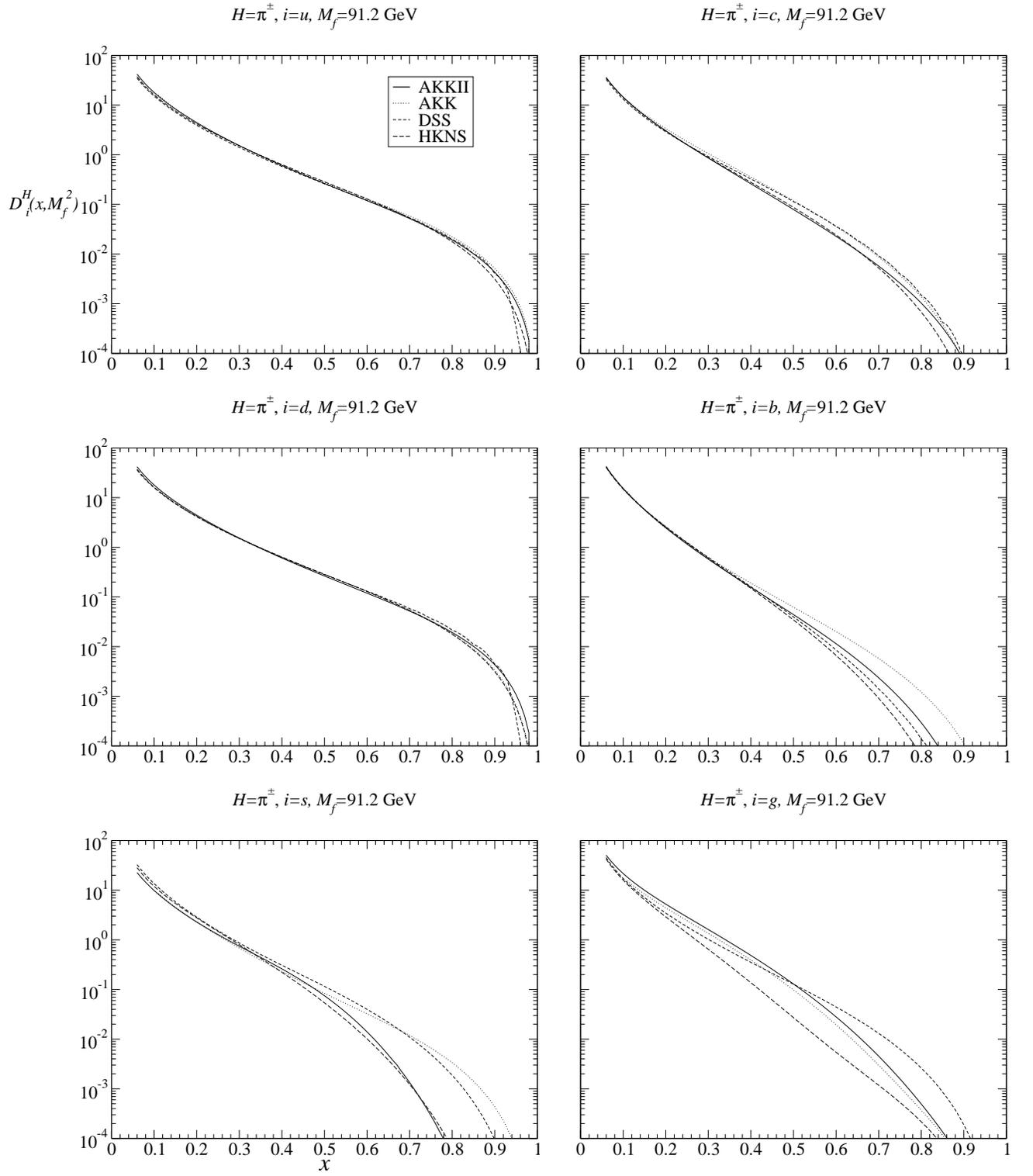}
\caption{The FFs for $\pi^\pm$ at $M_f=91.2$ GeV. \label{91Pion}}
\end{center}
\end{figure}

\begin{figure}
\begin{center}
\includegraphics[width=17cm]{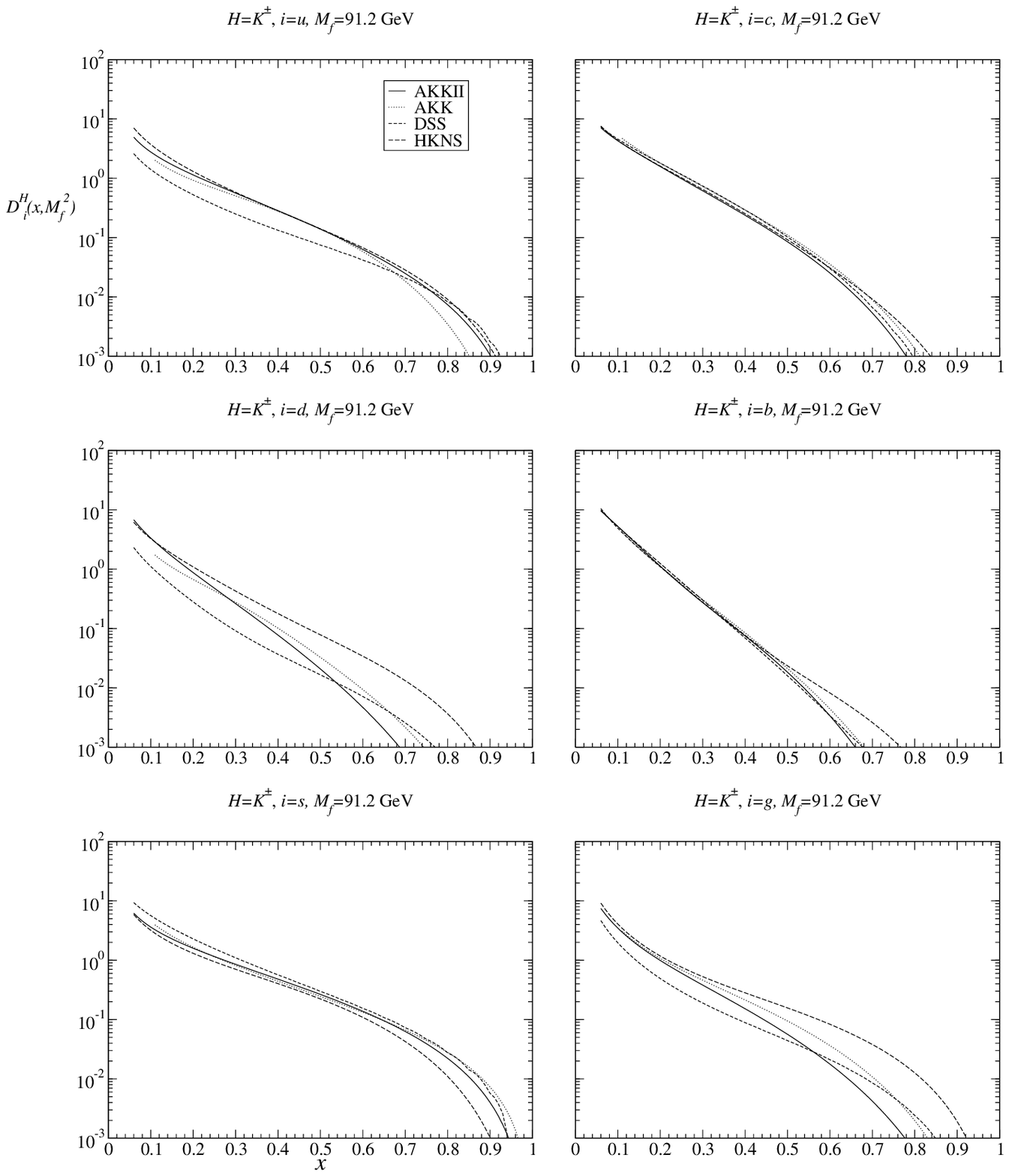}
\caption{The FFs for $K^\pm$ at $M_f=91.2$ GeV. \label{91Kaon}}
\end{center}
\end{figure}

\begin{figure}
\begin{center}
\includegraphics[width=17cm]{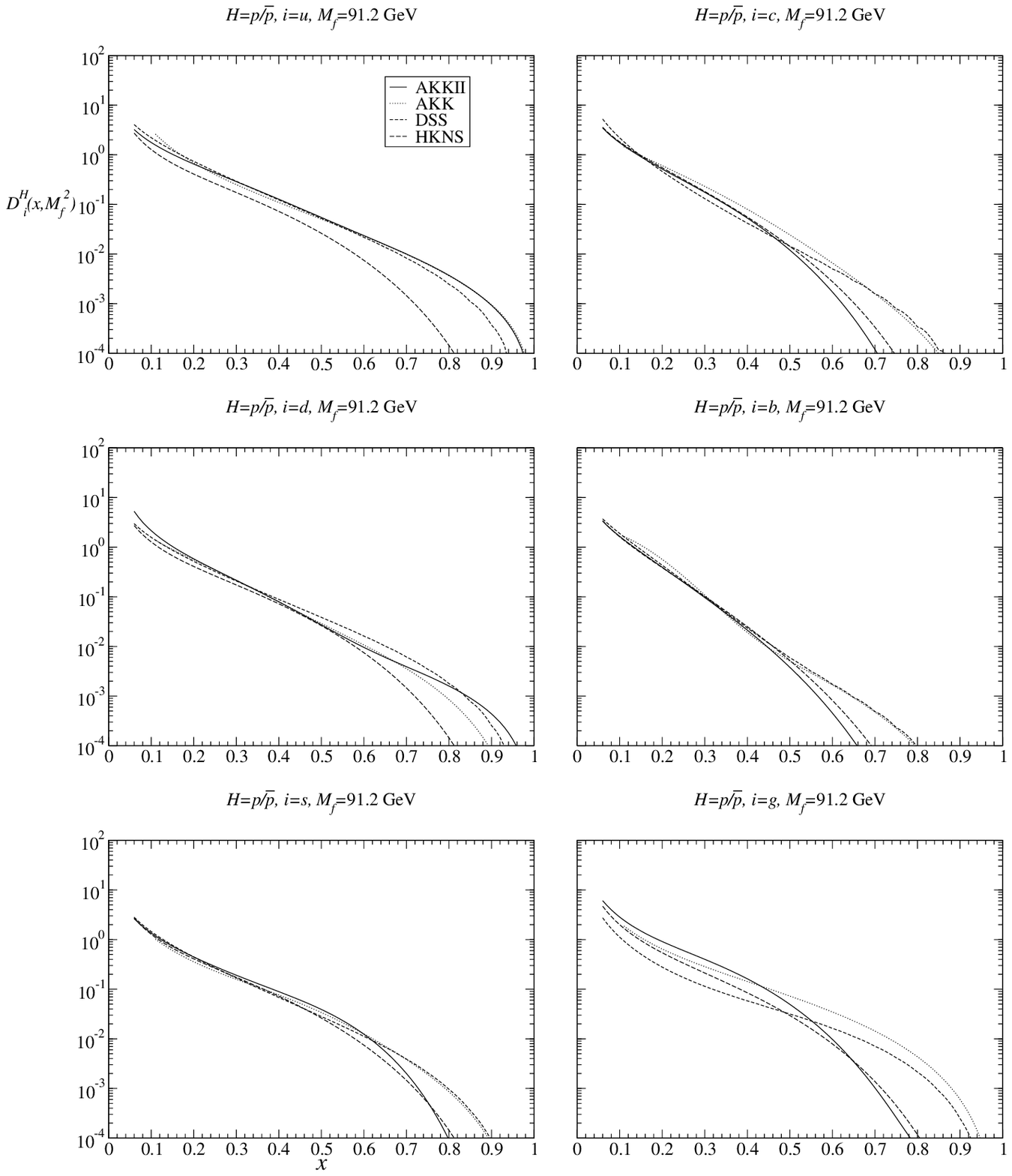}
\caption{The FFs for $p/\overline{p}$ at $M_f=91.2$ GeV. \label{91Proton}}
\end{center}
\end{figure}

\begin{figure}
\begin{center}
\includegraphics[width=17cm]{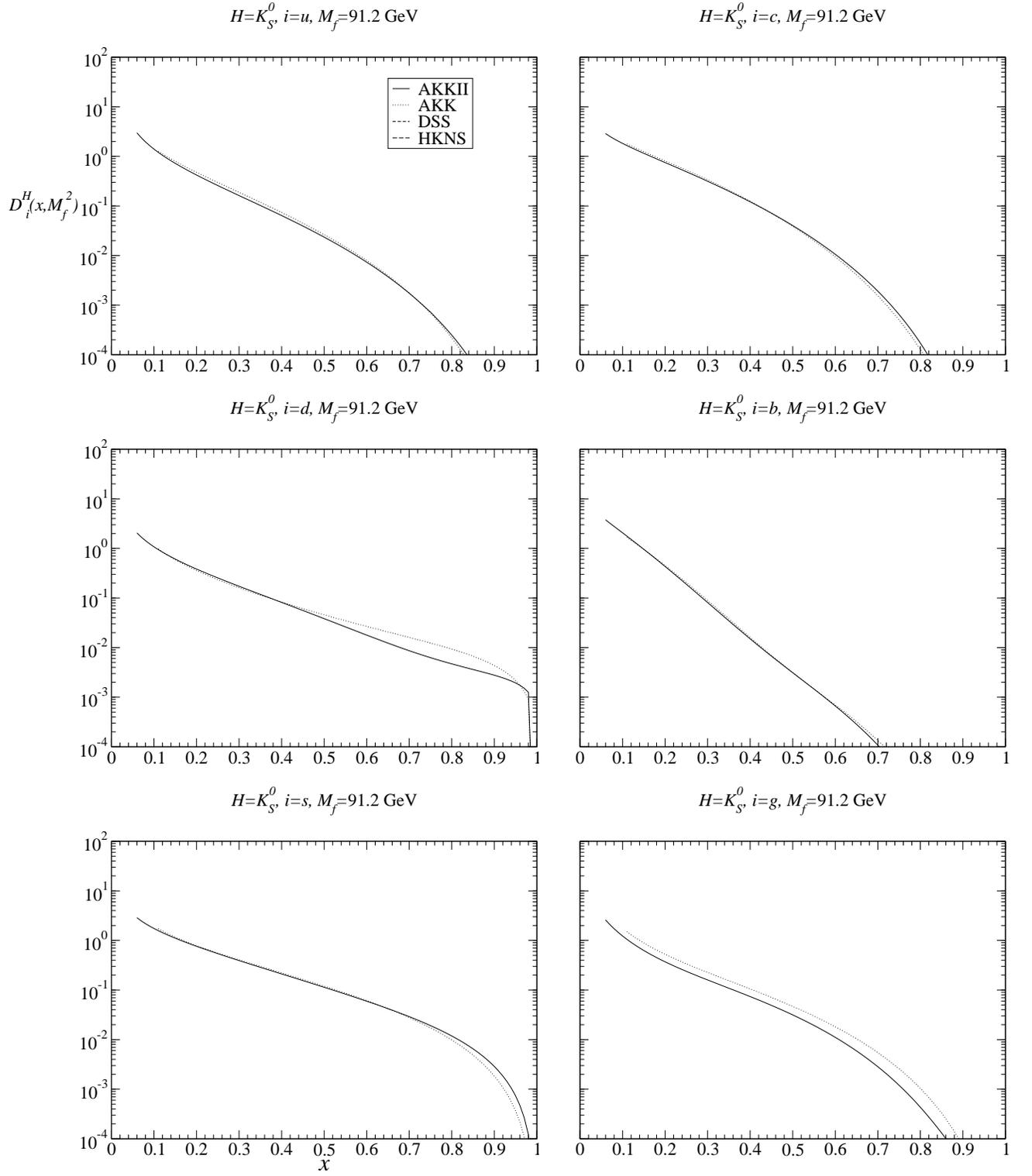}
\caption{The FFs for $K^0_S$ at $M_f=91.2$ GeV. \label{91K0S}}
\end{center}
\end{figure}

\begin{figure}
\begin{center}
\includegraphics[width=17cm]{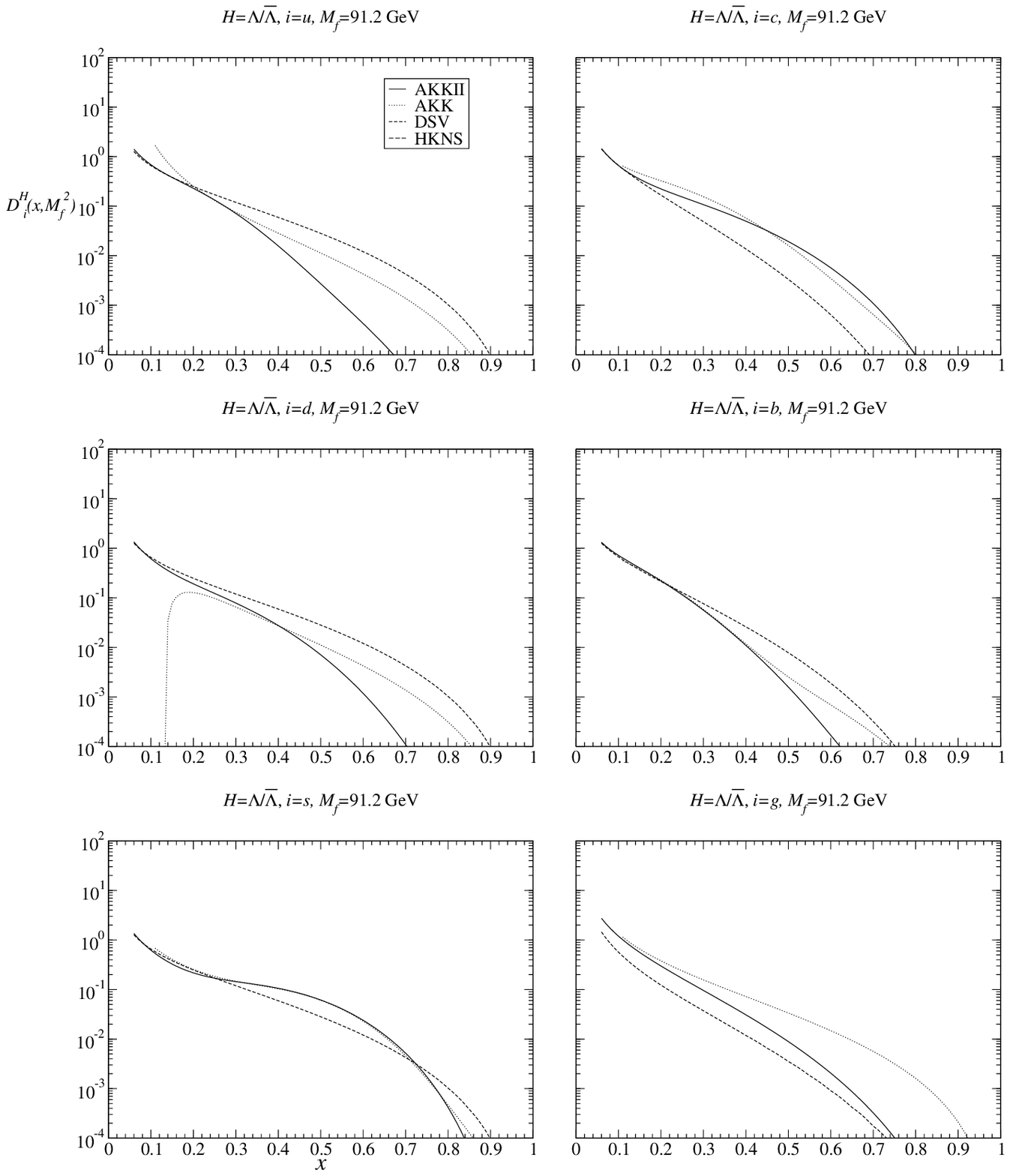}
\caption{The FFs for $\Lambda/\overline{\Lambda}$ at $M_f=91.2$ GeV. \label{91Lambda}}
\end{center}
\end{figure}

\begin{figure}
\begin{center}
\includegraphics[width=8.5cm]{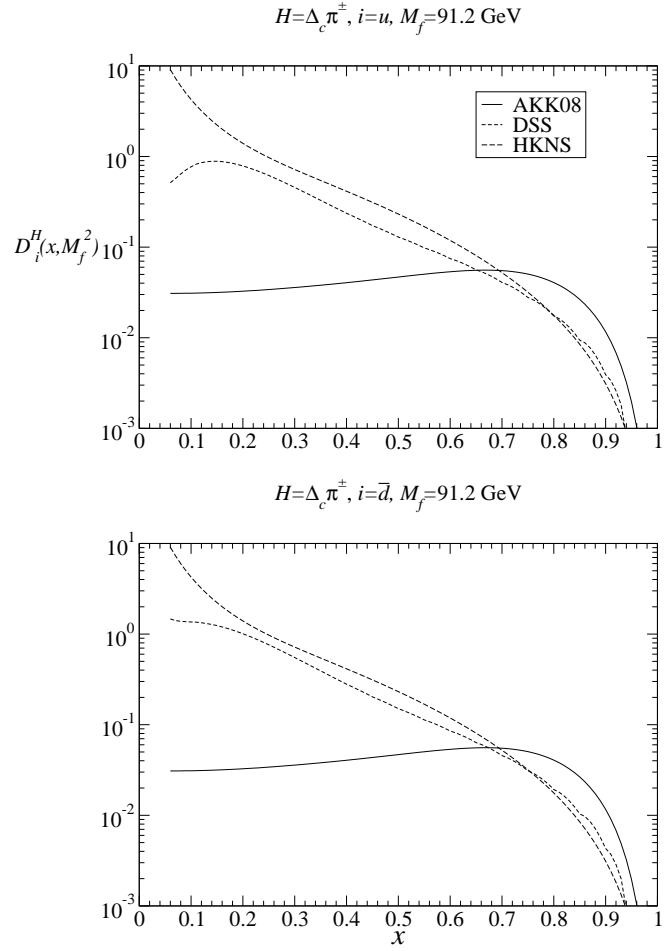}
\caption{The quark FFs for $\Delta_c \pi^\pm$ at $M_f=91.2$ GeV. \label{91PionVal}}
\end{center}
\end{figure}

\begin{figure}
\begin{center}
\includegraphics[angle=-90,width=8.7cm]{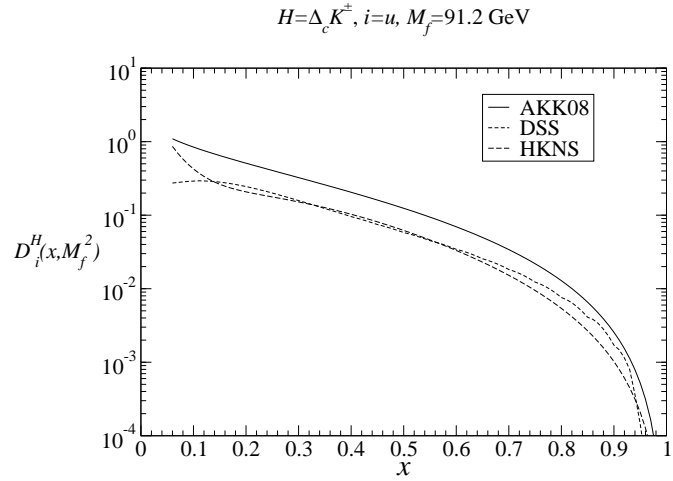}
\caption{The $u$ quark FF for $\Delta_c K^\pm$ at $M_f=91.2$ GeV. \label{up91KaonVal}}
\end{center}
\end{figure}

\begin{figure}[h!]
\begin{center}
\includegraphics[width=8.5cm]{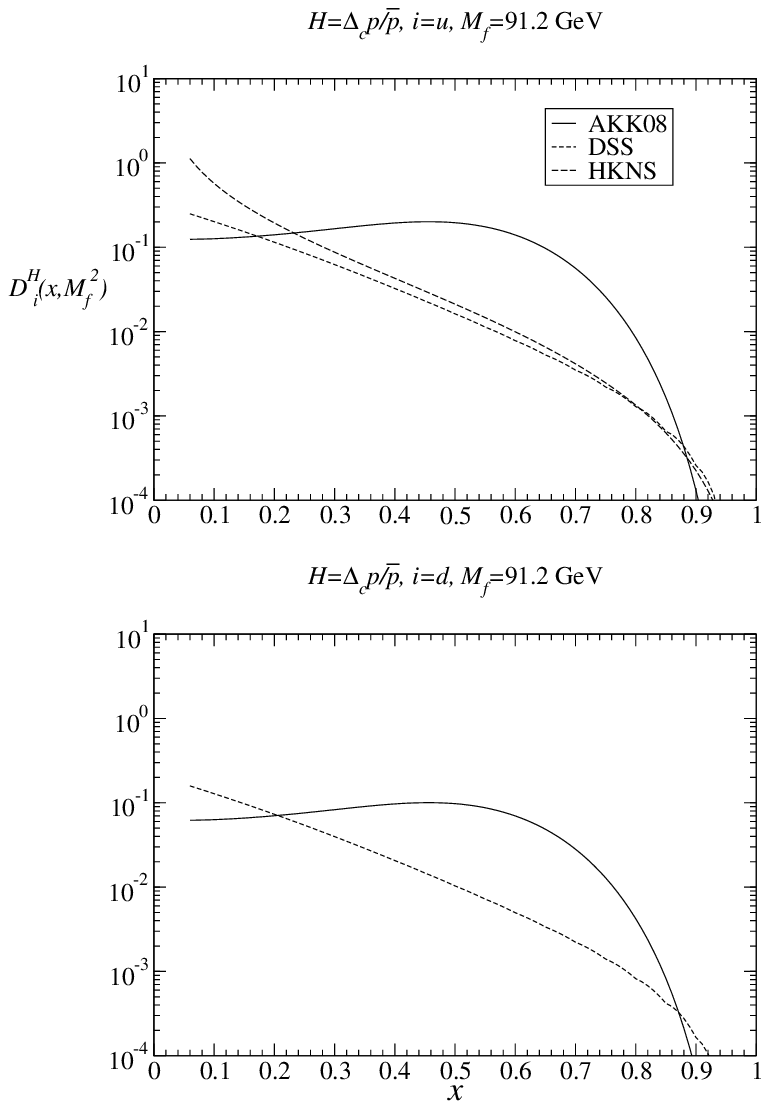}
\caption{The quark FFs for $\Delta_c p/\overline{p}$ at $M_f=91.2$ GeV. \label{91ProtonVal}}
\end{center}
\end{figure}

To illustrate the various features of the theoretical calculations, we include a number of figures.
First we study $pp(\overline{p})$ reactions:
For charge-sign unidentified hadrons,
we show the comparisons of the calculation with the experimental data in Fig.\ \ref{pp}, 
the effects of hadron mass and the relative theoretical errors in Figs.\ \ref{ppBRAHMS} and \ref{ppSTAR}, 
and the calculation using other FF sets in Figs.\ \ref{ppBRAHMS_ffc} and \ref{ppSTAR_ffc}.
For the charge-sign asymmetries, these features are respectively shown
in Fig.\ \ref{ppVal}, 
in Figs.\ \ref{ppValBRAHMS} and \ref{ppValSTAR} 
and in Figs.\ \ref{ppValBRAHMS_ffc} and \ref{ppValSTAR_ffc}.
Finally, the contributions to the production from the fragmentations of the sea and valence partons
of the initial protons at RHIC for charge-sign unidentified hadrons 
are shown in Figs.\ \ref{ppValenceRatioBRAHMS} and \ref{ppValenceRatioSTAR}
and for the charge-sign asymmetries in Figs.\ \ref{ppValenceRatioVal_BRAHMS} and \ref{ppValenceRatioVal_STAR}.
Next we study $e^+ e^-$ reactions:
We show the comparisons of the calculation with some of the experimental
data for a representative range of $\sqrt{s}$ in Fig.\ \ref{ee}, the 
effects of hadron mass, large $x$ resummation and the relative theoretical errors
in Figs.\ \ref{TASSO14_Z-qq} and \ref{OPAL_Z-qq},
and the calculation using other FF sets compared with 
data at $\sqrt{s}=91.2$ GeV, at $\sqrt{s}=14$ GeV and at various values of $\sqrt{s}$
in Fig.\ \ref{ee_ffc}, in Fig.\ \ref{TASSO14_Z-qq_ffc} and in Fig.\ \ref{qqProton_ffc} respectively.
Finally, the calculations using the various FF sets compared with the OPAL tagging probabilities are shown
in Figs.\ \ref{OPAL-eta_eta-Pion_ffc}--\ref{OPAL91-eta-qqLambda_ffc} .
Lastly, we show the various FFs at $\sqrt{s}=91.2$ GeV directly:
the charge-sign unidentified FFs in Figs.\ \ref{91Pion}--\ref{91Lambda}
and the charge-sign asymmetry FFs in Figs.\ \ref{91PionVal}--\ref{91ProtonVal}.

For inclusive $\pi^\pm$ production in $pp(\bar{p})$ collisions (Fig.\ \ref{pp}), the 
calculation agrees well with the PHENIX and STAR data. Although the calculation undershoots
the BRAHMS data for both rapidity ranges, this is not so 
serious when the systematic effects are taken into account since the magnitudes of the fitted $\lambda_K$
(see Table \ref{PionResults}) are less than 2, as discussed earlier.
This undershoot is more serious for the $p/\bar{p}$ fit.
The FF sets for each of $K^0_S$ and $K^\pm$ fit the data for both $K^0_S$ and $K^\pm$ well, 
which is consistent with SU(2) isospin invariance.
The calculation with the $\Lambda/\overline{\Lambda}$ FF set agrees reasonably well 
with the CDF data at the larger $p_T$ values only, and significantly undershoots the STAR data.
As expected, hadron mass effects are negligible in $\pi^\pm$ production data regardless of the
rapidity (Figs.\ \ref{ppBRAHMS} and \ref{ppSTAR}). A weak effect is seen in the slightly
heavier $K^\pm/K_S^0$ production data, which
does not depend on rapidity but which is largest at the smaller $p_T$ values as expected. The suppression due to
hadron mass effects is strongest for the heavier baryons $p/\bar{p}$ and $\Lambda/\overline{\Lambda}$, and is slightly 
larger at higher rapidity.
We also perform the calculations using the AKK \cite{Albino:2005me,Albino:2005mv}, DSS \cite{deFlorian:2007aj},
DSV \cite{deFlorian:1997zj} (for $\Lambda/\overline{\Lambda}$)
and HKNS \cite{Hirai:2007cx} FF sets (Figs.\ \ref{ppBRAHMS_ffc} and \ref{ppSTAR_ffc}). 
In these cases the hadron mass is set to zero, as was done in those fits.
It is expected that all FFs including the gluon FF for $\pi^\pm$ production are reasonably
well constrained since the $e^+ e^- \rightarrow \pi^\pm +X$ data are fairly precise
and lead to good agreement with $pp$ reaction data.
The differences in the calculations using different FF sets are then most likely due to theoretical 
errors from different choices for the various (NNLO) ambiguities
in the calculations, such as the calculation of $\alpha_s(\mu^2)$,
which explains why the calculations for $\pi^\pm$ at low and high rapidity 
are similar relative to the theoretical error,
and also why the calculation using the previous AKK set is close to the one using this paper's FF set.
However, for $K^\pm$ the HKNS calculation deviates from the rest at high rapidity.
The calculation with the previous AKK set for $p/\overline{p}$ 
at high rapidity deviates from the others but gives good agreement with the BRAHMS data.
However, the other FF sets agree better with the STAR data at lower rapidity.
For $K^0_S$, the description of the data is better with the fit of this paper than with the previous AKK one. 
However, for $\Lambda/\overline{\Lambda}$, the situation is the opposite, and in fact there is a strong 
discrepancy between theory and data when using this paper's and the DSV FF sets, implying
a possible inconsistency between the $pp$ and $e^+ e^-$ reaction data for $\Lambda/\overline{\Lambda}$ production. 
On the other hand, rather good agreement is obtained when using the previous AKK FF set, because
the gluon FF for $\Lambda/\overline{\Lambda}$ at the initial scale in 
Ref.\ \cite{Albino:2005mv} was fixed to 1/3 that for the AKK proton for this purpose.
Ultimately, a determination of the error on this prediction from the experimental errors,
including correlation effects, on the FFs would better determine whether an inconsistency really exists.

The BRAHMS data for which $3.25<y<3.35$ (Fig.\ \ref{ppVal}) provide most of the 
constraint on the charge-sign asymmetry FFs, 
while the constraints from the STAR data are rather poor.
The description of the $\Delta_c p/\bar{p}$ data from BRAHMS for which $3.25<y<3.35$ is particularly poor.
The $\Delta_c \pi^\pm$ data are much less precise 
than the $\pi^\pm$ data, which is due
to the similar yields of $\pi^+$ and $\pi^-$ relative to the experimental error. 
The $\Delta_c K^\pm$ and $\Delta_c p/\bar{p}$ data do not suffer this problem as much,
particularly the $\Delta_c p/\bar{p}$ data, since the yields for each charge-sign, 
particularly in the case of $p/\bar{p}$, are significantly different.
For the higher rapidity data at BRAHMS, the theoretical error is slightly 
lower for $\Delta_c \pi^\pm$ (Fig.\ \ref{ppValBRAHMS}) than for $\pi^\pm$ (Fig.\ \ref{ppBRAHMS}),
and lower to a greater degree for $\Delta_c K^\pm$ than for $K^\pm$, and 
for $\Delta_c p/\overline{p}$ than for $p/\overline{p}$,
which may result from some cancellation of the theoretical error
between the cross sections for each charge-sign, whose calculations are similar. 
However, the theoretical errors for $\Delta_c \pi^\pm$ (Fig.\ \ref{ppValSTAR}) and
$\pi^\pm$ (Fig.\ \ref{ppSTAR}) data at STAR are similar,
while the theoretical errors for $\Delta_c p/\overline{p}$ are larger than 
for $p/\overline{p}$, which may be due partly to 
(hidden) theoretical errors at low rapidity.
Note that, perhaps for the same reason, some cancellation of hadron mass dependence also occurs
for $\Delta_c p/\overline{p}$ relative to $p/\overline{p}$ at high and low rapidity.
For the higher rapidity data (Fig.\ \ref{ppValBRAHMS_ffc}) the calculations with this paper's FF sets are somewhat higher than
the others (at higher $p_T$ for $\Delta_c \pi^\pm$), presumably due to larger experimental errors in the data fitted to.
This occurs to a much greater degree for $\Delta_c p/\overline{p}$ at low rapidity,
probably because, in contrast to the other FF set fits, we impose additional constraints on the $\Delta_c p/\overline{p}$ FFs
with the BRAHMS data for which $2.9<y<3$, 
with which the calculations using the other FF sets disagree strongly.
The calculations for $\Delta_c \pi^\pm$ for STAR kinematics using this paper's FF set and the DSS FF sets are similar,
and somewhat different to the calculations using the HKNS FF set (Fig.\ \ref{ppValSTAR_ffc}).
Note that in this latter set the charge-sign asymmetry FFs were constrained using theoretical assumptions only. 

For $\pi^\pm$, the fact that charge-sign asymmetry is observed at BRAHMS and not at STAR is explained 
by the greater importance of the charge-sign 
asymmetric fragmentation from the initial protons' valence $u$ and $d$ quarks
(Figs.\ \ref{ppValenceRatioBRAHMS} and \ref{ppValenceRatioSTAR}).
The charge-sign asymmetry at STAR should become visible at sufficiently large $p_T$.
The excess of fragmentation from valence $u$ quarks over valence $d$ implies an excess of
$\pi^+$ over $\pi^-$, because then the first term in
Eq.\ (\ref{FfromFhatandDwithindicesforval}), which is positive, is larger in magnitude than the second,
which is negative (Figs.\ \ref{ppValenceRatioVal_BRAHMS} and \ref{ppValenceRatioVal_STAR}).
(If the magnitude of the ``$d+\bar{d}$''
contribution to the excess of $\pi^+$ over $\pi^-$ were larger than that of the ``$u+\bar{u}$'', this
excess would become negative). The lower fractional
contribution from the initial protons' valence quark fragmentations to the cross section for the STAR data
implies that this charge-sign asymmetry increases with rapidity.
As expected from the discussion following Eq.\ (\ref{FfromFhatandDwithindicesdecomp}), 
the contribution to $K^\pm$ production from fragmentation of valence $d$ quarks is negligible. 
While the production of $K^+$ exceeds that of $K^-$, at low rapidity 
the small contribution from valence quark fragmentation
suggests that any charge-sign asymmetry is difficult to observe.
For $p/\overline{p}$ at high and low rapidity,
valence $u$ quark fragmentation dominates over $d$, more so than for $\pi^\pm$ as expected.
It is clear that the excess of $p$ over $\bar{p}$ (relative to the $p/\bar{p}$ production)
increases with $p_T$ and rapidity.
At high rapidity, the valence $u$ and $d$ quarks contribute to the excess of $p$ over $\bar{p}$ as expected.
However, at low rapidity the contribution from the fragmentation of 
valence $d$ quarks to $\bar{p}$ exceeds that to $p$. This is inconsistent with our expectations.
Note that there may be larger hidden theoretical errors at lower rapidities. 
This was also the reason given earlier in this Section for the larger theoretical 
error for $\Delta_c p/\overline{p}$ than for $p/\overline{p}$. 
In addition, the propagated experimental errors on the FFs may still allow for this contribution
to the charge-sign asymmetry to have the opposite sign.
This is suggested by the fact that the cross section is much smaller at these lower rapidities.

We now turn to inclusive production in $e^+ e^-$ collisions.
Good agreement is found with the OPAL data at $\sqrt{s}=91.2$ GeV \cite{Akers:1994ez} for all particles 
(Fig.\ \ref{ee}), as indicated by the $\chi^2_{\rm DF}$ values
in Tables \ref{PionResults}, \ref{KaonResults}, \ref{K0SResults} and \ref{LambdaResults}, except for $p/\overline{p}$, 
which is indicated by the high $\chi^2_{\rm DF}$ value in Table \ref{ProtonResults}.
Reasonable agreement with the TASSO data at $\sqrt{s}=34$ GeV is found for all particles.
Note for example that, for these data, significant disagreement is found for $\pi^\pm$ at the two largest $x$ values, despite 
the value of $\lambda_K$ being positive (see Table \ref{PionResults}), because in fact the
theory would otherwise overshoot the rest of these data, which furthermore are more accurate.
For all particles the calculation tends to undershoot the
less accurate TASSO data at $\sqrt{s}=14$ GeV.
However, looking at $\pi^\pm$ for example, the value 
for $\lambda_K$ for these data is strongly negative and its magnitude is not unacceptably large.
A fitted mass for $p/\bar{p}$ results in a strong 
suppression of the cross section for $x\lesssim 0.5$ (Fig.\ \ref{TASSO14_Z-qq}), implying 
that hadron mass effects cannot be neglected in this region.
On the other hand, hadron mass effects for $\pi^\pm$ are negligible.
In general, the effect of large $x$ resummation is to enhance the cross section at large $x$.
However, note that there is some suppression at small $x$.
Formally, this suppression is less important than the effect of resummation of small $x$
divergences arising from soft gluons.
The theoretical error both with and without resummation is similar (on a logarithmic scale), although
it must be remembered that the same choice in both cases 
for the variation in $k$ might not be appropriate.
At large $\sqrt{s}$ (Fig.\ \ref{OPAL_Z-qq}),
the mass effects for $p/\bar{p}$ are negligible as expected.
Resummation effects are also reduced at higher $\sqrt{s}$.
At $\sqrt{s}=91.2$ GeV, all FF sets give similar results at $x\lesssim 0.4$
(Fig.\ \ref{ee_ffc}), but there are of course some differences at large $x$ where the experimental uncertainties are larger.
This paper's fit and the calculation with the HKNS FF set for $\pi^\pm$ 
prefer the SLD data around $x\simeq 0.6$, but,
as for the other FF sets, the FF set of this paper tries to fit the DELPHI, OPAL and SLD data at $x\simeq 0.8$.
The result for $K^\pm$ at large $x$ tries to fit both the OPAL
and SLD data points at $x\simeq 0.8$.
For $p/\overline{p}$, the OPAL data at all $x$ values are clearly inconsistent 
with the others, but this does not affect the good agreement with the other data. 
However, the calculation with this paper's FF set overshoots the others at values of $x$ above those for the data.
For $\Lambda/\overline{\Lambda}$, the calculations are all significantly different at large $x$.
Note that the DELPHI data are inconsistent with the others at small $x$.
The agreement between the calculations for $\pi^\pm$ at lower $\sqrt{s}=14$ GeV 
(Fig.\ \ref{TASSO14_Z-qq_ffc}) is similar to that at $\sqrt{s}=91.2$ GeV.
Note that data on inclusive $\pi^\pm$ production is generally much more accurate and abundant than for other particles.
At intermediate $x$, slightly more disagreement among the calculations exists for $K^\pm$ for $\sqrt{s}=14$ GeV 
than for $\sqrt{s}=91.2$ GeV.
Strong disagreement among the calculations at smaller $x$ and $\sqrt{s}=14$ GeV for 
$p/\overline{p}$ is found. 
However, the calculation with this paper's FF set describes the smaller $x$ HRS
and TPC data at $\sqrt{s}=29$ GeV better (Fig.\ \ref{qqProton_ffc}), and also the
TOPAZ data at $\sqrt{s}=58$ GeV where the calculation also ``points to'' the data for which $x<0.05$.
For $K_S^0$, both this paper's FF set and the previous AKK FF set give similar results (Fig.\ \ref{TASSO14_Z-qq_ffc}).
The calculation with this paper's FF set for $\Lambda/\overline{\Lambda}$ are significantly different
to the others at the smaller $x$ values, 
although the agreement here with the TASSO data at $\sqrt{s}=14$ GeV is better.

Since the OPAL tagging probabilities are physical, they give a better 
indication than the FFs themselves do of how well the fragmentation
from the individual quarks is currently understood. 
(Nevertheless, for completeness we will also study the FFs themselves next.)
For $\pi^\pm$ (Fig.\ \ref{OPAL-eta_eta-Pion_ffc}), $u$ and $d$ quark fragmentations are well
constrained, and are consistent with the corresponding OPAL tagging probabilities,
while $s$ and $b$ quark fragmentations are badly constrained at large $x$. 
The $e^+ e^-$ reaction data constrain the fragmentation of these quarks (and the $d$ quark) less
than the others because of their smaller coupling to the electroweak boson.
Since the corresponding tagging probabilities have large errors and
untagged or light quark tagged data cannot constrain the difference between $s$ and $d$ quark fragmentations, the only
other constraints, in the case of this paper's and the DSS FF sets, come from the $pp$ reaction data, 
which must therefore be rather poor for the $s$ quark fragmentation 
where the two calculations disagree markedly.
However, additional constraints for $c$ and $b$ quark fragmentation are provided by light, $c$ and $b$
quark tagged cross section measurements from DELPHI, SLD and TPC (see Table \ref{PionResults}).
The calculations overshoot the measured tagging probabilities at the lower $x$ values, 
particularly the heavy quark ones, which may be due to
inconsistencies between the theoretical and experimental definitions
in this region as discussed in Ref.\ \cite{Albino:2005me}.
For $K^\pm$ (Fig.\ \ref{OPAL-eta_eta-Kaon_ffc}), $d$ quark fragmentation is poorly constrained, 
and unfortunately little more constraint is provided by the tagging probabilities. 
However, the other quark fragmentations are better constrained since 
their calculations using the different FF sets yield similar results.
Note that for both $\pi^\pm$ and $K^\pm$, the favoured $u$ quark fragmentation is rather well
constrained, while the unfavoured light quark fragmentation ($s$ and $d$ respectively) is not.
For $p/\overline{p}$ (Fig.\ \ref{OPAL-eta_eta-Proton_ffc}),
more discrepancies are found at large $x$, particularly for the heavy quarks although here the
calculations with this paper's and the DSS FF sets are relatively less separated.
Fragmentation to $p/\overline{p}$ from $u$ quarks is fairly well understood, and also from the 
other quarks except at the large $x$ values.
For $K^0_S$ (Fig.\ \ref{OPAL91-eta-K0S_ffc}),
the calculations using this paper's and the previous AKK FF sets are similar.
Good agreement is obtained with the (most accurate) $s$ and $b$ 
measured tagging probabilities, even at the smallest $x$.
For $\Lambda/\overline{\Lambda}$ (Fig.\ \ref{OPAL91-eta-qqLambda_ffc}), 
as for $K^0_S$ for all quarks, the calculation with this paper's FF set 
is close to that using the previous AKK FF set for $s$ and $c$ quark tagging, but otherwise
these and the calculation with the DSV FF set are quite different.

Finally, we compare the FFs at $M_f=91.2$ GeV (Figs.\ \ref{91Pion}--\ref{91Lambda}).
In general, the relative behaviour of the FFs at lower $M_f$ values is qualitatively similar.
For each parton,
this paper's FFs are typically lower than the others because the large $x$ resummation enhances the cross section.
Fragmentations from the gluon are poorly constrained, even those from fits where $pp$ reaction data is used 
(comparing this paper's gluon FFs with the corresponding DSS ones where applicable), 
because although these data depend more strongly on the gluon FF than the $e^+ e^-$ reaction data,
the former data are much more limited than the latter.
For $\pi^\pm$, $K^\pm$ and $K_S^0$, the relative behaviour of the quark FFs for each 
flavour are similar to their tagging probability counterparts discussed above. 
However, excluding HKNS, for $p/\overline{p}$ there is better agreement 
at large $x$ among the $u$ and $d$ quark FFs than among their tagging probabilities, the uncertainty in the latter case
presumably coming from the contributions of the other quark FFs. 
For $\Lambda/\overline{\Lambda}$, the FFs are rather similar at the lower $x$ values, but,
apart from the agreement between this paper's and the previous AKK FFs for the $s$ and $c$ quark at large $x$,
are otherwise markedly different.

For $\Delta_c \pi^\pm$, the similarity between the DSS and HKNS results for the $u$
and $\overline{d}$ (Fig.\ \ref{91PionVal}) quark FFs and 
the $u$ quark FF for $\Delta_c K^\pm$ (Fig.\ \ref{up91KaonVal}) 
and for $\Delta_c p/\overline{p}$ (Fig.\ \ref{91ProtonVal}) is to be expected since similar non perturbative assumptions 
were made for these fits while for the fits of this paper we have let the data decide.
For $\Delta_c \pi^\pm$, this paper's quark FFs are typically much smaller than the other FFs,
while the situation is the opposite for $K^\pm$ and even more so for $\Delta_c p/\overline{p}$.

\section{Conclusions}
\label{conc}

We have determined charge-sign unidentified FFs for $\pi^\pm$, $K^\pm$, $p/\overline{p}$, $K_S^0$ and $\Lambda/\overline{\Lambda}$, 
and FFs for the charge-sign asymmetries $\Delta_c \pi^\pm$, $\Delta_c K^\pm$ and $\Delta_c p/\overline{p}$
from $e^+ e^-$ and $pp(\overline{p})$ reaction data.
Relative to the previous AKK fits \cite{Albino:2005me,Albino:2005mv}, we added $e^+ e^-$ reaction data 
at smaller $x$ and lower $\sqrt{s}$, which provides stronger constraints on the gluon fragmentation
that enters at NLO.
To account for deviations from the FO calculation in this region, we included hadron mass effects in the calculation,
with the mass in the calculation for the $e^+ e^-$ reaction data left as a free parameter in the fit. 
For the baryons $p/\overline{p}$ and $\Lambda/\overline{\Lambda}$ we obtained excellent agreement with the true masses.
Therefore these particles are probably produced almost exclusively from direct partonic fragmentation, and therefore
may be key to our understanding of the partonic fragmentation process.
This is then a strong reason for the inconsistencies between the descriptions of $e^+ e^-$ and $pp$ reactions 
in baryon production to be resolved.
In particular, the description of the STAR data for $\Lambda/\overline{\Lambda}$ is poor,
while the contribution from the initial protons' valence $d$ quarks to the 
charge-sign asymmetry for $p/\overline{p}$ from STAR is negative.
However, good agreement with the BRAHMS data for $p/\overline{p}$ was found in general.
For the mesons $\pi^\pm$, $K^\pm$ and $K_S^0$ we obtain the correct order of magnitude.
The overshoot (undershoot) in the case of $\pi^\pm$ ($K^\pm$ and $K_S^0$) suggests that
a significant portion of the sample originates from the decay of heavier particles (complicated decay channels).
We implemented large $x$ resummation in $e^+ e^-$ reactions and in the DGLAP
evolution of the FFs, which for most particles leads to a significant improvement in the fit.
We included RHIC $pp$ reaction data for all particles and Tevatron $p\overline{p}$ reaction data from the CDF collaboration for
$K_S^0$ and $\Lambda/\overline{\Lambda}$ to improve the constraints on the gluon fragmentation, the quark flavour separation
and also, in the case of the RHIC data, to determine the charge-sign asymmetry FFs. 
Hadron mass effects were included in the calculation of these data as well. 
For both $e^+ e^-$ and $pp(\overline{p})$ reaction data, the normalization errors were accounted for 
by including them in the correlation matrix used for the calculation of $\chi^2$.
The corresponding $\lambda_K$ values were determined, and were typically in the
reasonable range $|\lambda_K|\lesssim 2$.
In order to prevent the large errors of the statistically lower quality charge-sign asymmetry data 
propagating to all FFs, we fitted the valence quark FFs on which they
depend separately from the other FF degrees of freedom. 
This avoids the need to make
unreliable non perturbative assumptions which will ultimately affect all fitted FFs in unclear ways.

In order to make predictions, our FFs sets over the fitted range $0.05<z<1$ and $M_0^2<M_f^2<100000$ GeV$^2$
can be obtained from the FORTRAN routines at \verb!http://www.desy.de/~simon/AKK.html!, 
which are calculated using cubic spline interpolation on a linear grid in $z$ 
and linear interpolation on a linear grid in $\ln M_f^2$. 
We note a number of crucial points relevant to the use of our FF sets.
Firstly, it is neither incorrect nor even inconsistent to use NLO FF sets such 
as ours in a LO calculation. The result will simply be to LO only.
Secondly, while our mass corrections are desirable, it is not necessary to include them in
calculations using our FF sets, which would then
be at least as accurate as cross sections calculated using other FF sets that were
extracted without the use of mass corrections.
Finally, the FF sets here were fitted at a low factorization scale $M_0=\sqrt{2}$ GeV, where the N$^2$LO
error on our NLO evolution may be quite sizable. It is therefore incorrect to evolve our FFs using
a different NLO evolution procedure (e.g.\ one in which the DGLAP equation is solved numerically).
Rather, the FFs supplied at the web site just given should be used.

\begin{acknowledgments}

The authors would like to thank Mark Heinz and Bedanga Mohanty for
the numerical values for the
measurements of $\pi^+$, $\pi^-$, $p$, $\overline{p}$, $K^0_S$ and $\Lambda/\overline{\Lambda}$ production in $pp$ collisions from
the STAR collaboration, and Ramiro Debbe for the numerical values for the
measurements of $\pi^+$, $\pi^-$, $K^+$, $K^-$, $p$ and $\overline{p}$ from the BRAHMS collaboration.
This work was supported in part by the Deutsche Forschungsgemeinschaft
through Grant No.\ KN~365/5-1 and by the Bundesministerium f\"ur Bildung und
Forschung through Grant No.\ 05~HT4GUA/4.

\end{acknowledgments}


\appendix
\section{Correlated errors}
\label{correrr}

In this appendix we assume symmetric errors.
In the absence of systematic effects,
the probability for observables $f_i$ to take values between $f_i^t$ and $f_i^t+df_i^t$, given
measured values $f_i^e\pm \sigma_i$, is proportional to $\exp[-\chi^2/2] df_i^t$, where
\be
\chi^2=\sum_i \left(\frac{f_i^t-f_i^e}{\sigma_i}\right)^2.
\ee
The $K$th source of systematic uncertainty will cause $f_i^e$ to be shifted to
$f_i^e+\lambda_K \sigma_i^K$, where the probability density in $\lambda_K$ is proportional to $\exp[-\lambda_K^2/2]$,
which defines each systematic uncertainty $\sigma_i^K$. Therefore, systematic effects modify $\chi^2$ to
\be
\chi^2=\sum_i \left(\frac{f_i^t-f_i^e-\sum_K \lambda_K \sigma_i^K}{\sigma_i}\right)^2+\sum_K \lambda_K^2.
\label{chi2intermsoflambdak}
\ee
The most likely values of the $\lambda_K$ occur at
\be
\frac{\partial \chi^2}{\partial \lambda_K}=0.
\ee
Solving these equations for the $\lambda_K$ gives
\be
\lambda_K=\sum_i \frac{f_i^t-f_i^e}{\sigma_i^2}
\left(\sigma_i^K-\sum_{jkL}\sigma_i^L\sigma_j^L \left(C^{-1}\right)_{jk}\sigma_k^K\right),
\ee
where the covariance matrix 
\be
C_{ij}=\sigma_i^2 \delta_{ij}+\sum_K \sigma_i^K \sigma_j^K,
\ee
and Eq.\ (\ref{chi2intermsoflambdak}) becomes
\be
\chi^2=\sum_{ij} (f_i^t-f_i^e)\left(C^{-1}\right)_{ij} (f_j^t-f_j^e).
\ee



\begin{table}
\caption{Summary of the fit for $\pi^\pm$ production. Details of the data used are shown to the left of the vertical
double line,
fit results to the right. The columns labeled ``Norm.'' and ``Shift'' respectively give the experimental normalization and
``fitted'' (i.e.\ using the fitted $\lambda_K$) normalization error(s) and as a percentage. In the column labeled
``Properties'', ``$l$ tagged'' means the light quarks are tagged. The BRAHMS data have
5 sources of normalization error, whose values are shown for $p_T$ above (below) 3 GeV.
The normalization error
of $11.7\%$ on the STAR data follows from the total cross section measurement $\sigma =30 \pm 3.5$mb.
\label{PionResults}}
\begin{center}
\begin{tabular}{|c|c|c|c|c||c|c|c|}
\hline 
\multirow{2}{*}{Collaboration} & \multirow{2}{*}{Properties}
& \multirow{2}{*}{\vspace{0.3cm} $\sqrt{s}$} & \multirow{2}{*}{\vspace{0.3cm} \#} & \multirow{2}{*}{\vspace{0.3cm} Norm.\ } & 
\multirow{2}{*}{$\chi^2_{\rm DF}$} & \multirow{2}{*}{$\lambda_K$} & \multirow{2}{*}{\vspace{0.3cm} Shift} \\
& & (GeV) & data & (\%) & & & (\%) \\
\hline \hline
\input{figs_PionTable}
\hline
\end{tabular}
\end{center}
\end{table}

\begin{table}
\caption{As in Table \ref{PionResults}, but for the charge-sign asymmetry $\Delta_c \pi^\pm$.
\label{chargedpionvalresults}}
\begin{center}
\begin{tabular}{|c|c|c|c|c||c|c|c|}
\hline 
\multirow{2}{*}{Collaboration} & \multirow{2}{*}{Properties}
& \multirow{2}{*}{\vspace{0.3cm} $\sqrt{s}$} & \multirow{2}{*}{\vspace{0.3cm} \#} & \multirow{2}{*}{\vspace{0.3cm} Norm.\ } & 
\multirow{2}{*}{$\chi^2_{\rm DF}$} & \multirow{2}{*}{$\lambda_K$} & \multirow{2}{*}{\vspace{0.3cm} Shift} \\
& & (GeV) & data & (\%) & & & (\%) \\
\hline \hline
\input{figs_PionValTable}
\hline
\end{tabular}
\end{center}
\end{table}

\begin{table}
\caption{As in Table \ref{PionResults}, but for $K^\pm$ production.
\label{KaonResults}}
\begin{center}
\begin{tabular}{|c|c|c|c|c||c|c|c|}
\hline
\multirow{2}{*}{Collaboration} & \multirow{2}{*}{Properties}
& \multirow{2}{*}{\vspace{0.3cm} $\sqrt{s}$} & \multirow{2}{*}{\vspace{0.3cm} \#} & \multirow{2}{*}{\vspace{0.3cm} Norm.\ } & 
\multirow{2}{*}{$\chi^2_{\rm DF}$} & \multirow{2}{*}{$\lambda_K$} & \multirow{2}{*}{\vspace{0.3cm} Shift} \\
& & (GeV) & data & (\%) & & & (\%) \\
\hline \hline 
\input{figs_KaonTable}
\hline
\end{tabular}
\end{center}
\end{table}

\begin{table}
\caption{As in Table \ref{KaonResults}, but for the charge-sign asymmetry $\Delta_c K^\pm$.
\label{chargedkaonvalresults}}
\begin{center}
\begin{tabular}{|c|c|c|c|c||c|c|c|}
\hline 
\multirow{2}{*}{Collaboration} & \multirow{2}{*}{Properties}
& \multirow{2}{*}{\vspace{0.3cm} $\sqrt{s}$} & \multirow{2}{*}{\vspace{0.3cm} \#} & \multirow{2}{*}{\vspace{0.3cm} Norm.\ } & 
\multirow{2}{*}{$\chi^2_{\rm DF}$} & \multirow{2}{*}{$\lambda_K$} & \multirow{2}{*}{\vspace{0.3cm} Shift} \\
& & (GeV) & data & (\%) & & & (\%) \\
\hline \hline
\input{figs_KaonValTable}
\hline
\end{tabular}
\end{center}
\end{table}

\begin{table}
\caption{As in Table \ref{PionResults}, but for the fit for $p/\bar{p}$ production.
\label{ProtonResults}}
\begin{center}
\begin{tabular}{|c|c|c|c|c||c|c|c|}
\hline
\multirow{2}{*}{Collaboration} & \multirow{2}{*}{Properties}
& \multirow{2}{*}{\vspace{0.3cm} $\sqrt{s}$} & \multirow{2}{*}{\vspace{0.3cm} \#} & \multirow{2}{*}{\vspace{0.3cm} Norm.\ } & 
\multirow{2}{*}{$\chi^2_{\rm DF}$} & \multirow{2}{*}{$\lambda_K$} & \multirow{2}{*}{\vspace{0.3cm} Shift} \\
& & (GeV) & data & (\%) & & & (\%) \\
\hline \hline
\input{figs_ProtonTable}
\hline
\end{tabular}
\end{center}
\end{table}

\begin{table}
\caption{As in Table \ref{ProtonResults}, but for the charge-sign asymmetry $\Delta_c p/\bar{p}$.
\label{chargedprotonvalresults}}
\begin{center}
\begin{tabular}{|c|c|c|c|c||c|c|c|}
\hline 
\multirow{2}{*}{Collaboration} & \multirow{2}{*}{Properties}
& \multirow{2}{*}{\vspace{0.3cm} $\sqrt{s}$} & \multirow{2}{*}{\vspace{0.3cm} \#} & \multirow{2}{*}{\vspace{0.3cm} Norm.\ } & 
\multirow{2}{*}{$\chi^2_{\rm DF}$} & \multirow{2}{*}{$\lambda_K$} & \multirow{2}{*}{\vspace{0.3cm} Shift} \\
& & (GeV) & data & (\%) & & & (\%) \\
\hline \hline
\input{figs_ProtonValTable}
\hline
\end{tabular}
\end{center}
\end{table}

\begin{table}
\caption{As in Table \ref{PionResults}, but for $K_S^0$ production.
The CDF data are normalized by the unknown total cross section, which is therefore fitted to obtain the result $\sigma=13.7$ mb.
\label{K0SResults}}
\begin{center}
\begin{tabular}{|c|c|c|c|c||c|c|c|}
\hline
\multirow{2}{*}{Collaboration} & \multirow{2}{*}{Properties}
& \multirow{2}{*}{\vspace{0.3cm} $\sqrt{s}$} & \multirow{2}{*}{\vspace{0.3cm} \#} & \multirow{2}{*}{\vspace{0.3cm} Norm.\ } & 
\multirow{2}{*}{$\chi^2_{\rm DF}$} & \multirow{2}{*}{$\lambda_K$} & \multirow{2}{*}{\vspace{0.3cm} Shift} \\
& & (GeV) & data & (\%) & & & (\%) \\
\hline \hline
\input{figs_K0STable}
\hline
\end{tabular}
\end{center}
\end{table}

\begin{table}
\caption{As in Table \ref{PionResults}, but for $\Lambda/\overline{\Lambda}$ production.
The CDF data are normalized by the unknown total cross section, which is therefore fitted to obtain the result $\sigma=4.5$ mb.
\label{LambdaResults}}
\begin{center}
\begin{tabular}{|c|c|c|c|c||c|c|c|}
\hline
\multirow{2}{*}{Collaboration} & \multirow{2}{*}{Properties}
& \multirow{2}{*}{\vspace{0.3cm} $\sqrt{s}$} & \multirow{2}{*}{\vspace{0.3cm} \#} & \multirow{2}{*}{\vspace{0.3cm} Norm.\ } & 
\multirow{2}{*}{$\chi^2_{\rm DF}$} & \multirow{2}{*}{$\lambda_K$} & \multirow{2}{*}{\vspace{0.3cm} Shift} \\
& & (GeV) & data & (\%) & & & (\%) \\
\hline \hline
\input{figs_LambdaTable}
\hline
\end{tabular}
\end{center}
\end{table}

\clearpage
\clearpage
\clearpage

\end{document}

%% file: figs_chi2summary
$\pi^\pm$ &  518.7 &  519.0 \\
\hline
$K^\pm$ &  416.6 &  439.4 \\
\hline
$p/\bar{p}$ &  525.2 &  538.0 \\
\hline
$K^0_S$ &  317.2 &  318.7 \\
\hline
$\Lambda/\overline{\Lambda}$ &  273.1 &  325.7 \\
\hline

%% file: figs_hadmasssummary
$\pi^\pm$ &  154.6 &  139.6 \\
\hline
$K^\pm$ &  337.0 &  493.7 \\
\hline
$p/\bar{p}$ &  948.8 &  938.3 \\
\hline
$K^0_S$ &  343.0 &  497.6 \\
\hline
$\Lambda/\overline{\Lambda}$ & 1127.0 & 1115.7 \\
\hline

%% file: figs_paramtable
$N_g$ &    247.80 &     16.11 &  16155.68 &      1.64 &     26.92 \\ \hline
$a_g$ &      1.93 &      2.13 &      7.26 &      0.84 &      4.49 \\ \hline
$b_g$ &      6.14 &      3.28 &      9.07 &      4.11 &      5.18 \\ \hline
$c_g$ &      0.96 &      0.78 &      2.04 &      1.18 &      3.58 \\ \hline
$d_g$ &     -0.53 &      2.26 &     -0.43 &     -0.07 &     -1.31 \\ \hline \hline
$N_u$ &      0.32 &      1.66 &      0.49 &   3781.89 &      0.60 \\ \hline
$a_u$ &     -2.07 &      0.22 &     -0.05 &      4.68 &     -0.27 \\ \hline
$b_u$ &      0.96 &      3.55 &      1.84 &     16.79 &      2.25 \\ \hline
$c_u$ &     -0.81 &      0.50 &     -0.24 &      2.34 &      0.01 \\ \hline
$d_u$ &      2.91 &     -1.74 &     -0.01 &     -0.26 &     -2.67 \\ \hline \hline
$N_d$ & $=N_u$  &      3.10 &      0.03 &    121.78 &      0.71 \\ \hline
$a_d$ & $=a_u$  &     -0.29 &     -2.61 &      3.89 &     -0.62 \\ \hline
$b_d$ & $=b_u$  &      6.71 &      0.69 &      9.68 &      3.32 \\ \hline
$c_d$ & $=c_u$  &     -0.07 &     -0.91 &      1.67 &      2.56 \\ \hline
$d_d$ & $=d_u$  &      5.52 &      0.46 &      0.26 &     19.80 \\ \hline \hline
$N_s$ & 152607.12 &      0.82 &   3574.00 &    659.46 &      3.65 \\ \hline
$a_s$ &      7.34 &     -0.04 &     10.57 &      6.31 &      0.00 \\ \hline
$b_s$ &     12.29 &      1.62 &     16.87 &      6.80 &      4.69 \\ \hline
$c_s$ & 0 (fixed) &      1.16 &     39.06 &      1.53 &      0.18 \\ \hline
$d_s$ & 0 (fixed) &      0.06 &     -6.55 &      0.45 &     -3.67 \\ \hline \hline
$N_c$ &      0.33 &     12.06 &     43.30 &      6.82 &      6.68 \\ \hline
$a_c$ &     -2.05 &      0.99 &      2.35 &      2.19 &      0.43 \\ \hline
$b_c$ &      2.61 &      4.77 &      9.36 &      5.87 &      5.29 \\ \hline
$c_c$ &     -0.88 &      5.45 &     15.04 &      0.92 &      0.78 \\ \hline
$d_c$ &      2.13 &      6.52 &     13.74 &     -0.35 &     -0.07 \\ \hline \hline
$N_b$ &      1.25 &     15.72 &      6.81 &     17.23 &     35.20 \\ \hline
$a_b$ &     -0.45 &      0.96 &      0.48 &      1.32 &      0.60 \\ \hline
$b_b$ &      4.37 &      7.94 &     11.89 &     12.17 &     18.91 \\ \hline
$c_b$ &     17.48 &     21.05 &      0.43 &      0.84 &      1.45 \\ \hline
$d_b$ &     10.79 &     11.38 &      0.00 &     -0.02 &     -5.59 \\ \hline

%% file: figs_paramValtable
$N_u$ &    153.26 &      0.73 &  52301.42 \\ \hline
$a_u$ &     13.00 & 0 (fixed) &     11.06 \\ \hline
$b_u$ &      2.51 &      1.45 &      6.74 \\ \hline \hline
$N_d$ & $=N_u$ & 0 (fixed) & $=0.5 N_u$ \\ \hline
$a_d$ & $=a_u$ & 0 (fixed) & $=a_u$ \\ \hline
$b_d$ & $=b_u$ & 0 (fixed) & $=b_u$ \\ \hline

%% file: figs_PionTable
TASSO \cite{Brandelik:1980iy}    & untagged &   12  & 5  & 20  &
 0.56 &   0.26 &     5.3 \\
\hline
TASSO \cite{Althoff:1982dh}      & untagged &   14  & 10 & 8.5 &
 0.91 &  -1.24 &   -10.5 \\
\hline
TASSO \cite{Althoff:1982dh}      & untagged &   22  & 1  & 6.3 &
 0.00 &   0.05 &     0.3 \\
\hline
HRS \cite{Derrick:1985wd}        & untagged &   29  & 6  &    &
 1.37 &  &  \\
\hline
TPC \cite{Aihara:1986mv}         & l tagged &   29  & 9  &    &
 0.30 &  &  \\
\hline
TPC \cite{Aihara:1986mv}         & c tagged &   29  & 9  &    &
 0.68 &  &  \\
\hline
TPC \cite{Aihara:1986mv}         & b tagged &   29  & 9  &    &
 1.23 &  &  \\
\hline
TPC \cite{Aihara:1988fc}         & untagged &   29  & 27 &    &
 1.09 &  &  \\
\hline
TASSO \cite{Brandelik:1980iy}    & untagged &   30  & 4  & 20  &
 0.61 &   0.72 &    14.5 \\
\hline
TASSO \cite{Braunschweig:1988hv} & untagged &   34  & 10 & 6   &
 1.11 &   0.61 &     3.7 \\
\hline
TASSO \cite{Braunschweig:1988hv} & untagged &   44  & 7  & 6   &
 2.01 &   0.67 &     4.0 \\
\hline
TOPAZ \cite{Itoh:1994kb}         & untagged &   58  & 8  &    &
 0.85 &  &  \\
\hline
ALEPH \cite{Buskulic:1994ft}     & untagged & 91.2  & 22 & 3   &
 0.57 &  -0.55 &    -1.6 \\
\hline
DELPHI \cite{Abreu:1998vq}       & l tagged & 91.2  & 17 &    &
 1.76 &  &  \\
\hline
DELPHI \cite{Abreu:1998vq}       & b tagged & 91.2  & 17 &    &
 1.76 &  &  \\
\hline
DELPHI \cite{Abreu:1998vq}       & untagged & 91.2  & 17 &    &
 1.76 &  &  \\
\hline
OPAL \cite{Abbiendi:1999ry}      & u tagged & 91.2  & 5  &    &
 5.78 &  &  \\
\hline
OPAL \cite{Abbiendi:1999ry}      & d tagged & 91.2  & 5  &    &
 5.43 &  &  \\
\hline
OPAL \cite{Abbiendi:1999ry}      & s tagged & 91.2  & 5  &    &
 4.43 &  &  \\
\hline
OPAL \cite{Abbiendi:1999ry}      & c tagged & 91.2  & 5  &    &
 7.25 &  &  \\
\hline
OPAL \cite{Abbiendi:1999ry}      & b tagged & 91.2  & 5  &    &
10.92 &  &  \\
\hline
OPAL \cite{Akers:1994ez}         & untagged & 91.2  & 20 &    &
 1.26 &  &  \\
\hline
SLD \cite{Abe:2003iy}            & l tagged & 91.2  & 28 &    &
 0.79 &  &  \\
\hline
SLD \cite{Abe:2003iy}            & c tagged & 91.2  & 28 &    &
 0.78 &  &  \\
\hline
SLD \cite{Abe:2003iy}            & b tagged & 91.2  & 28 &    &
 0.71 &  &  \\
\hline
SLD \cite{Abe:2003iy}            & untagged & 91.2  & 28 &    &
 0.57 &  &  \\
\hline
DELPHI \cite{Abreu:2000gw}       & untagged & 189   & 3  &    &
 4.67 &  &  \\
\hline
\multirow{2}{*}{\vspace{-0.6cm} BRAHMS \cite{Arsene:2007jd}}  &
\multirow{2}{*}{\vspace{0.3cm} $y\in [2.9,3]$} & \multirow{2}{*}{200}
& 8  & \multirow{2}{*}{\vspace{0.3cm} 11,7,8(13),} &
 0.90 &  -1.67,  -1.06,  -1.20,  -0.30,  -0.14
&   -18.4,    -7.4,    -9.6,    -0.6,    -0.1  \\
\cline{2-2} \cline{4-4} \cline{6-8} & $y\in [3.25,3.35]$ &
& 7  & \multirow{2}{*}{\vspace{0.3cm} 2,1(3)} & 
 2.83 &  -1.91,  -1.21,  -1.81,  -0.35,  -0.34
&   -21.0,    -8.5,   -14.5,    -0.7,    -0.3  \\
\hline
PHENIX \cite{Adler:2003pb} ($\pi^0$) & $|\eta|<0.35$ & 200  & 13  & 9.7 &
 0.48 &  -0.01 &    -0.1 \\
\hline
STAR \cite{Adams:2006uz} ($\pi^0$) & $\eta=3.3$ & 200   & 4  & 16   &
 0.72 &  -0.58 &    -9.4 \\
\hline
STAR \cite{Adams:2006uz} ($\pi^0$) & $\eta=3.8$ & 200   & 2  & 16   &
 0.54 &  -0.27 &    -4.4 \\
\hline
STAR \cite{Adams:2006nd} & $|y|<0.5$ & 200 & 10 & 11.7 &
 0.48 &   0.08 &     0.9 \\
\hline \hline
Total      &      &     & 382 &      &   1.36 &       &  \\

%% file: figs_PionValTable
\multirow{2}{*}{\vspace{-0.6cm} BRAHMS \cite{Arsene:2007jd}}  &
\multirow{2}{*}{\vspace{0.3cm} $y\in [2.9,3]$} & \multirow{2}{*}{200}
& 8  & \multirow{2}{*}{\vspace{0.3cm} 11,7,8(13),} &
 0.66 &  -0.59,  -0.38,  -0.48,  -0.11,  -0.07
&    -6.5,    -2.6,    -3.8,    -0.2,    -0.1  \\
\cline{2-2} \cline{4-4} \cline{6-8} & $y\in [3.25,3.35]$ &
& 7  & \multirow{2}{*}{\vspace{0.3cm} 2,1(3)} & 
 0.47 &  -0.39,  -0.25,  -0.25,  -0.07,  -0.02
&    -4.3,    -1.7,    -2.0,    -0.1,     0.0  \\
\hline
STAR \cite{Adams:2006nd} & $|y|<0.5$ & 200 & 10 & 11.7 &
 0.07 &  -0.05 &    -0.5 \\
\hline \hline
Total      &      &     & 25 &      &   0.37 &       &  \\

%% file: figs_KaonTable
TASSO \cite{Brandelik:1980iy}    & untagged &   12  & 3  & 20  &
 0.92 &  -0.58 &   -11.6 \\
\hline
TASSO \cite{Althoff:1982dh}      & untagged &   14  & 9 & 8.5 &
 2.41 &   0.21 &     1.8 \\
\hline
TASSO \cite{Althoff:1982dh}      & untagged &   22  & 7  & 6.3 &
 1.21 &  -1.55 &    -9.8 \\
\hline
HRS \cite{Derrick:1985wd}        & untagged &   29  & 7  &    &
 1.13 &  &  \\
\hline
MARKII \cite{Schellman:1984yz}         & untagged &   29  & 2  & 12  &
 0.44 &  &  \\
\hline
TPC \cite{Aihara:1988fc}         & untagged &   29  & 26  &    &
 0.00 &  &  \\
\hline
TASSO \cite{Brandelik:1980iy}    & untagged &   30  & 2  & 20  &
 0.48 &  -0.42 &    -8.5 \\
\hline
TASSO \cite{Braunschweig:1988hv} & untagged &   34  & 5 & 6   &
 0.20 &  -0.36 &    -2.2 \\
\hline
TOPAZ \cite{Itoh:1994kb}         & untagged &   58  & 5  &    &
 0.13 &  &  \\
\hline
ALEPH \cite{Buskulic:1994ft,Barate:1996fi}     & untagged & 91.2  & 18 & 3   &
 0.56 &   0.37 &     1.1 \\
\hline
DELPHI \cite{Abreu:1998vq}       & l tagged & 91.2  & 17 &    &
 0.77 &  &  \\
\hline
DELPHI \cite{Abreu:1998vq}       & b tagged & 91.2  & 17 &    &
 0.77 &  &  \\
\hline
DELPHI \cite{Abreu:1998vq}       & untagged & 91.2  & 17 &    &
 0.77 &  &  \\
\hline
OPAL \cite{Abbiendi:1999ry}      & u tagged & 91.2  & 5  &    &
 1.11 &  &  \\
\hline
OPAL \cite{Abbiendi:1999ry}      & d tagged & 91.2  & 5  &    &
 0.78 &  &  \\
\hline
OPAL \cite{Abbiendi:1999ry}      & s tagged & 91.2  & 5  &    &
 1.97 &  &  \\
\hline
OPAL \cite{Abbiendi:1999ry}      & c tagged & 91.2  & 5  &    &
 5.71 &  &  \\
\hline
OPAL \cite{Abbiendi:1999ry}      & b tagged & 91.2  & 5  &    &
 8.97 &  &  \\
\hline
OPAL \cite{Akers:1994ez}         & untagged & 91.2  & 10 &    &
 0.47 &  &  \\
\hline
SLD \cite{Abe:2003iy}            & l tagged & 91.2  & 28 &    &
 1.82 &  &  \\
\hline
SLD \cite{Abe:2003iy}            & c tagged & 91.2  & 28 &    &
 0.96 &  &  \\
\hline
SLD \cite{Abe:2003iy}            & b tagged & 91.2  & 28 &    &
 1.93 &  &  \\
\hline
SLD \cite{Abe:2003iy}            & untagged & 91.2  & 28 &    &
 0.46 &  &  \\
\hline
DELPHI \cite{Abreu:2000gw}       & untagged & 189   & 3  &    &
 1.10 &  &  \\
\hline
\multirow{2}{*}{\vspace{-0.6cm} BRAHMS \cite{Arsene:2007jd}}  &
\multirow{2}{*}{\vspace{0.3cm} $y\in [2.9,3]$} & \multirow{2}{*}{200}
& 8  & \multirow{2}{*}{\vspace{0.3cm} 11,7,8(13),} &
 1.14 &  -1.25,  -0.80,  -0.35,  -0.23,   0.00
&   -13.8,    -5.6,    -2.8,    -0.5,     0.0  \\
\cline{2-2} \cline{4-4} \cline{6-8} & $y\in [3.25,3.35]$ &
& 6  & \multirow{2}{*}{\vspace{0.3cm} 2,1(3)} & 
 2.93 &  -0.78,  -0.49,  -1.45,  -0.14,  -0.25
&    -8.5,    -3.5,   -11.6,    -0.3,    -0.2  \\
\hline
CDF \cite{Acosta:2005pk} ($K_S^0$) & $|\eta|<1$ & 630 & 37 & 10 &
 0.59 &  -2.14 &   -21.4 \\
\hline \hline
STAR \cite{Adams:2006nd} ($K_S^0$) & $|y|<0.5$ & 200 & 9 & 11.7 &
 1.04 &  -1.70 &   -19.8 \\
\hline \hline
Total      &      &     & 346 &      &   1.20 &       &  \\

%% file: figs_KaonValTable
\multirow{2}{*}{\vspace{-0.6cm} BRAHMS \cite{Arsene:2007jd}}  &
\multirow{2}{*}{\vspace{0.3cm} $y\in [2.9,3]$} & \multirow{2}{*}{200}
& 8  & \multirow{2}{*}{\vspace{0.3cm} 11,7,8(13),} &
 1.15 &  -0.59,  -0.38,   0.03,  -0.11,   0.04
&    -6.5,    -2.6,     0.2,    -0.2,     0.0  \\
\cline{2-2} \cline{4-4} \cline{6-8} & $y\in [3.25,3.35]$ &
& 6  & \multirow{2}{*}{\vspace{0.3cm} 2,1(3)} & 
 1.07 &  -1.13,  -0.72,  -1.02,  -0.21,  -0.14
&   -12.4,    -5.0,    -8.1,    -0.4,    -0.1  \\
\hline
Total      &      &     & 14 &      &   1.12 &       &  \\

%% file: figs_ProtonTable
TASSO \cite{Brandelik:1980iy}    & untagged &   12  & 3  & 20  &
 0.49 &  -0.31 &    -6.3 \\
\hline
TASSO \cite{Althoff:1982dh}      & untagged &   14  & 9 & 8.5 &
 2.30 &  -0.61 &    -5.2 \\
\hline
TASSO \cite{Althoff:1982dh}      & untagged &   22  & 9  & 6.3 &
 1.36 &  -0.99 &    -6.2 \\
\hline
HRS \cite{Derrick:1985wd}        & untagged &   29  & 7  &    &
 4.31 &  &  \\
\hline
TPC \cite{Aihara:1988fc}         & untagged &   29  & 20  &    &
 1.08 &  &  \\
\hline
TASSO \cite{Brandelik:1980iy}    & untagged &   30  & 3  & 20  &
 0.52 &  -0.82 &   -16.5 \\
\hline
JADE \cite{Bartel:1981sw}    & untagged &   34  & 2  & 14  &
 6.07 &  -0.82 &   -16.5 \\
\hline
TASSO \cite{Braunschweig:1988hv} & untagged &   34  & 7 & 6   &
 1.12 &  -1.63 &    -9.8 \\
\hline
TOPAZ \cite{Itoh:1994kb} & untagged &   58  & 5 &    &
 0.29 &  &  \\
\hline
ALEPH \cite{Buskulic:1994ft,Barate:1996fi}     & untagged & 91.2  & 18 & 3   &
 0.63 &  -1.78 &    -5.3 \\
\hline
DELPHI \cite{Abreu:1998vq}       & l tagged & 91.2  & 17 &    &
 0.21 &  &  \\
\hline
DELPHI \cite{Abreu:1998vq}       & b tagged & 91.2  & 17 &    &
 1.00 &  &  \\
\hline
DELPHI \cite{Abreu:1998vq}       & untagged & 91.2  & 17 &    &
 0.26 &  &  \\
\hline
OPAL \cite{Abbiendi:1999ry}      & u tagged & 91.2  & 5  &    &
 2.14 &  &  \\
\hline
OPAL \cite{Abbiendi:1999ry}      & d tagged & 91.2  & 5  &    &
 2.48 &  &  \\
\hline
OPAL \cite{Abbiendi:1999ry}      & s tagged & 91.2  & 5  &    &
 3.03 &  &  \\
\hline
OPAL \cite{Abbiendi:1999ry}      & c tagged & 91.2  & 5  &    &
 2.67 &  &  \\
\hline
OPAL \cite{Abbiendi:1999ry}      & b tagged & 91.2  & 5  &    &
 1.51 &  &  \\
\hline
OPAL \cite{Akers:1994ez}         & untagged & 91.2  & 10 &    &
 8.78 &  &  \\
\hline
SLD \cite{Abe:2003iy}            & l tagged & 91.2  & 29 &    &
 1.52 &  &  \\
\hline
SLD \cite{Abe:2003iy}            & c tagged & 91.2  & 29 &    &
 1.51 &  &  \\
\hline
SLD \cite{Abe:2003iy}            & b tagged & 91.2  & 29 &    &
 1.77 &  &  \\
\hline
SLD \cite{Abe:2003iy}            & untagged & 91.2  & 29 &    &
 0.47 &  &  \\
\hline
DELPHI \cite{Abreu:2000gw}       & untagged & 189   & 3  &    &
 2.01 &  &  \\
\hline
\multirow{2}{*}{\vspace{-0.6cm} BRAHMS \cite{Arsene:2007jd}}  &
\multirow{2}{*}{\vspace{0.3cm} $y\in [2.9,3]$} & \multirow{2}{*}{200}
& 7  & \multirow{2}{*}{\vspace{0.3cm} 11,7,8(13),} &
 2.42 &  -2.69,  -1.71,  -1.70,  -0.49,  -0.14
&   -29.6,   -12.0,   -13.6,    -1.0,    -0.1  \\
\cline{2-2} \cline{4-4} \cline{6-8} & $y\in [3.25,3.35]$ &
& 5  & \multirow{2}{*}{\vspace{0.3cm} 2,1(3)} & 
 5.33 &  -3.68,  -2.34,  -2.51,  -0.67,  -0.27
&   -40.4,   -16.4,   -20.0,    -1.3,    -0.3  \\
\hline
STAR \cite{Adams:2006nd} & $|y|<0.5$ & 200 & 8 & 11.7 &
 3.01 &  -1.94 &   -22.6 \\
\hline \hline
Total      &      &     & 309 &      &   1.89 &       &  \\

%% file: figs_ProtonValTable
\multirow{2}{*}{\vspace{-0.6cm} BRAHMS \cite{Arsene:2007jd}}  &
\multirow{2}{*}{\vspace{0.3cm} $y\in [2.9,3]$} & \multirow{2}{*}{200}
& 7  & \multirow{2}{*}{\vspace{0.3cm} 11,7,8(13),} &
 1.83 &  -1.57,  -1.00,  -0.58,  -0.29,   0.08
&   -17.3,    -7.0,    -4.6,    -0.6,     0.1  \\
\cline{2-2} \cline{4-4} \cline{6-8} & $y\in [3.25,3.35]$ &
& 5  & \multirow{2}{*}{\vspace{0.3cm} 2,1(3)} & 
 3.77 &  -3.10,  -1.97,  -2.12,  -0.56,  -0.23
&   -34.1,   -13.8,   -16.9,    -1.1,    -0.2  \\
\hline
STAR \cite{Adams:2006nd} & $|y|<0.5$ & 200 & 8 & 11.7 &
 1.51 &  -0.16 &    -1.8 \\
\hline \hline
Total      &      &     & 20 &      &   2.19 &       &  \\

%% file: figs_K0STable
TASSO \cite{Althoff:1984iz}    & untagged &   14  & 8  & 15  &
 0.53 &  -1.32 &   -19.8 \\
\hline
TASSO \cite{Braunschweig:1989wg}      & untagged &   14.8  & 8 &  &
 0.60 &  &  \\
\hline
TASSO \cite{Braunschweig:1989wg}      & untagged &   21.5  & 5 &  &
 0.44 &  &  \\
\hline
TASSO \cite{Althoff:1984iz}      & untagged &   22  & 5  & 15 &
 0.79 &  -0.08 &    -1.3 \\
\hline
HRS \cite{Derrick:1985wd}        & untagged &   29  & 12  &    &
 2.61 &  &  \\
\hline
MARK II \cite{Schellman:1984yz}      & untagged &   29  & 17  & 12 &
 0.76 &   0.26 &     3.1 \\
\hline
TPC \cite{Aihara:1984mk}         & untagged &   29  & 7  &    &
 0.41 &  &  \\
\hline
TASSO \cite{Brandelik:1981ta} & untagged &   33.3  & 7 & 15   &
 0.55 &   0.26 &     3.9 \\
\hline
TASSO \cite{Althoff:1984iz} & untagged &   34  & 13 & 15   &
 0.93 &  -0.05 &    -0.8 \\
\hline
TASSO \cite{Braunschweig:1989wg} & untagged &   34.5  & 13 &    &
 1.72 &  &  \\
\hline
CELLO \cite{Behrend:1989ae} & untagged &   35  & 9 &    &
 0.46 &  &  \\
\hline
TASSO \cite{Braunschweig:1989wg} & untagged &   35  & 13 &    &
 1.78 &  &  \\
\hline
TASSO \cite{Braunschweig:1989wg} & untagged &   42.6  & 13 &    &
 0.88 &  &  \\
\hline
TOPAZ \cite{Itoh:1994kb}         & untagged &   58  & 4  &    &
 0.05 &  &  \\
\hline
ALEPH \cite{Barate:1996fi}     & untagged & 91.2  & 16 & 2   &
 0.44 &  -1.52 &    -3.0 \\
\hline
DELPHI \cite{Abreu:1994rg}       & untagged & 91.2  & 13 &    &
 0.55 &  &  \\
\hline
OPAL \cite{Abbiendi:1999ry}      & u tagged & 91.2  & 5  &    &
 0.98 &  &  \\
\hline
OPAL \cite{Abbiendi:1999ry}      & d tagged & 91.2  & 5  &    &
 0.63 &  &  \\
\hline
OPAL \cite{Abbiendi:1999ry}      & s tagged & 91.2  & 5  &    &
 0.47 &  &  \\
\hline
OPAL \cite{Abbiendi:1999ry}      & c tagged & 91.2  & 5  &    &
 1.54 &  &  \\
\hline
OPAL \cite{Abbiendi:1999ry}      & b tagged & 91.2  & 5  &    &
 1.98 &  &  \\
\hline
OPAL \cite{Abbiendi:2000cv}         & untagged & 91.2  & 16 &  6  &
 0.46 &  -0.40 &    -2.4 \\
\hline
SLD \cite{Abe:1998zs}            & l tagged & 91.2  & 9 &    &
 0.71 &  &  \\
\hline
SLD \cite{Abe:1998zs}            & c tagged & 91.2  & 9 &    &
 0.91 &  &  \\
\hline
SLD \cite{Abe:1998zs}            & b tagged & 91.2  & 9 &    &
 1.59 &  &  \\
\hline
SLD \cite{Abe:1998zs}            & untagged & 91.2  & 9 &    &
 1.17 &  &  \\
\hline
DELPHI \cite{Abreu:2000gw}       & untagged & 183   & 2  &    &
11.39 &  &  \\
\hline
DELPHI \cite{Abreu:2000gw}       & untagged & 189   & 3  &    &
 3.81 &  &  \\
\hline
\multirow{2}{*}{\vspace{-0.6cm} BRAHMS \cite{Arsene:2007jd}} ($K^\pm$) &
\multirow{2}{*}{\vspace{0.3cm} $y\in [2.9,3]$} & \multirow{2}{*}{200}
& 8  & \multirow{2}{*}{\vspace{0.3cm} 11,7,8(13),} &
 1.37 &  -1.24,  -0.79,   0.00,  -0.23,   0.07
&   -13.6,    -5.5,     0.0,    -0.5,     0.1  \\
\cline{2-2} \cline{4-4} \cline{6-8} & $y\in [3.25,3.35]$ &
& 6  & \multirow{2}{*}{\vspace{0.3cm} 2,1(3)} & 
 1.81 &  -0.55,  -0.35,  -0.91,  -0.10,  -0.15
&    -6.1,    -2.5,    -7.3,    -0.2,    -0.2  \\
\hline
CDF \cite{Acosta:2005pk} & $|\eta|<1$ & 630 & 48 &  &
 0.59 &  -2.19 &   -21.9 \\
\hline \hline
STAR \cite{Adams:2006nd} & $|y|<0.5$ & 200   & 9  & 11.7   &
 1.06 &  -1.58 &   -18.5 \\
\hline
Total      &      &     & 323 &      &   1.03 &       &  \\

%% file: figs_LambdaTable
TASSO \cite{Althoff:1984iz}      & untagged &   14  & 3 & 20 &
 0.17 &   0.05 &     1.1 \\
\hline
TASSO \cite{Althoff:1984iz}      & untagged &   22  & 4  & 20 &
 0.43 &  -0.60 &   -12.0 \\
\hline
HRS \cite{Baringer:1986jd}  & untagged &   29  & 12  &    &
 0.93 &  &  \\
\hline
MARK II \cite{delaVaissiere:1984xg}  & untagged &   29  & 15  &    &
 0.71 &  &  \\
\hline
TASSO \cite{Brandelik:1981ta}    & untagged &   33.3  & 6  & 15  &
 0.73 &  -1.36 &   -20.5 \\
\hline
TASSO \cite{Althoff:1984iz} & untagged &   34  & 6 & 20   &
 0.48 &  -0.96 &   -19.2 \\
\hline
TASSO \cite{Braunschweig:1988wh} & untagged &   34.8  & 9 & 9   &
 1.88 &   0.26 &     2.3 \\
\hline
CELLO \cite{Behrend:1989ae}         & untagged &   35  & 7  &    &
 1.02 &  &  \\
\hline
TASSO \cite{Braunschweig:1988hv} & untagged &   42.1  & 4  & 9   &
 0.22 &  -0.27 &    -2.4 \\
\hline
ALEPH \cite{Barate:1996fi}     & untagged & 91.2  & 16 & 4   &
 0.34 &  -1.35 &    -5.4 \\
\hline
DELPHI \cite{Abreu:1993mm}       & untagged & 91.2  & 7 &    &
 1.33 &  &  \\
\hline
OPAL \cite{Abbiendi:1999ry}      & u tagged & 91.2  & 5  &    &
 0.22 &  &  \\
\hline
OPAL \cite{Abbiendi:1999ry}      & d tagged & 91.2  & 5  &    &
 0.20 &  &  \\
\hline
OPAL \cite{Abbiendi:1999ry}      & s tagged & 91.2  & 5  &    &
 0.43 &  &  \\
\hline
OPAL \cite{Abbiendi:1999ry}      & c tagged & 91.2  & 5  &    &
 1.20 &  &  \\
\hline
OPAL \cite{Abbiendi:1999ry}      & b tagged & 91.2  & 5  &    &
 0.32 &  &  \\
\hline
OPAL \cite{Alexander:1996qj}         & untagged & 91.2  & 12 &    &
 2.61 &  &  \\
\hline
SLD \cite{Abe:1998zs}            & l tagged & 91.2  & 4 &    &
 3.40 &  &  \\
\hline
SLD \cite{Abe:1998zs}            & c tagged & 91.2  & 4 &    &
 1.50 &  &  \\
\hline
SLD \cite{Abe:1998zs}            & b tagged & 91.2  & 4 &    &
 2.70 &  &  \\
\hline
SLD \cite{Abe:1998zs}            & untagged & 91.2  & 9 &    &
 0.67 &  &  \\
\hline
DELPHI \cite{Abreu:2000gw}       & untagged & 183   & 3  &    &
 6.42 &  &  \\
\hline
DELPHI \cite{Abreu:2000gw}       & untagged & 189   & 3  &    &
 6.11 &  &  \\
\hline
CDF \cite{Acosta:2005pk} & $|\eta|<1$ & 630 & 34 & 10 &
 1.27 &  -3.18 &   -31.8 \\
\hline \hline
STAR \cite{Abelev:2006cs} & $|y|<0.5$ & 200 & 9 & 11.7 &
 5.84 &  -6.81 &   -79.5 \\
\hline
Total      &      &     & 188 &      &   1.45 &       &  \\